\documentclass[onecolumn,prd,tightenlines,nofootinbib,amsmath,amssymb,amsfonts,superscriptaddress]{revtex4}
\usepackage[pass,letterpaper]{geometry}

\usepackage{hyperref}
\usepackage{lipsum}
\usepackage{graphicx}
\usepackage{dsfont}
\usepackage{subeqnarray}
\usepackage{yfonts}

\usepackage{slashed}
\usepackage{xcolor}
\usepackage[normalem]{ulem}

\def\signofmetric{0}

\newcommand{\be}{\begin{equation}}
\newcommand{\ee}{\end{equation}}

\newcommand{\bea}{\begin{subeqnarray}}
\newcommand{\eea}{\end{subeqnarray}}
\newcommand{\beanosub}{\begin{eqnarray}}
\newcommand{\eeanosub}{\end{eqnarray}}

\newcommand{\CF}[1]{}

\newcommand{\dd}{{\rm d}}
\usepackage{bm}
\newcommand{\gr}[1]{{\bm #1}}

\newcommand{\com}[1]{}

\newcommand{\hatM}{{\lambda}}

\newcommand{\vol}{{\cal V}}

\newcommand{\ii}{{\rm i}}

\newcommand{\deltarel}{\underline{\delta}}
\newcommand{\antipar}[1]{{\bar{#1}}}

\if0\signofmetric
\def\sgnzz{+}
\def\sgnzi{\pm}
\def\sgnii{-}

\def\sgnslash{}
\def\sgnminusslash{-}
\fi

\if1\signofmetric
\def\sgnzz{-}
\def\sgnzi{\mp}
\def\sgnii{+}

\def\sgnslash{-}
\def\sgnminusslash{}
\fi

\begin{document}

\title{Kinetic theory of fermions in curved spacetime}

\author{Christian Fidler}
\email{christian.fidler@uclouvain.be }
\affiliation{Catholic University of Louvain - Center for Cosmology, Particle Physics and Phenomenology (CP3)
2, Chemin du Cyclotron, B-1348 Louvain-la-Neuve, Belgium}

\author{Cyril Pitrou}
\email{pitrou@iap.fr}
\affiliation{Institut d'Astrophysique de Paris, CNRS-UMR 7095, UPMC - Paris VI, Sorbonne
  Universit\'es, 98 bis Bd Arago, 75014 Paris, France}

\date{\today}


\begin{abstract}
We build a statistical description of fermions, taking into account the spin degree of freedom in addition to the momentum of particles, and we detail its use in the context of the kinetic theory of gases of fermions particles. We show that the one-particle distribution function needed to write a Liouville equation is a spinor valued operator. The degrees of freedom of this function are covariantly described by an intensity function and by a polarisation vector which are parallel transported by free streaming. \com{These properties are essentially analogous to the description of polarised light in geometric optics.} Collisions are described on the microscopic level and lead to a Boltzmann equation for this operator. We apply our formalism to the case of weak interactions, which at low energies can be considered as a contact interaction between fermions, allowing us to discuss the structure of the collision term for a few typical weak-interaction mediated reactions. In particular we find for massive particles that a dipolar distribution of velocities in the interacting species is necessary to generate linear polarisation, as opposed to the case of photons for which linear polarisation is generated from the quadrupolar distribution of velocities.
\end{abstract}

\maketitle



\section*{Introduction}

Out of equilibrium evolution of gases is well described in the context
of kinetic theory. This description is based on a distribution
function $f(t,\gr{x},\gr{p})$, which describes the density of
particles in phase space, that is the probability of finding a particle at the spatial coordinate $\gr{x}$ with the
momentum $\gr{p}$ at a given time $t$. This description can be generalised to
the relativistic context and is used in curved space-time when
including gravitational effects~\cite{Stewart1971,Bernstein1988}.
If collisions can be ignored and gas particles are free streaming, the evolution of the
distribution functions is dictated by a Liouville equation. However,
when interactions, between the same species or among different species, cannot
be ignored, a collision term must be added and this leads to a
Boltzmann equation dictating the evolution of the distribution
function. 

In many circumstances, this description needs to be approximated
to reduce the computational cost. If interactions are very strong, the
gas is driven toward local thermal equilibrium and behaves effectively
as a fluid with pressure. Thus, when decomposing the directional dependence in the momentum, one needs only
the monopole which is related to the energy density of the fluid, and the
dipole which is related to the fluid velocity. If interactions are not
strong enough to support equilibrium, but the mean free path is still much smaller than the
typical scales of variation of the distribution function, we can use
an approximate description, either as a viscous fluid approximation
in the usual context of non-perfect fluids, or using e.g. a tight-coupling
expansion~\cite{Ma:1995ey,Hu:1995en,Doran:2005ep,Pitrou:2010ai,CyrRacine:2010bk,Blas:2011rf}
in the context of cosmic microwave background (CMB) in the early
Universe before the electron decoupling

Among the relativistic contexts in which the kinetic theory cannot be
avoided, we find {\it i) } the big-bang nucleosynthesis
\cite{Dicus1982,Bernstein1988,Bernstein:1988ad,Serpico:2004gx,Iocco2008,Grohs:2015tfy}
during which the neutrinos are highly relativistic
and gradually decouple from the other species; {\it ii)} supernovae explosions to treat the radiative
transfer of neutrinos~\cite{Kotake:2012nd,Peres:2013pua}; {\it iii)}
the CMB in the decoupling phase and in the transparent matter dominated era.

In this last case, it happens that the description is not complete if we ignore the polarisation state for the gas of
photons since a quadrupolar distribution of velocities induces
linear polarisation. Indeed, when the coupling to electrons ends to be efficient around recombination, photons develop a quadrupolar
distribution of velocities and this leads to the generation of linear
polarisation. In the case of photons,
the description of polarization is based on a tensor-valued distribution function $f_{\mu\nu}(t,\gr{x},\gr{p})$,
which is doubly projected orthogonally to the observer and to the momentum~\cite{Tsagas:2007yx,Pitrou2008}. While the description of polarisation in the kinetic theory is well established for photons, this is not the case for fermions and there are several open questions. First with which
type of distribution function can we describe the various polarisation
states for a gas of fermions? What are the conditions required to
generate or erase polarisation in a gas of fermions? Under which circumstances polarisation can be ignored and a description based on a simple distribution function is sufficient?

\subsection*{Main features of the formalism}

In this article we aim at answering these questions by building the
kinetic theory for fermions, allowing for the description of
polarisation.  We build a meaningful procedure to obtain a classical description out of a quantum system. This is even more crucial for a gas of fermions, which are a set of excitations of an underlying quantum field, since there is no classical version of the field as in the case of bosons, e.g. in the case of photons where the classical field is described by classical electrodynamics. The fundamental step is to consider the number operator $N_{rs}(\gr{p},\gr{p}')
\equiv 
a_r^\dagger(\gr{p}) a_s(\gr{p}') $ in the quantum
system, where the indices of the creation and annihilation operators $a^\dagger_r$ and $a_s$ refer
to helicity. Its expectation value takes a particularly simple form in the case of a homogeneous system since this imposes
\be\label{AverageNrs}
\langle N_{rs}(\gr{p},\gr{p}')\rangle =   (2\pi)^3 2 p^0 \delta^3(\gr{p}-\gr{p}')  f_{rs} (\gr{p})\,.
\ee
We can show that the function $f_{rs} (\gr{p})$ describes correctly
the statistics of momenta and in order to obtain a classical limit we only need
to consider that the results obtained in the homogeneous case are
valid locally, that is we consider a function $f_{rs}
(\gr{x},t,\gr{p})$. 

An operator in spinor space for fermion particles can then be formed by
contraction with the plane-wave solutions of the Dirac equation $u_s(\gr{p})$
\be\label{IntroOperator}
\gr{F}(\gr{x},t,\gr{p}) \equiv  \sum_{rs} u_s(\gr{p}) f_{rs} (\gr{x},t,\gr{p})
\bar u_r(\gr{p})\,,
\ee
and we show that it is a central object in the computation of the
collision term of the Boltzmann equation. The method is slightly
different but essentially similar for fermion antiparticles. In full generality an operator in spinor space has $16$ degrees of
freedom, but given its construction based on the number operator, we have only four independent degrees of freedom in the operator $\gr{F}(\gr{x},t,\gr{p})$ (corresponding to the two helicity states and their correlation). We show hereafter that these degrees
of freedom are recovered in a covariant and observer independent
manner when decomposing the operator (\ref{IntroOperator}) as
\be
\gr{F} = \frac{1}{2}\left(I \sgnzz \gamma^5 \gamma^\mu {\cal Q}_\mu \right)\left(m\sgnzz \slashed{p}\right)\,.
\ee
One degree of freedom corresponds to the total intensity $I(\gr{x},t,\gr{p})$ while the three remaining
degrees correspond to the state of polarisation and are
covariantly contained in a vector ${\cal Q}^\mu(\gr{x},t,\gr{p})$ with
the transverse property $p_\mu {\cal Q}^\mu=0$. 

These degrees of freedom are parallel transported by free streaming and 
thus allow for a description by a simple Liouville equation. More
importantly, they are independent from the observer even though a
choice of observer is needed for the construction of helicity states; they are true functions of the phase space. We further show that the collision term for fermion interactions, such as weak interactions, is naturally expressed in these covariant quantities.

\subsection*{Outline}

We first summarize the description of fermion quantum fields in
\S~\ref{SecQFT}. We then build in \S~\ref{SecQDF} the distribution
function $\gr{F}(\gr{x},t,\gr{p})$ which is an operator in spinor
space and we show how its covariant components $I(\gr{x},t,\gr{p})$
and ${\cal Q}^\mu(\gr{x},t,\gr{p})$ are extracted. The transformation
properties of these quantities are investigated in \S
\ref{SecTProperties} and we show that they are observer
independent. Hence they are well suited to describe the free streaming
of the distribution function and we derive the Liouville equation that
they satisfy in \S~\ref{SecLiouville}. We then turn to the effect of
collisions in \S~\ref{SecBoltzmann} where we derive the general
Boltzmann equation for fermions. This is applied to the specific case
of low energy weak interactions in \S~\ref{SecWeak}. We detail the
form of the collision terms obtained for standard reactions in
\S~\ref{SecStandard} so as to recover standard results when
polarisation can be ignored. The case of vector bosons, including photons, is summarized in appendix
\ref{AppBosons}. Finally, technical details are gathered in appendix
\ref{AppCommute} for commutation rules of quantum creation and
annihilation operators, appendix \ref{TraceTechnologie} for traces of
Dirac gamma matrices, and appendix \ref{AppTwoComponents} for
two-component spinors and their relation to Dirac spinors.

\subsection*{Notation}

An inertial frame is defined
by a tetrad field, that is by a timelike vector field $e_0$ and three
spacelike vector fields $e_i$, together with the associated co-tetrad
$e^0,e^i$. Latin indices such as $i,j,\dots$ indicate
spatial components in the tetrad basis.  A four-vector is written as $V^\mu$ where Greek indices $\mu,\nu,\dots$ denote components in the tetrad basis. \com{In particular, the components of the tetrad vectors and co-vectors in the tetrad basis are by definition $[e_\mu]^\nu = \delta_\mu^\nu$ and $[e^\mu]_\nu = \delta^\mu_\nu$. \CF{I find the previous confusing.}} If gravity can be ignored, that is in the context of special relativity, the inertial frame is global. However, when including the effect of gravity in the context of general relativity, the inertial frame is local and one must employ general coordinates whose indices are labelled by $\alpha,\beta,\dots$. For a given vector this implies $V^\alpha = V^\mu [e_\mu]^\alpha$. 

The momentum vector $p^\mu$ will often simply be denoted as $p$ and its spatial components $p^i$ allow to
build the spatial momentum $\gr{p} = p^i e_i$. More generally, we reserve boldface notation to spatial vectors. The energy associated with the momentum is given by the time component
$E = p^0$. When a quantity depends on the spatial momentum, we use indifferently $\gr{p}$ or $p$ when no ambiguity can arise. The (special) relativistic (and Lorentz covariant) integration measure is defined as
\be
[\dd p] \equiv \frac{\dd^3 \gr{p}}{(2\pi)^3 2 p^0}
\ee
and its associated (special) relativistic Dirac function is defined
accordingly as
\be
\deltarel(p-p') = (2\pi)^3 2 p^0 \delta^3(\gr{p}-\gr{p}') \,,
\ee
such that $\int [\dd p] \deltarel(p-p') =1$.
\if0\signofmetric
Our metric convention follows particle physics standards\footnote{A compiled version with the opposite convention, which is often employed in the cosmological context, can be downloaded at \href{http://www2.iap.fr/users/pitrou/publi/CFCPCosmoMetric.pdf}{{\tt http://www2.iap.fr/users/pitrou/publi/CFCPCosmoMetric.pdf}} or recompiled from the source file by changing a boolean value in the preambule.}. 
\else
Our metric convention follows the standard notation employed in cosmology.
\fi
In the tetrad basis, the metric $g_{\mu\nu}$ reduces to the Minkowski metric
\be
\eta = {\rm diag}(\sgnzz,\sgnii,\sgnii,\sgnii)\,.
\ee

The Levi-Civita tensor is fully antisymmetric and in the tetrad basis
all its components are deduced from the choice
\be
\epsilon_{0123} = -\epsilon^{0123} = 1\,.
\ee
We identify the time-like vector of a tetrad $e_0$ with the velocity $u$ of an observer and its spatial Levi-Civita is obtained from $\epsilon_{ijk}\equiv u^\mu\epsilon_{\mu ijk} $, and it is such that $\epsilon_{123}={\epsilon^{12}}_3 = 1$.

\section{Description of fermions in quantum field theory}\label{SecQFT}

\subsection{Quantum fields}

Let us consider a species of fermions $a$ (e.g. electrons or neutrinos) and the
corresponding antiparticle species $\antipar{a}$. The associated
quantum field is decomposed as
\be\label{DefQuantumField}
\psi = \sum_{s=\pm\tfrac{1}{2}} \int [\dd p]\left[ {\rm e}^{\sgnslash \ii p \cdot x }
\antipar{a}^\dagger_s(p) v_s(p)+ {\rm e}^{\sgnminusslash \ii p \cdot x }
a_s(p) u_s(p)\right]\,.
\ee
In this expression, the creation and annihilation operators of the particles and antiparticles satisfy the anti-commutation rules 
\be\label{AntiCommuteRule}
\{a_r(p), a^\dagger_s(p')\} = \delta^{\rm K}_{rs}
\deltarel(p-p')\,,\qquad \{\antipar{a}_r(p), \antipar{a}^\dagger_s(p')\}
= \delta^{\rm K}_{rs} \deltarel(p-p')
\ee
with all other anti-commutators vanishing and where $\delta^{\rm K}$
is the Kronecker function.  Furthermore, the $u_s$ and $v_s$ are plane-wave solutions of the Dirac equation
\be\label{EqDirac}
(\ii \gamma^\mu \partial_\mu \sgnii m) \psi = 0\,,
\ee
where the Dirac matrices $\gamma^\mu$ satisfy the Clifford algebra
(for more details, see Appendix \ref{AppClifford})
\be\label{DefClifford}
\{\gamma^\mu,\gamma^\nu\} =  \sgnslash 2 g^{\mu\nu} \mathds{1}\,. 
\ee
More precisely, the $u_s$ ($v_s$) are the positive (negative)
frequency solutions and satisfy
\be\label{eq:particledef}
(\slashed{p} \sgnii m) u_s(p) = 0\,,\qquad (\slashed{p}  \sgnzz  m) v_s(p) = 0\,,
\ee
with the standard Dirac slashed notation $\slashed{p} \equiv p_\mu \gamma^\mu$.

\subsection{Helicity basis}

The creation and annihilation operators employed to build the field $\psi$ in Eq.~(\ref{DefQuantumField}) refer to a complete set of solutions to the Dirac equation and we are free to choose the basis characterising this Hilbert space. In the following we will employ helicity states $s = \pm\tfrac{1}{2}$, with left-helical states labelled as $-1/2$ and right-helical ones as $+1/2$. We first introduce the matrix 
\be
\gamma^5 \equiv\frac{\ii}{4 !} \epsilon_{\mu\nu\rho\lambda}\gamma^\mu
\gamma^\nu \gamma^\rho \gamma^\lambda = \ii \gamma^0 \gamma^1 \gamma^2
\gamma^3 \,,
\ee
which satisfies $\{\gamma^5,\gamma^\mu\}=0$ and $(\gamma^5)^2=\mathds{1}$, from which the projectors on the left and right chiral parts are built
\be
P_- \equiv \frac{1}{2}(\mathds{1}-\gamma^5) \,,\qquad P_+ \equiv \frac{1}{2}(\mathds{1}+ \gamma^5) \,.
\ee
Let us consider a momentum $p$ with energy $E=p^0$, and whose spatial direction unit vector $\hat{\gr{p}}\equiv \gr{p}/|\gr{p}|$ is decomposed into spherical coordinates as $\hat{p}^i = \left({\rm sin}(\theta){\rm sin}(\phi),{\rm sin}(\theta){\rm cos}(\phi),{\rm cos}(\theta)\right)$. We define left- and right-chiral states $u_{\rm L} = P_- u$ and $u_{\rm R} = P_+ u $. From Eq.~(\ref{eq:particledef}), specifying the particle and antiparticle solutions of the Dirac equation, we find
\bea\label{eq:particledef2}
&\left(E\gamma^0-\gr{p} \cdot {\bm \gamma}\right) u_{s,{\rm R}} = m u_{s,{\rm L}} \qquad & 
\left(E\gamma^0-\gr{p}\cdot {\bm \gamma} \right)u_{s,{\rm L}} = m u_{s,{\rm R}} \\ 
&\left(E\gamma^0-\gr{p}\cdot {\bm \gamma} \right)v_{s,{\rm R}} = -m v_{s,{\rm L}}  \qquad &
 \left(E\gamma^0-\gr{p}\cdot {\bm \gamma} \right)v_{s,{\rm L}} = -m v_{s,{\rm R}}\,,
\eea
enforcing a connection between the left- and right-chiral components of particles and antiparticles for massive fermions. Helicity states, decribing the spin related to the spatial momentum of the particle, are defined by
\be\label{eq:helicitydef}
-\hat{\gr{p}}\cdot {\bm \gamma}\gamma^5 u_s = 2 s \gamma^0 u_s \qquad -\hat{\gr{p}}\cdot {\bm \gamma}\gamma^5 v_s = - 2s \gamma^0 v_s\,.
\ee
For helicity states Eqs.~(\ref{eq:particledef2}) thus simplify to
\bea\label{eq:particledefhelical}
&\left(E+ 2s |\gr{p}|\right)  u_{s,{\rm R}} = m \gamma^0 u_{s,{\rm L}} \qquad & 
\left(E- 2s |\gr{p}|\right)  u_{s,{\rm L}} = m \gamma^0 u_{s,{\rm R}} \\ 
&\left(E- 2s |\gr{p}|\right) v_{s,{\rm R}} = -m \gamma^0 v_{s,{\rm L}}  \qquad &
 \left(E+ 2s |\gr{p}|\right)v_{s,{\rm L}} = -m  \gamma^0 v_{s,{\rm R}}\,.
\eea
In the helicity basis for the Dirac matrices Eq.~(\ref{eq:helicitydef}) reads
\be
\com{\left(\begin{array}{cc}\hat{\gr{p}}\cdot {\bm \sigma}&0\\
0&\hat{\gr{p}} \cdot {\bm \sigma}\end{array}\right)}\hat{\gr{p}} \cdot{\bm \sigma}  u_{s,\rm{L/R}} = 2s u_{s,{\rm L/R}} \qquad 
\hat{\gr{p}}\cdot {\bm \sigma} v_{s,\rm{L/R}} = - 2s  v_{s,{\rm L/R}}\,,
\ee
and the left- and right-chiral parts are individually solved by the well known chiral 2-spinors of quantum-mechanics
\be
\chi_{\tfrac{1}{2}}= \left(\begin{array}{c}  {\rm e}^{-\ii \phi/2} {\rm
                            cos}(\theta/2) \\ {\rm e}^{\ii\phi/2}{\rm
                            sin}(\theta/2)\end{array}\right) \qquad \chi_{-\tfrac{1}{2}}= \left(\begin{array}{c} - {\rm e}^{-\ii\phi/2}{\rm
                            sin}(\theta/2) \\  {\rm e}^{\ii\phi/2}{\rm cos}(\theta/2) \end{array}\right) \,.
\ee
The full Dirac spinors are then obtained by combining the left- and right-chiral parts under consideration of Eq.~(\ref{eq:particledefhelical}) and their normalisation
\begin{eqnarray}\label{usvseasy}
u_{s} = \left(\begin{array}{c} \sqrt{E+2s|\vec{p}|}\chi_s \\ \sqrt{E-2s|\vec{p}|}\chi_s \end{array}\right) \qquad 
v_{s} = \left(\begin{array}{c} 2s\sqrt{E-2s|\vec{p}|}\chi_{-s} \\ -2s\sqrt{E+2s|\vec{p}|}\chi_{-s} \end{array}\right)\,.
\end{eqnarray}
In Appendix~\ref{AppTwoComponents} we detail how these solutions are also expressed with the two-components spinor formalism.

\section{Quantum distribution function}\label{SecQDF}

\subsection{Number operator and distribution function}\label{SecNrs}

Let us define the number operator 
\be\label{defNrs}
N_{rs}(p,p') \equiv a^\dagger_r(p) a_s(p')\,.
\ee
If we consider a homogenous system described by the state $|\Psi\rangle$, then its average in this quantum state must be
diagonal and of the form
\be\label{deffab}
\langle \Psi | N_{rs}(p,p')| \Psi \rangle \equiv  
\deltarel(p-p') f_{rs} (p)\,.
\ee
This can be understood when considering the Wigner
functional~\cite{Nagirner:2001xd}. To that purpose, let us
introduce the average and difference of momentum
\be
\gr{K} \equiv \gr{p}'-\gr{p}\qquad \overline{\gr{p}} \equiv \frac{1}{2}(\gr{p}'+\gr{p})\,.
\ee
Then the Wigner transform is defined as
\be
\langle N_{rs}(\bar p,\gr{x})\rangle \equiv \int \frac{\dd^3
  \gr{K}}{(2\pi)^3 2 K^0}\langle \Psi | N_{rs}(p,p')| \Psi \rangle  {\rm
  e}^{\ii \gr{K} \cdot \gr{x}}\,.
\ee
Considering a homogeneous system in the sense of the Wigner functional means that $\langle
N_{rs}(\bar p,\gr{x})\rangle$ does not depend on $\gr{x}$ and in that
case it is equal to the function  $ f_{rs} (\bar p)$ defined in Eq. (\ref{deffab}). 

It is then straightforward to show that this
should be identified with the one-particle distribution function of a
homogeneous system. Indeed the total occupation operator is obtained from a sum over all possible
momenta of the diagonal part as
\be
N_{rs} \equiv \int [\dd p] N_{rs}(p,p)\,.
\ee
Its average in a given quantum state is 
\be
\langle \Psi | N_{rs}| \Psi \rangle  = \vol \int \frac{\dd^3 \gr{p}}{(2\pi)^3} f_{rs}(p)\,,
\ee
where we introduced the total volume $ (2\pi)^3 \delta^3(0)= \vol$. This corresponds exactly to the definition of a classical one-particle distribution function.
By construction $N_{rs}$ and $f_{rs}$ are Hermitian, that is
\be\label{HermitianProperty}
N^\star_{rs}(p,p') =N_{sr}(p',p) \qquad \Rightarrow \qquad f^\star_{rs}(p) =f_{sr}(p)\,.
\ee
For antiparticles, the construction is equivalent but with the
operators of antiparticles instead.

\subsection{Molecular chaos and the Boltzmann approximation}

In principle, when considering an interacting system, the one-particle
distribution function is not enough to describe it statistically,
because $n$-particle correlation functions are generated by
collisions. In order to obtain a description only in terms of a one-particle
distribution function, we must assume that the connected part of the
$n$-particle functions vanishes and thus that $n$-particle functions
are expressed only in terms of one-particle functions, corresponding to the assumption of molecular chaos. The $n$-particle number operators for species $a$ are defined as
\be
N^{(n)}_{r_1 \dots r_n s_1 \dots s_n} \equiv a^\dagger_{r_1}\dots a^\dagger_{r_n}a_{s_1} \dots a_{s_n}\,,\qquad\Rightarrow \qquad N^{(2)}_{r_1 r_2 s_1 s_2} \equiv a^\dagger_{r_1}a^\dagger_{r_2}a_{s_1}a_{s_2}\,,
\ee
where for simplicity we omit to write explicitly the momentum dependence of the operators in this section. Their expectation value in a quantum state are then expressed as\footnote{For bosons we remove all minus signs, that is the factor $(-1)^{n+1}\epsilon(\sigma)$  so we would for instance get
$
\langle N^{(2)}_{r_1 r_2 s_1 s_2} \rangle = \langle N^{(1)}_{r_1 s_1}
\rangle \langle N^{(1)}_{r_2 s_2} \rangle + \langle N^{(1)}_{r_1 s_2}
\rangle \langle N^{(1)}_{r_2 s_1} \rangle
$.
}
\be\label{Chaos}
\langle N^{(n)}_{r_1 \dots r_n s_1 \dots s_n} \rangle = \sum_{\sigma \in S_n}(-1)^{n+1} \epsilon(\sigma) \langle N^{(1)}_{r_1 s_{\sigma(1)}} \rangle \dots \langle N^{(1)}_{r_n s_{\sigma(n)}}\rangle\quad \Rightarrow \quad
\langle N^{(2)}_{r_1 r_2 s_1 s_2} \rangle =   \langle N^{(1)}_{r_1 s_2}
\rangle \langle N^{(1)}_{r_2 s_1} \rangle -\langle N^{(1)}_{r_1 s_1}
\rangle \langle N^{(1)}_{r_2 s_2} \rangle\,,
\ee
where the sum is on the group of permutation $S_n$ and $\epsilon(\sigma)$ is the signature of the permutation.
\com{\langle N^{(3)}_{r_1 r_2 r_3 s_1 s_2 s_3} \rangle &=& -\langle N^{(1)}_{r_1 s_1} \rangle (\langle N^{(1)}_{r_2 s_2} \rangle\langle N^{(1)}_{r_3 s_3}\rangle-\langle N^{(1)}_{r_2 s_3} \rangle\langle N^{(1)}_{r_3 s_2}\rangle) + \langle N^{(1)}_{r_1 s_2} \rangle (\langle N^{(1)}_{r_2 s_1} \rangle\langle N^{(1)}_{r_3 s_3}\rangle-\langle N^{(1)}_{r_2 s_3} \rangle\langle N^{(1)}_{r_3 s_1}\rangle) \nonumber\\
&&+\langle N^{(1)}_{r_1 s_3} \rangle (\langle N^{(1)}_{r_3 s_1} \rangle\langle N^{(1)}_{r_2 s_2}\rangle-\langle N^{(1)}_{r_3 s_2} \rangle\langle N^{(1)}_{r_2 s_1}\rangle) \,.
\eea}
This approximation is exactly similar to the Boltzmann approximation of the BBGKY hierarchy. In practice this assumption of molecular chaos is used to obtain the following property for the expectation in a quantum state of a product of one-particle number operators
\be\label{MagicChaos}
\langle N^{(1)}_{r_1 s_1}\dots N^{(1)}_{r_n s_n} \rangle = \sum_{\sigma \in S_n}\langle \{N^{(1)}_{r_1 s_{\sigma(1)}}\}\rangle \dots \langle \{N^{(1)}_{r_n s_{\sigma(n)}}\}\rangle \,\qquad \{N^{(1)}_{r_a s_b}\}= \begin{cases}N^{(1)}_{r_a s_b}\quad a\leq b\\\widehat{N}^{(1)}_{r_a s_b}\quad a > b\end{cases}\,,
\ee
where $\widehat{N}_{rs}$ is the Pauli blocking operator defined in Eq.~(\ref{Magic4}). The expectation value of a product of one-particle number operators is the sum of products of expectation values of all possible pairings between creation and annihilation operators. For each pair, if the indices $(r_a,s_b)$ correspond to operators which were initially in the creation-annihilation order, that is with $a\leq b$ (resp. annihilation-creation order, that is $a>b$) we use $N_{r_a s_b}$ (resp. $\widehat{N}_{r_a s_b}$). For instance the expectation value for a product of two one-particle number operators is simply
\be
\langle N^{(1)}_{r_1 s_1}N^{(1)}_{r_2 s_2} \rangle  = \langle N^{(1)}_{r_1 s_1}\rangle \langle N^{(1)}_{r_2 s_2} \rangle + \langle N^{(1)}_{r_1 s_2}\rangle \langle \widehat{N}^{(1)}_{r_2 s_1} \rangle\,.
\ee
\com{\bea
\langle N^{(1)}_{r_1 s_1}N^{(1)}_{r_2 s_2}N^{(1)}_{r_3 s_3} \rangle  &=& \langle N^{(1)}_{r_1 s_1}\rangle (\langle N^{(1)}_{r_2 s_2} \rangle \langle N^{(1)}_{r_3 s_3} \rangle + \langle N^{(1)}_{r_2 s_3}\rangle \langle \widehat{N}^{(1)}_{r_3 s_2} \rangle) + \langle N^{(1)}_{r_1 s_2}\rangle (\langle N^{(1)}_{r_2 s_2} \rangle \langle N^{(1)}_{r_3 s_3} \rangle + \langle N^{(1)}_{r_2 s_3}\rangle \langle \widehat{N}^{(1)}_{r_3 s_1} \rangle) \nonumber\\
&&+\langle N^{(1)}_{r_1 s_3}\rangle (\langle \widehat{N}^{(1)}_{r_3 s_2} \rangle \langle \widehat{N}^{(1)}_{r_2 s_1} \rangle + \langle N^{(1)}_{r_2 s_2}\rangle \langle \widehat{N}^{(1)}_{r_3 s_1} \rangle) \,,
\eea}
Finally, we also assume that species are uncorrelated such that the expectation value for operators of various species is the product of expectation values of the operators of each species. For instance for two species $a$ and $b$ we assume $\langle a_r^\dagger a_s b_p^\dagger b_q \rangle=\langle a_r^\dagger a_s \rangle \langle b_p^\dagger b_q \rangle$.

\subsection{Spinor valued distribution function}\label{SecSpinorValued}

In order to extract the information contained in $f_{rs}(p)$ in a
covariant form, we build operators in spinor space following
\be\label{DefFparticles}
F_{\mathfrak{a}}^{\,\,\,\mathfrak{b}}(p) \equiv \begin{cases} \sum \limits_{rs} 
f_{rs}(p) u_{s,\mathfrak{a}}(p)  \bar u^{\mathfrak{b}}_r(p)\qquad {\rm
  particles},\\
\sum \limits_{rs} 
f_{rs}(p) v_{r,\mathfrak{a}}(p) \bar v^{\mathfrak{b}}_s(p) \qquad {\rm
antiparticles}\,.
\end{cases}
\ee
For the sake of clarity we use a notation where components of operators in spinor space and Dirac spinors are explicit and are denoted by indices of the type $\mathfrak{a},\mathfrak{b},\dots$. For instance
the identity operator is ${\delta_{\mathfrak{a}}}^{\mathfrak{b}}$ and the plane waves solutions are written $u_{s,\mathfrak{a}}$ and
$v_{s,\mathfrak{a}}$. 

A spinor valued operator has $16$ degrees of freedom and we thus need a $16$-dimensional basis to decompose operators in spinor space. 
The set 
\be
{\cal O} \equiv\{\mathds{1},\gamma^\mu,\Sigma^{\mu\nu},\gamma^\mu
\gamma^5,\gamma^5\}
\ee 
is a complete basis for the space of operators in spinor space, where
we define the matrices 
\be\label{DefSigma}
\Sigma^{\mu\nu} \equiv \frac{\ii}{4}[\gamma^\mu,\gamma^\nu]\,,\qquad \widetilde{\Sigma}^{\mu\nu} \equiv
\frac{1}{2}{\epsilon^{\mu\nu}}_{\alpha\beta}\Sigma^{\alpha\beta} =\ii \gamma^5 \Sigma^{\mu\nu} \,.
\ee

The operators are orthogonal and we find
for any two different operators $X$ and $Y$(${X_\mathfrak{a}}^\mathfrak{b}$ and
${Y_\mathfrak{a}}^\mathfrak{b}$ in spinor components) in the set ${\cal
  O}$ that ${\rm  Tr}[X \cdot Y]={X_\mathfrak{a}}^\mathfrak{b} {Y_\mathfrak{b}}^\mathfrak{a}
=0$.
Using this property, any operator can be decomposed onto
this basis with the help of the Fierz identity~\cite{Nishi:2004st} (see also Eq. G.1.99
of \cite{BibleSpinors} taking into account a factor $2$ difference in the definition of $\Sigma^{\mu\nu}$)
\be\label{MyFierz}
\delta_{\mathfrak{a}}^\mathfrak{b} \delta_{\mathfrak{c}}^{\mathfrak{d}} = \frac 1 4 \left[
  \delta_{\mathfrak{a}}^\mathfrak{d} \delta_{\mathfrak{c}}^{\mathfrak{b}}
  +(\gamma^{5})_{\mathfrak{a}}^{\,\,\mathfrak{d}}
  (\gamma^{5})_{\mathfrak{c}}^{\,\,\mathfrak{b}} \sgnzz
  (\gamma^{\mu})_{\mathfrak{a}}^{\,\,\mathfrak{d}}
  (\gamma_{\mu})_{\mathfrak{c}}^{\,\,\mathfrak{b}} \sgnii (\gamma^\mu
  \gamma^5)_{\mathfrak{a}}^{\,\,\mathfrak{d}} (\gamma_\mu
  \gamma^5)_{\mathfrak{c}}^{\,\,\mathfrak{b}} + 2 (\Sigma^{\mu
    \nu})_{\mathfrak{a}}^{\,\,\mathfrak{d}} (\Sigma_{\mu\nu})_{\mathfrak{c}}^{\,\,\mathfrak{b}}
\right] = \sum \limits_{X \in {\cal O}} c_X X_{\mathfrak{a}}^{\,\,\mathfrak{d}} X_{\mathfrak{c}}^{\,\,\mathfrak{b}}\,.
\ee
The last equality defines the coefficients $c_X$ of the expansion which, by
construction, satisfy $c_X  = 1/{\rm Tr}[X\cdot X]$.
Note also that this identity can be used with the matrices $\widetilde{\Sigma}^{\mu\nu}$
instead of $\Sigma^{\mu\nu}$ by employing $ (\Sigma^{\mu
    \nu})_{\mathfrak{a}}^{\,\,\mathfrak{d}}
  (\Sigma_{\mu\nu})_{\mathfrak{c}}^{\,\,\mathfrak{b}} =-  (\widetilde\Sigma^{\mu
    \nu})_{\mathfrak{a}}^{\,\,\mathfrak{d}} (\widetilde
  \Sigma_{\mu\nu})_{\mathfrak{c}}^{\,\,\mathfrak{b}}$, and in the
  following this proves to be more convenient.

Any operator $O$ in spinor space is decomposed on the basis ${\cal O}$ thanks to Eq. (\ref{MyFierz}) as
\be\label{DecompositionFormula}
{O_\mathfrak{a}}^\mathfrak{b} = \sum_{c_X \in {\cal O}} c_X {\rm Tr}[O\cdot X] {X_\mathfrak{a}}^\mathfrak{b}\,.
\ee
In particular, any bilinear tensor product of the form $u_r(p)
\bar{u}_s(p)$ or $v_r(p) \bar{v}_s(p)$ [with the standard notation
$\bar u_s = u_s^\dagger \gamma^0$, see Eq. (\ref{RigorousDiracConjugate})], is a spinor-space operator and can be decomposed as 
\be\label{urus}
u_{r,\mathfrak{a}}(p) \bar{u}_{s}^\mathfrak{b}(p) = \sum \limits_{X
  \in {\cal O}} c_X
[\bar{u}_{s}(p) X u_r(p) ] {X_\mathfrak{a}}^\mathfrak{b}
= \sum \limits_{X  \in {\cal O}} c_X [\bar{u}_{s}^\mathfrak{c}(p) {X_\mathfrak{c}}^\mathfrak{d} u_{r,\mathfrak{d}}(p) ] {X_\mathfrak{a}}^\mathfrak{b}  \,.
\ee
Hence, we only need to compute the $\bar{u}_{s}(p) X u_r(p) $ and
$\bar{v}_{s}(p) X v_r(p) $ (these expressions are computed in App. \ref{SecuXu}) for all
$X \in {\cal O}$ to decompose the operator
${F_\mathfrak{a}}^\mathfrak{b}$, defined in Eq. (\ref{DefFparticles}), into
\be\label{Fspinor}
 F_{\mathfrak{a}}^{\,\,\,\mathfrak{b}} = \begin{cases}\sum \limits_{X
     \in {\cal O}} c_X \sum \limits_{rs} f_{rs} (\bar u_r X u_s)X_{\mathfrak{a}}^{\,\,\mathfrak{b}} \com{= \sum
\limits_{X \in {\cal O}} c_X \sum_{ab }\left({}^ X{U_{ab}}  f_{ab} \right)
X_{\mathfrak{a}}^{\,\,\mathfrak{b}}}\qquad {\rm particles},\\
\sum \limits_{X \in {\cal O}} c_X \sum \limits_{rs} f_{sr} \,(\bar v_r X v_s) X_{\mathfrak{a}}^{\,\,\mathfrak{b}} \com{= \sum
\limits_{X \in {\cal O}} c_X \sum_{ab }\left( {}^ XV_{ab} f_{ba} \right)X_{\mathfrak{a}}^{\,\,\mathfrak{b}}}\qquad {\rm antiparticles}\,.
\end{cases}
\ee
Note how the indices are in reverse order for antiparticles ($f_{sr}$ instead of $f_{rs}$).

\subsubsection{Covariant components of the spinor valued distribution function}\label{SecCovariantF}

\begin{figure}[htb!]
    \includegraphics[scale=0.395,angle=0]{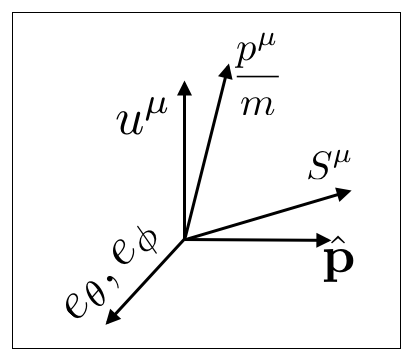}
\includegraphics[scale=0.4,angle=0]{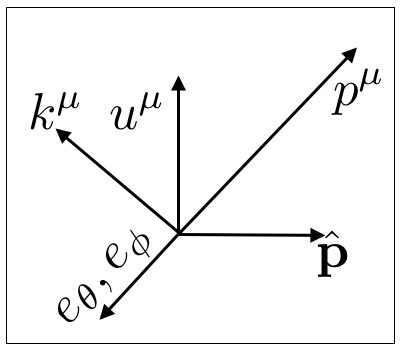}
     \caption{Left: In the massive case, $p^\mu/m$ and $e_\theta,e_\phi,S^\mu$ form an
       orthonormal basis. Right: In the massless case, $k^\mu,p^\mu$
       are null vectors orthogonal to $e_\theta,e_\phi$ such that
       $k^\mu p_\mu = \sgnslash 1$. In both cases, the polarisation
       basis is formed by ${\epsilon}^\mu_{\pm} \equiv (e^\mu_\theta \mp \ii
e^\mu_\phi)/\sqrt{2}$}
\label{fig1} 
  \end{figure}

While $f_{rs}(p)$ has only four degrees of freedom,
${F_\mathfrak{a}}^\mathfrak{b}$ has 16 components, and in the
decomposition this must be made manifest. To that purpose let us define a
set of three unit spatial vectors. 
For a given observer with four-velocity $u^\alpha$ which is chosen to be aligned with the time-like tetrad vector $e_0$, we can define a spatial momentum
$\gr{p}$ and its spatial direction $\hat{\gr{p}}\equiv \gr{p}/|\gr{p}|$. In spherical coordinates the momentum direction is given by $\theta,\phi$ and defines a radial unit vector. We then also consider
the usual basis in spherical coordinates $e_\theta$ and
$e_\phi$, which are purely spatial unit vectors. In tetrad components
these are given by
\be\label{Explicitsphericalbasis}
\hat p^i= \left( \begin{array}{c}\cos \phi \sin \theta\\\sin \phi \sin
                   \theta\\ \cos \theta \end{array} \right)\,,\qquad
               e_\theta^i =\left( \begin{array}{c}\cos \phi \cos \theta\\\sin \phi \cos \theta\\ -\sin \theta \end{array} \right) \,,\qquad
               e_\phi^i =\left( \begin{array}{c}-\sin \phi\\\cos \phi\\ 0\end{array} \right)\,.
\ee
Let us introduce the helicity vector
\be\label{defSmu}
S^\mu(u^\nu,p^\nu) = -\frac{m}{|\gr{p}|} u^\mu + \frac{E}{m|\gr{p}|}p^\mu\,, \qquad E
= \sgnslash u_\mu p^\mu \,,\qquad |\gr{p}| = \sqrt{E^2-m^2}\,,
\ee
which is a unit vector in the direction of the spatial
momentum that is transverse to $p^\mu$ in the sense $S^\mu p_\mu =0$ and is thus
spacelike. Since the space of vectors
orthogonal to $p^\mu$ is three-dimensional, the transverse property is
not enough to specify the helicity vector and the definition
(\ref{defSmu}) depends explicitly on the observer which is used to define the spatial part of the
momentum. When no ambiguity can arise we write simply $S^\mu$. In
components the helicity vector is given by
\be\label{DefS}
S^0 = m^{-1}|\gr{p}|\,,\qquad S^i = m^{-1} E \hat{p}^i\,.
\ee
Geometrically (see Fig. \ref{fig1}), the helicity vector corresponds to the spatial direction unit vector
$\hat{\gr{p}}$ boosted in its direction by the same boost needed to
obtain $p^\mu$ from $u^\mu$. 

Finally, we define the polarisation basis
\be\label{Explicitpolarbasis}
{\epsilon}^\mu_{\pm} \equiv\frac{1}{\sqrt{2}}(e^\mu_\theta \mp \ii
e^\mu_\phi)\,,\qquad \Rightarrow\quad {\epsilon}^0_{\pm} =0\,,\qquad {\epsilon}^i_{\pm}=\frac{1}{\sqrt{2}}\left( \begin{array}{c}
\text{cos}(\theta)\text{cos}(\phi) \pm i
\text{sin}(\phi)\\
\text{cos}(\theta)\text{sin}(\phi) \mp i
\text{cos}(\phi)\\
-\text{sin}(\theta) \end{array} \right)\,.
\ee

The set of vectors $S^\mu,\epsilon_\pm^\mu$, and $p^\mu/m$ constitute an adapted orthonormal basis, with which the expressions for the operators of the type (\ref{urus}) take a simple form. From
Eqs. (\ref{baruGu}), and noting $m \mathds{1}$ simply as $m$, we first recover immediately the known decompositions
\begin{subeqnarray}\label{Closure}
u_s \bar{u}_s &=& \frac 1 2 \left(\mathds{1} \sgnzz 2 s \gamma_5
  \slashed{S}\right) \left(\sgnslash \slashed{p} + m \right) = \frac 1 2 \left(\sgnslash \slashed{p} + m \right)\left(\mathds{1} \sgnzz 2 s \gamma_5
  \slashed{S}\right)  \,,\quad\qquad
\sum_s u_s \bar{u}_s = \sgnslash \slashed{p} + m\slabel{Closureuu}\\
v_s \bar{v}_s &=& \frac 1 2 \left(\mathds{1} \sgnzz 2 s \gamma_5
  \slashed{S}\right) \left(\sgnslash \slashed{p} - m \right) = \frac 1 2 \left(\sgnslash \slashed{p} - m \right)\left(\mathds{1} \sgnzz 2 s \gamma_5
  \slashed{S}\right)  \,,\quad\qquad
\sum_s v_s \bar{v}_s = \sgnslash \slashed{p} - m \slabel{Closurevv}\,.
\end{subeqnarray}
Note however that in the massless limit, $m S^\mu \to p^\mu$
where $p^\mu$ is becomes a null vector. The previous decompositions in the massless case take the simpler forms
\begin{subeqnarray}\label{Closuremassless}
u_s \bar{u}_s &=& \sgnslash \frac 1 2 \left(\mathds{1} + 2 s \gamma_5 \right) \slashed{p} =\sgnslash \frac 1 2 \slashed{p}  \left(\mathds{1} - 2 s \gamma_5 \right)  \,,\quad\qquad
\sum_s u_s \bar{u}_s = \sgnslash \slashed{p}\\
v_s \bar{v}_s &=& \sgnslash \frac 1 2 \left(\mathds{1} - 2 s \gamma_5 \right) \slashed{p} =\sgnslash \frac 1 2 \slashed{p}  \left(\mathds{1} + 2 s \gamma_5 \right)  \,,\quad\qquad
\sum_s v_s \bar{v}_s = \sgnslash \slashed{p}\,.
\end{subeqnarray}
Additionally we also find corresponding decompositions when the helicities are different, the so-called Bouchiat-Michel formulae~\cite{1958NucPh...5..416B} (see
also App. H.4 of Ref. \cite{BibleSpinors}), and using the polarisation basis (\ref{Explicitpolarbasis}) these take the simple forms  
\bea\label{uruscov}
u_s \bar u_r &=& \sqrt{2} p^\mu
\epsilon_{r-s}^\nu\widetilde{\Sigma}_{\mu\nu} \sgnzz
\frac{m}{\sqrt{2}} \gamma^5 \gamma_\mu \epsilon_{r-s}^\mu=
\frac{1}{\sqrt{2}} \gamma^5\slashed{\epsilon}_{r-s}\left(\slashed{p}\sgnzz
  m\right)\qquad  {\rm if}\qquad  r \neq s \\
v_s \bar v_r &= & \sqrt{2} p^\mu
\epsilon_{s-r}^\nu \widetilde{\Sigma}_{\mu\nu} \sgnii \frac{m}{\sqrt{2}}\gamma^5 \gamma_\mu \epsilon_{s-r}^\mu=
\frac{1}{\sqrt{2}} \gamma^5\slashed{\epsilon}_{s-r}\left(\slashed{p}\sgnii
  m\right)\qquad  {\rm if}\qquad  r \neq s \,.
\eea

We are now in position to decompose the spinor space operator ${F_\mathfrak{a}}^\mathfrak{b}$. Let us define\footnote{We use the obvious abuse of notation $f_{++}$ for e.g. $f_{+\tfrac{1}{2}\,+\tfrac{1}{2}}$.}
\be\label{DefIVQ}
I\equiv f_{++}+f_{--}\,,\qquad V=f_{++}-f_{--}\,,\qquad Q_\pm\equiv \sqrt{2}f_{\pm\mp} \,,
\ee
together with
\be\label{DefcalQ}
Q^\mu \equiv Q_+ \epsilon_+^\mu + Q_- \epsilon_-^\mu\,,\qquad {\cal
  Q}^\mu \equiv Q^\mu+V S^\mu\,.
\ee
$I$ is the total intensity, $V$ the circular polarisation, and $Q^\mu$
is the purely linear polarisation vector. ${\cal Q}^\mu$ is the
total polarisation vector, taking into account both circular and linear polarisation. By construction the total polarisation ${\cal Q}^\mu$ is transverse to the momentum (${\cal Q}^\mu p_\mu=0$). The linear polarisation $Q^\mu$ is transverse both to the momentum and to the observer velocity $u^\mu$, that is it is a purely spatial vector.

\begin{itemize}
\item In the massive case, using  Eqs. (\ref{Closure}) and (\ref{uruscov}) in Eq. (\ref{Fspinor}), with again the short notation $M$ for $M \mathds{1}$, the operator $\gr{F}$ is decomposed on the basis of operators ${\cal O}$ as
\be\label{BigDecompositionGood}
\gr{F} = \frac{I}{2}\left(M\sgnzz \slashed{p}\right)\sgnzz\frac{M}{2} \gamma^5\gamma_\mu {\cal Q}^\mu  +p^\mu \tilde \Sigma_{\mu\nu} {\cal Q}^\nu\,.
\ee
We have written this decomposition in a form which is valid for both
particles and antiparticles, by introducing the notation 
\be
M=\begin{cases}
m, \qquad {\rm particle}\\
-m, \,\,\quad {\rm antiparticle}.
\end{cases}
\ee
Using $\gamma^5 \slashed{\cal Q} \slashed{p} = \slashed{p} \gamma^5 \slashed{\cal Q} = 2 p^\mu \widetilde{\Sigma}_{\mu\nu} {\cal Q}^\nu$ the decomposition can also be written (with the notation $I$ for $I \mathds{1}$)
\be\label{BigDecompositionGood2}
\gr{F} = \frac{1}{2}\left(I \sgnzz \gamma^5 \slashed{\cal Q}\right)\left(M\sgnzz \slashed{p}\right) = \frac{1}{2}\left(M\sgnzz \slashed{p}\right)\left(I \sgnzz \gamma^5 \slashed{\cal Q}\right)\,.
\ee

\item In the massless limit, we obtain the simpler decomposition
\be\label{BigDecompositionGoodMassless}
\gr{F} =\sgnslash\frac{1}{2}(I+ \hatM V \gamma^5)\slashed{p}+p^\mu \widetilde{\Sigma}_{\mu\nu} Q^\nu\,,\qquad \hatM \equiv \begin{cases}1\,\,\,\,\,\quad{\rm particles}\\-1\quad{\rm antiparticles}\,.\end{cases}
\ee
Note that the linear polarisation $Q^\mu$ and the circular polarisation $V$ enter separately, and not as a total polarisation vector ${\cal Q}^\mu$ as is the case in the massive case. Using $\gamma^5 \slashed{Q} \slashed{p} = \slashed{p} \gamma^5 \slashed{Q} = 2 p^\mu \widetilde{\Sigma}_{\mu\nu} {Q}^\nu$ it can also be rewritten as
\be\label{BigDecompositionGoodMassless2}
\gr{F} =\sgnslash\frac{1}{2}(I+ \hatM V \gamma^5 \sgnzz \gamma^5 \slashed{Q})\slashed{p} = \sgnslash\frac{1}{2} \slashed{p} (I- \hatM V \gamma^5 \sgnzz \gamma^5 \slashed{Q})\,.
\ee
The decompositions (\ref{BigDecompositionGoodMassless}) or (\ref{BigDecompositionGoodMassless2}) match the ones obtained in Refs. \cite{Vlasenko:2013fja} \cite{Blaschke:2016xxt} which focused on the massless case. 
\end{itemize}

\subsubsection{Properties of the decomposition in covariant components}\label{SecPropertiesDecomposition}

\begin{itemize}
\item The decomposition of an antiparticle is the same as the one of
  its particle counterpart, except for the replacement $m \to
  -m$. However it is usually admitted that the rule to go from the
  particle to the antiparticle description is $p \to -p$. In the
  decomposition (\ref{BigDecompositionGood}),  we note that the
  replacement $m \to -m$ is equivalent to the replacement $p \to -p$
  up to an overall minus sign. We detail this point in \S~\ref{SecChargeConjugation}. 

\item In the massless case, the particle and antiparticle operators
  differ only by the replacement $V \to -V$. 
\item All operators $X$ appearing in the decomposition (\ref{BigDecompositionGood}) satisfy the property $X^\dagger =
  \gamma^0 \cdot X \cdot \gamma^0$ or more rigorously $X^\dagger = A
  \cdot X \cdot A^{-1}$ (see App. \ref{AppClifford} and Eq. G.1.22 of
  Ref.~\cite{BibleSpinors}). Note that this is not the case for the operator $\gamma^5$ which does not appear. Hence, the operator $\gr{F}$ satisfies the same property
\be\label{HermitianSpinor}
\gr{F}^\dagger = \gamma^0  \cdot \gr{F} \cdot \gamma^0\,.
\ee
This property is a consequence of the Hermiticity  of the distribution function [see Eq. (\ref{HermitianProperty})].

\item In the decomposition of the massless case
  (\ref{BigDecompositionGoodMassless}), we remark that any extra term
  proportional to $p^\mu$ can be added to $Q^\mu$ without altering the
  decomposition due to the antisymmetry of
  $\widetilde{\Sigma}_{\mu\nu}$, and without altering the transverse
  property $p^\mu {Q}_\mu=0$. Indeed ${\cal Q}^\mu$ has three degrees of freedom
  since it is built from $V,Q_+,Q_-$ and this is reflected by the
  transverse property ${\cal Q}^\mu p_\mu = 0$, but $Q^\mu$ contains only the
two degrees of freedom corresponding to $Q_-,Q_+$ and this leads to
this ambiguity. 
This is identical to the gauge freedom of the
massless photon which is remaining even in the Lorentz gauge $k_\mu A^\mu = 0$ since it has only two physical degrees
of freedom and not three as in the massive case (Proca theory). The solution to this problem is exactly the same, and by construction  of $Q^\mu$ we have demanded an extra condition which is observer dependent, $Q_\mu u^\mu = 0$. This means that for
this observer the linear polarisation vector $Q^\mu$ must be purely
spatial. This condition is analogous to the observer dependent Coulomb
gauge choice ($A^0=0$) required to fully fix the gauge
potential. However, the physical results do not depend on this
Coulomb-type gauge choice since it does not alter the decomposition (\ref{BigDecompositionGoodMassless}). To summarize, linear polarisation in the massless case is the coset 
\be\label{DefCoset}
[ Q^\mu ] \equiv \{ Q^\mu + \alpha p^\mu,\,\,\, \alpha \in \mathds{R}\}
\ee
whose preferred representative element for a given observer is the one defined in Eq. (\ref{DefcalQ}), which is the only one purely spatial for that observer.

\item Except in the massless case, the total polarisation vector is a spin-$1$ representation of $SO(3)\simeq SU(2)$ and the intensity is a spin-$0$ representation. Indeed, when forming the number operator (\ref{defNrs}), and thus $f_{rs}$, we are building the tensor product of spin-$1/2$ representations and what we have achieved is a decomposition of the reducible representation $\gr{2}\otimes \gr{2}$ in irreducible components $\gr{3} \oplus \gr{1}$, where we have denoted $\gr{1},\gr{2},\gr{3}$ the spin-$0,1/2,1$ representations of $SU(2)$. In the massless case, the little group of the Lorentz group (see Ref. \cite{Weinberg1}) is not $SO(3)$ but $SO(2)\simeq U(1)$. Hence the decomposition in irreducible representations is of the form $\gr{2}_1 \oplus \gr{1} \oplus \gr{1}$ where the purely linear polarisation is in the spin-$1$ representation of $SO(2)$ (noted $\gr{2}_1$) and circular polarisation is in the representation $\gr{1}$. See Appendix~\ref{AppBosons} for the decomposition in irreducible components of the distribution function for massless vector bosons which obeys the same logic.
\end{itemize}

\subsubsection{Extraction of covariant components}

The covariant components can be obtained by multiplying $\gr{F}$ with the appropriate operator and taking the trace, using the Fierz identity  (\ref{MyFierz}). In the massive case we find
\bea\label{ExtractCovParts1}
I &=& \frac{1}{2 M}{\rm Tr}[\gr{F}] = \sgnslash\frac{1}{2 m^2}{\rm Tr}[\gr{F} \cdot \slashed{p}]=\frac{1}{4 m^2}{\rm Tr}[\gr{F} \cdot (M \sgnzz \slashed{p})]\\
{\cal Q}^\mu &=& \frac{1}{2 M}{\rm Tr}[\gr{F} \cdot \gamma^\mu \gamma^5] = \sgnslash\frac{1}{2 m^2}{\rm Tr}[\gr{F} \cdot \gamma^\mu \gamma^5\slashed{p}]=\frac{1}{4 m^2}{\rm Tr}[\gr{F} \cdot \gamma^\mu \gamma^5(M \sgnzz \slashed{p})]=\frac{1}{4 m^2}{\rm Tr}[\gr{F} \cdot (M \sgnzz \slashed{p})\gamma^\mu \gamma^5]\,.
\eea
In the last equality we have used $\{\slashed{p},\gamma^\mu \gamma^5\} = \sgnslash 2 \gamma^5 p^\mu$ and ${\rm Tr}[\gr{F}\cdot \gamma^5]=0$.

In the massless case, we must first introduce a future directed null
vector $k^\mu$ (see Fig. \ref{fig1}) such that 
\be\label{Defkmu}
k^\mu p_\mu = \sgnslash 1\,,\qquad k_\mu \epsilon_\pm^\mu=0\,,
\ee 
that is which lies in the plane spanned by the observer four-velocity $u^\mu$ and $p^\mu$. The covariant parts are then obtained from
\bea\label{ExtractCovParts2}
I &=& \sgnslash\frac{1}{2}{\rm Tr}[\gr{F} \cdot \slashed{k}]\slabel{ExtractIMassless}\\
\hatM V &=& \sgnslash\frac{1}{2}{\rm Tr}[\gr{F} \cdot \slashed{k}\gamma^5]\\
{Q}^\mu &=& \sgnslash\frac{1}{2}{\rm Tr}[\gr{F} \cdot \gamma^\mu \gamma^5\slashed{k}]=\sgnslash\frac{1}{2}{\rm Tr}[\gr{F} \cdot \slashed{k}\gamma^\mu \gamma^5]\,.
\eea
Note that given the decomposition in the massless case (\ref{BigDecompositionGoodMassless}), the linear polarisation extracted with the previous expressions is automatically transverse to the observer four-velocity $u^\mu$ and to the momentum $p^\mu$. That is, introducing a symmetric screen projector ${\cal H}_{\mu\nu}$ such that ${{\cal H}_{\mu}}^\sigma {\cal H}_{\sigma\nu} = {\cal H}_{\mu\nu}$ and ${\cal H}_{\mu\nu}p^\nu={\cal H}_{\mu\nu}u^\nu=0$, which is built as
\be\label{DefScreenH}
{\cal H^\mu}_\nu \equiv \delta_\nu^\mu \sgnii p^\mu k_\nu \sgnii k^\mu p_\nu= \sgnminusslash \epsilon_+^\mu \epsilon_{-\,\nu} \sgnii \epsilon_-^\mu \epsilon_{+\,\nu}\,,
\ee
it satisfies automatically ${{\cal H}^\mu}_\nu Q^\nu = Q^\mu$.


\section{Transformation properties and observer independence}\label{SecTProperties}

\subsection{The definition of observer independence}\label{SecDiscussionObserverDependence}

In order to discuss the observer dependence of the covariant components, we must first specify the definition of the observer. So far we have performed our computations in special relativity in a Minkowski space-time, that is with a global inertial frame. However, we intend to embed this space-time into the general relativistic manifold and define our observer by a fully relativistic trajectory.

From the equivalence principle, the Minkowski space becomes in the context of general relativity the local tangent space, and it is connected to the manifold by the tetrads $[e_\mu]^\alpha$ carrying a Lorentz-index inside the brackets and a general relativistic index on the outside. The tetrads are not unique and at each point in space-time a group of tetrads exist, corresponding to the Lorentz symmetry of special relativity. 

In the following, we will choose the unique tetrad corresponding to the observer velocity and its spatial orientation (such that the observer velocity in tetrad indices is purely temporal ($u^\alpha = [e_0]^\alpha$), i.e. the observer is not moving in its own frame). The corresponding local Minkowski space-time is thus identified directly with our relativistic observer and Lorentz transformations therefore represent a change of observer. 

We split the general relativistic 4-momentum $P^\alpha$~\footnote{For clarity we use the notation $p^\mu$ for the components of the momentum in the local tetrad basis $e^\mu$ and $P^\alpha$ for the components of the same momentum in a general coordinates basis, that is we use both a different symbol and a different type of index.} into a covariant energy and spatial momentum according to 
\be
P^\alpha =  p^\mu [e_\mu]^\alpha = E [e_0]^\alpha + p^i [e_i]^\alpha\,.
\ee
$E$ and $p^i$ are by construction invariant under general coordinate transformations, but not under a change of the observer. Each observer will measure a different energy or spatial momentum corresponding to its own velocity and the orientation of the spatial vectors in its tetrad.
\com{Since we identify the observer directly with our local Minkowski space time it implies that the energy and spatial momentum are \emph{not} covariant under Lorentz transformations. The momentum in general relativity is connected to the special relativistic momentum by 
\be
P^\mu = [e_a]^\mu p^a =  E [e_0]^\mu - p^i [e_i]^\mu \,,
\ee
implying that the energy $E$ corresponds to the zero-component and the spatial momentum $p^i$ to the spatial-components of the 4-momentum. This is true in any Minkowski space-time.} 

We will call a quantity observer independent if under a change of observer and the corresponding Lorentz transformation it transforms in a given representation of the Lorentz group. This implies that the same object will be described differently by two observers, but this difference is entirely related to the local tetrad basis of the observer and not the properties of the observer itself. One example is the local 4-momentum $p^\mu$, transforming as a gneneric 4-vector, or the mass of particles $m$, which is a Lorentz scalar. 

On the other hand a quantity is observer-dependent if it does not follow the usual Lorentz transformations. This is for example the case for the local observer velocity $u^\mu = \delta_0^\mu$, which by definition takes this value for each observer and thus does not transform as a vector. In particular this is also the case for the helicity $S^\mu$, which directly depends on the observer-dependant $u^\mu$. The choice of observer therefore affects the helicity states and the same quantum system may be described both as left- or right-handed depending on the observer. We conclude that the distribution functions $f_{rs}$, the energy $E$ and the spatial momentum $p^i$ are observer-dependent.

\com{
\subsection{Observer independence of ${F_\mathfrak{a}}^\mathfrak{b}$ }

The field $\psi$ is observer independent and does transform in the usual spinor-representation. It is constructed from the operators $a_{s}$ which carry a free Hilbert space index $s$. We choose to utilise the helicity basis $s= \pm\tfrac{1}{2}$ introducing an explicit observer dependence in this sub-space. However, after summing over the entire Hilbert space, using the completeness relation the resulting field is independent from this choice and therefore observer independent. 

We can build $\psi$ for two different observers, the new observer being obtained through a Lorentz transformation $\Lambda$ from the old one. This corresponds to the passive transformation point of view. For the new observer, the coordinates of a given point are $\tilde x = \Lambda^{-1} x$ where $x$ are the coordinates of the same point for the old observer. The field spinor components seen by the new observer are related to the ones seen by the old observer thanks to 
\be
\widetilde{\psi}_\mathfrak{a} (\tilde x) = {{D}_\mathfrak{a}}^\mathfrak{b}(\Lambda^{-1}) \psi_\mathfrak{b} (x)\qquad \Rightarrow \qquad\widetilde{\psi}_\mathfrak{a} (x) = {{D}_\mathfrak{a}}^\mathfrak{b}(\Lambda^{-1}) \psi_\mathfrak{b} (\Lambda x)\,,
\ee
where ${D_\mathfrak{a}}^\mathfrak{b}(\Lambda)$ is the representation of the Lorentz transformation on the spinor space whose explicit expression is given in Appendix \ref{AppSUdeux}.

However, the Lorentz transformation is also represented in the Hilbert space by $U(\Lambda)$ and the field seen by the new observer can be built as
\be
U(\Lambda) \psi_\mathfrak{a} (x) U(\Lambda^{-1}) = \widetilde{\psi}_\mathfrak{a}(x)\,.
\ee
Combining these two ways of obtaining the transformed field, we deduce that the field must satisfy
\be
U(\Lambda) \psi_\mathfrak{a} (x) U(\Lambda^{-1}) ={D_\mathfrak{a}}^\mathfrak{b}(\Lambda^{-1}) \psi_\mathfrak{b} (\Lambda x)\,.
\ee
Using the decomposition (\ref{DefQuantumField}) for both $\psi_\mathfrak{a} (x)$ and $\psi_\mathfrak{b} (\Lambda x)$ in this expression, and using $[\dd \,(\Lambda p) ] = [\dd p]$, we deduce that
\com{We directly build the fields in both coordinate systems, with both observers defining the momentum $\gr{p}$ and the helicity $s$ based on their own coordinates
\begin{eqnarray}
\psi_\mathfrak{a}(x) &=& \sum_{s=\pm\tfrac{1}{2}} \int [\dd p]\left[ {\rm e}^{\sgnslash \ii p \cdot x }
\antipar{a}^\dagger_s(p) v_{s,\mathfrak{a}}(p)+ {\rm e}^{\sgnminusslash \ii p \cdot x }
a_s(p) u_{s,\mathfrak{a}}(p)\right] \\
\widetilde{\psi}_\mathfrak{a}(\tilde{x}) &=& \sum_{s=\pm\tfrac{1}{2}} \int [\dd \tilde{p}]\left[ {\rm e}^{\sgnslash \ii \tilde{p} \cdot \tilde{x} }
\antipar{a}^\dagger_s(\tilde{p}) v_{s,\mathfrak{a}}(\gr{\tilde{p}})+ {\rm e}^{\sgnminusslash \ii \tilde{p} \cdot \tilde{x} }
a_s(\tilde{p}) u_{s,\mathfrak{a}}(\tilde{p})\right] \\ \nonumber
&=& \sum_{s=\pm\tfrac{1}{2}} \int [\dd p]\left[ {\rm e}^{\sgnslash \ii p \cdot x }
\antipar{a}^\dagger_s(\Lambda p) v_{s,\mathfrak{a}}(\Lambda p)+ {\rm e}^{\sgnminusslash \ii p \cdot x }
a_s(\Lambda p) u_{s,\mathfrak{a}}(\Lambda p)\right]\,,
\end{eqnarray}
where in the last line we have used that $\tilde p = \Lambda p$, the scalar product $p \cdot x$ is independent of the coordinates and that the integration element $[\dd p]$ is Lorentz invariant. 
Combined with Eq.~(\ref{eq:trafofields}) this implies that }
\bea\label{Trulea}
\sum_{s} a_s(\Lambda p) u_{s,\mathfrak{a}}(\Lambda p) &=& \sum_{s} {D_\mathfrak{a}}^\mathfrak{b}(\Lambda)  \,U(\Lambda)a_s(p) U(\Lambda^{-1}) u_{s,\mathfrak{b}}(p) \\
\sum_{s} \bar a^\dagger_s(\Lambda p) v_{s,\mathfrak{a}}(\Lambda p) &=& \sum_{s} {D_\mathfrak{a}}^\mathfrak{b}(\Lambda)  \,U(\Lambda)\bar a^\dagger_s(p) U(\Lambda^{-1}) v_{s,\mathfrak{b}}(p) \,.
\eea
These equations mean, that even though the helicity states (defined by $a_s(p)$ and $a_s(\Lambda p)$) are observer dependent, after summing them with the corresponding spinors $u_{s,\mathfrak{a}}$ the combination is observer independent and transforms as a regular Dirac spinor valued operator.

The conjugated field $\bar{\psi}^{\mathfrak{a}}$ provides the analogous relations
\bea\label{Truleadagger}
\sum_s a^{\dagger}_s(\Lambda p) \bar u_s^\mathfrak{a}(\Lambda p) &=& \sum_s
\,U(\Lambda)a^{\dagger}_s(p) U(\Lambda^{-1})\bar u_s^\mathfrak{b}(p) \,{D^{}_\mathfrak{b}}^\mathfrak{a}(\Lambda^{-1})  \\
\sum_s \bar a_s(\Lambda p) \bar v_s^\mathfrak{a}(\Lambda p) &=& \sum_s
\,U(\Lambda)\bar a_s(p) U(\Lambda^{-1})\bar v_s^\mathfrak{b}(p) \,{D^{}_\mathfrak{b}}^\mathfrak{a}(\Lambda^{-1})  \,,
\eea
or they can be deduced using the property (\ref{MagicSigma}). We have seen that the distribution functions $f_{rs}$ are observer dependent. However, using the above equations, the tensor ${F_\mathfrak{a}}^\mathfrak{b}$ built from the distribution functions turns out to be entirely observer independent, for both particles and antiparticles. Indeed, under the change of observer, the quantum state $|\Psi\rangle$ observed by the initial observer is seen as $\widetilde{|\Psi\rangle} = U(\Lambda^{-1})|\Psi\rangle$ for the new observer, and a momentum $p$ seen by the old observer is now seen as $\tilde p \equiv \Lambda^{-1} p$. From Eqs. (\ref{Trulea}) and (\ref{Truleadagger}) the spinor valued operator ${\widetilde{F}_\mathfrak{a}}^{\,\,\mathfrak{b}}$ defined by the new observer is related to the former by
\be \label{WonderfulTransformation}
{\widetilde{F}_\mathfrak{a}}^{\,\,\mathfrak{b}}(\tilde p) = {D_\mathfrak{a}}^{\mathfrak{a}'}(\Lambda^{-1}){F_{\mathfrak{a}'}}^{\mathfrak{b}'}( p){D_{\mathfrak{b}'}}^\mathfrak{b}(\Lambda)\,.
\ee
This is the expected (passive) transformation rule for spin valued tensors in momentum space, and according to the definition of
\S~\ref{SecDiscussionObserverDependence}, this proves that ${F_\mathfrak{a}}^\mathfrak{b}$ is observer independent.
}

\subsection{Observer independence of ${F_\mathfrak{a}}^\mathfrak{b}$ }

The field $\psi$ is observer independent by construction and does transform in the usual spinor-representation.
\com{I suggest to remove the following because it looks like our previous blurry understanding.  It is constructed from the operators $a_{s}$ which carry a Hilbert space label $s$. We choose to utilise the helicity basis $s= \pm\tfrac{1}{2}$ and since helicity is observer dependent we introduce an explicit observer dependency in this sub-space. However, after summing over the entire Hilbert space and employing the completeness relation the resulting field is independent from this choice. }
We can build $\psi$ for two different observers connected by the passive Lorentz transformation $\Lambda$, which thus acts actively on the coordinates $\tilde x = \Lambda x$, $\tilde x$ being the coordinates of a given point for the new observer and $x$ the coordinates of the same point for the original one. In addition to the coordinates change, each observer also defines his own Hilbert space basis and therefore describes quantum states and operators differently. We take this into account via the unitary transformation $U(\Lambda)$ and we find the usual transformation rules
\be \label{eq:trafofields}
\widetilde{\psi}_{\mathfrak{a}}(\tilde{x}) = {D_\mathfrak{a}}^\mathfrak{b}(\Lambda) U(\Lambda)\psi_{\mathfrak{b}}(x)U(\Lambda^{-1})\,,
\ee
where ${D_\mathfrak{a}}^\mathfrak{b}(\Lambda)$ is the representation
of the Lorentz transformation on the spinor space whose explicit
expression is given in Appendix \ref{AppSUdeux}.
We directly build the fields in both coordinate systems, with the observers defining the momentum $\gr{p}$ and the helicity $s$ based on their own coordinates
\begin{eqnarray}
\psi_\mathfrak{a}(x) &=& \sum_{s=\pm\tfrac{1}{2}} \int [\dd p]\left[ {\rm e}^{\sgnslash \ii p \cdot x }
\antipar{a}^\dagger_s(p) v_{s,\mathfrak{a}}(p)+ {\rm e}^{\sgnminusslash \ii p \cdot x }
a_s(p) u_{s,\mathfrak{a}}(p)\right] \\
\widetilde{\psi}_\mathfrak{a}(\tilde{x}) &=& \sum_{s=\pm\tfrac{1}{2}} \int [\dd \tilde{p}]\left[ {\rm e}^{\sgnslash \ii \tilde{p} \cdot \tilde{x} }
\antipar{a}^\dagger_s(\tilde{p}) v_{s,\mathfrak{a}}(\gr{\tilde{p}})+ {\rm e}^{\sgnminusslash \ii \tilde{p} \cdot \tilde{x} }
a_s(\tilde{p}) u_{s,\mathfrak{a}}(\tilde{p})\right] \\ \nonumber
&=& \sum_{s=\pm\tfrac{1}{2}} \int [\dd p]\left[ {\rm e}^{\sgnslash \ii p \cdot x }
\antipar{a}^\dagger_s(\Lambda p) v_{s,\mathfrak{a}}(\Lambda p)+ {\rm e}^{\sgnminusslash \ii p \cdot x }
a_s(\Lambda p) u_{s,\mathfrak{a}}(\Lambda p)\right]\,,
\end{eqnarray}
where in the last line we have defined $\tilde p = \Lambda p$ and used that the scalar product $p \cdot x$ is independent of the coordinates and that the integration element $[\dd p]$ is Lorentz invariant. 
Combined with Eq.~(\ref{eq:trafofields}) this implies that 
\bea\label{Trulea}
\sum_{s} a_s(\Lambda p) u_{s,\mathfrak{a}}(\Lambda p) &=& \sum_{s} {D_\mathfrak{a}}^\mathfrak{b}(\Lambda)  \,U(\Lambda)a_s(p) U(\Lambda^{-1}) u_{s,\mathfrak{b}}(p) \\
\sum_{s} \bar a^\dagger_s(\Lambda p) v_{s,\mathfrak{a}}(\Lambda p) &=& \sum_{s} {D_\mathfrak{a}}^\mathfrak{b}(\Lambda)  \,U(\Lambda)\bar a^\dagger_s(p) U(\Lambda^{-1}) v_{s,\mathfrak{b}}(p) \,.
\eea
These equations mean, that even though the helicity states (defined by $a_s(p)$ and $a_s(\Lambda p)$) are observer dependent, after summing them with the corresponding spinors $u_{s,\mathfrak{a}}$ the combination is observer independent and transforms as a regular Dirac spinor valued operator.

The conjugated field $\bar{\psi}^{\mathfrak{a}}$ provides the analogous relations
\bea\label{Truleadagger}
\sum_s a^{\dagger}_s(\Lambda p) \bar u_s^\mathfrak{a}(\Lambda p) &=& \sum_s
\,U(\Lambda)a^{\dagger}_s(p) U(\Lambda^{-1})\bar u_s^\mathfrak{b}(p) \,{D^{}_\mathfrak{b}}^\mathfrak{a}(\Lambda^{-1})  \\
\sum_s \bar a_s(\Lambda p) \bar v_s^\mathfrak{a}(\Lambda p) &=& \sum_s
\,U(\Lambda)\bar a_s(p) U(\Lambda^{-1})\bar v_s^\mathfrak{b}(p) \,{D^{}_\mathfrak{b}}^\mathfrak{a}(\Lambda^{-1})  \,,
\eea
or they can be deduced using the property (\ref{MagicSigma}).

We deduce that a change of coordinates is equivalent to a change in the Hilbert basis
\begin{eqnarray}
\psi_\mathfrak{a}(\Lambda x) &=& \sum_{s=\pm\tfrac{1}{2}} \int [\dd p]\left[ {\rm e}^{\sgnslash \ii p \Lambda x }
\antipar{a}^\dagger_s(p) v_{s,\mathfrak{a}}(p)+ {\rm e}^{\sgnminusslash \ii p \Lambda x }
a_s(p) u_{s,\mathfrak{a}}(p)\right]  \\ \nonumber
&=& \sum_{s=\pm\tfrac{1}{2}} \int [\dd \tilde{p}]\left[ {\rm e}^{\sgnslash \ii \tilde{p} \cdot x }
\antipar{a}^\dagger_s(\Lambda \tilde{p}) v_{s,\mathfrak{a}}(\Lambda \tilde{p})+ {\rm e}^{\sgnminusslash \ii \tilde{p} \cdot x }
a_s(\Lambda \tilde{p}) u_{s,\mathfrak{a}}(\Lambda \tilde{p})\right] \\ \nonumber
&=& {D_\mathfrak{a}}^\mathfrak{b}(\Lambda)  \,U(\Lambda)  \sum_{s=\pm\tfrac{1}{2}} \int [\dd \tilde{p}]\left[ {\rm e}^{\sgnslash \ii \tilde{p} \cdot x }
\antipar{a}^\dagger_s(\tilde{p}) v_{s,\mathfrak{b}}(\tilde{p})+ {\rm e}^{\sgnminusslash \ii \tilde{p} \cdot x }
a_s( \tilde{p}) u_{s,\mathfrak{b}}( \tilde{p})\right]\, U^{-1}(\Lambda) \\ \nonumber
&=&{D_\mathfrak{a}}^\mathfrak{b}(\Lambda)  \,U(\Lambda) \psi_\mathfrak{b}(x) U^{-1}(\Lambda) \,.
\end{eqnarray}
So we find that the field must statisfy 
\be
U(\Lambda) \psi_\mathfrak{a} (x) U(\Lambda^{-1}) ={D_\mathfrak{a}}^\mathfrak{b}(\Lambda^{-1}) \psi_\mathfrak{b} (\Lambda x)\,.
\ee
While the distribution functions $f_{rs}$ are observer dependent, the tensor ${F_\mathfrak{a}}^\mathfrak{b}$ built from the distribution functions turns out to be entirely observer independent, for both particles and antiparticles. Indeed, under the change of observer, the quantum state $|\Psi\rangle$ seen by the initial observer is seen as $\widetilde{|\Psi\rangle} = U(\Lambda)|\Psi\rangle$ for the new observer, and a momentum $p$ seen by the old observer is now seen as $\tilde p \equiv \Lambda p$. From Eqs. (\ref{Trulea}) and (\ref{Truleadagger}) the spinor valued operator ${\widetilde{F}_\mathfrak{a}}^{\,\,\mathfrak{b}}$ defined by the new observer is related to the former by
\be \label{WonderfulTransformation}
{\widetilde{F}_\mathfrak{a}}^{\,\,\mathfrak{b}}(\tilde p) = {D_\mathfrak{a}}^{\mathfrak{a}'}(\Lambda){F_{\mathfrak{a}'}}^{\mathfrak{b}'}( p){D_{\mathfrak{b}'}}^\mathfrak{b}(\Lambda^{-1})\,.
\ee
This is the expected transformation rule for spinor valued tensors in momentum space, and according to the definition of
\S~\ref{SecDiscussionObserverDependence}, this proves that ${F_\mathfrak{a}}^\mathfrak{b}$ is observer independent.

\subsection{Transformation properties of the covariant components}

The Dirac matrices satisfy the transformation rules
\be
D(\Lambda) \gamma^\mu D(\Lambda^{-1}) = \gamma^\nu {\Lambda_\nu}^\mu\,\quad \Rightarrow \quad  D(\Lambda^{-1}) \gamma^\mu D(\Lambda) = {\Lambda^\mu}_\nu \gamma^\nu\,.
\ee
For the proper orthochronous Lorentz group $SO^+(1,3)$ this implies
that $\gamma^5$ is invariant. Combining this property with the transformation rule
(\ref{WonderfulTransformation}) and the decomposition
(\ref{BigDecompositionGood}) we deduce that the covariant components
transform under a coordinate transformation $\Lambda \in SO^+(1,3)$ as
\be
\widetilde{I}(\tilde p ) = I(p)\,,\qquad \widetilde{\cal Q}^\mu(\tilde p) = {\Lambda^\mu}_\nu
{\cal Q}^\nu(p)\,.
\ee
This means that they transform exactly as a scalar and vector field and according to the discussion of \S~\ref{SecDiscussionObserverDependence}, they are therefore observer independent. 

The same analysis can be carried in the massless case using the decomposition (\ref{BigDecompositionGood2}) and we deduce that the covariant components transform as
\be
\widetilde{I}(\tilde p ) = I(p)\,,\qquad\widetilde{V}(\tilde p ) = V(p)\,,\qquad \widetilde{Q}^\mu(\tilde p) = {\widetilde{\cal H}^\mu}_{\,\,\sigma}{\Lambda^\sigma}_\nu
{Q}^\nu(p)\,.
\ee
The screen projector [see Def. (\ref{DefScreenH})] associated with the new observer and the new momentum, ${\widetilde{\cal H}^\mu}_{\,\,\sigma}$, ensures that the linear polarisation remains spatial for the new observer according to the discussion of \S~\ref{SecPropertiesDecomposition}. Hence in the massless case, the linear polarisation part is not strictly observer independent, but since this dependence introduced by the screen projector is there only as the result of a choice to remove a non physical degree of freedom, we can still conclude that in that sense the covariant components are observer independent. More rigorously, the coset of linear polarisation [see definition (\ref{DefCoset})] is observer independent and only the special choice of its representative element is observer dependent and we should rather write the transformation rule of linear polarisation cosets which is $\left[\widetilde{Q}^\mu(\tilde p) \right]=\left[ {\Lambda^\mu}_\nu{Q}^\nu(p)\right]$, for which the observer independence is manifest.

\subsection{Discussion}

The observer independence is important as it allows us to build a statistical description of the fluid without the need to specify an observer first. This will be particularly useful for deriving simple transport equations in general relativity.   

The scalar $I$ describes the total intensity of the field and is observer independent since the local number of particles is identical for each observer. The information of the polarisation of the fluid is contained in the observer independant vector ${\cal Q}^\mu$.

On the other hand the parameters $V$ and $Q_\pm$, describing individually the circular and linear polarisations are not observer independent. The circular polarisation $V$, for example, changes if the observer is boosted and overtakes the measured particle. We have defined
\be
 {\cal Q}^\mu = Q^\mu + V S^\mu\,,
\ee
where ${\cal Q}^\mu$ combines multiple observer dependant quantities
into one observer independent vector. In the example of the observer
overtaking a particle we change all left-helical $-\tfrac{1}{2}$
states into right-helical $+\tfrac{1}{2}$ states. This means that the
boosted observer will find $\widetilde{V} = - V$. At the same time the
vector $S^\mu$ is also observer dependent and the new observer will
define the spatial momentum of the particles with the opposite
sign. Therefore the combination $V S^\mu$ is invariant under this
boost. At the same time the off-diagonal distributions are swapped:
$\widetilde{f}_{+-} = f_{-+}$. However these are combined with the polarisation vectors $\epsilon_{\pm}$ to form $Q^\mu$, which are also interchanged for the new observer, leading to $Q^\mu$ being invariant. 

In a more general case $Q^\mu$ and $V$ cannot be disentangled in an observer independent manner and there always exists a subset of observers, all related by boosts along the momentum direction and rotations around the momentum direction, that will perceive the field to be entirely circularly polarised without any linear polarisation. For this reason we will work with the observer independent polarisation vector ${\cal Q}^\mu$ and only refer to the circular and linear polarisations when we have specified an observer. 

In the massless case the situation simplifies. It is no longer possible to overtake the particles as they move at the speed of light in any coordinate system. This leads to both, the circular and linear polarisations $V$ and $Q^\mu$ [more rigorously the coset of linear polarisation (\ref{DefCoset})] to be individually observer independent.

\subsection{Discrete transformations}\label{SecDiscrete}

We can repeat the analysis with discrete transformations. We are most notably interested in parity and charge conjugations which are needed to discuss weak interactions.

\subsubsection{Parity transformation}

For parity transformation, the unitary operator ${\cal P}$
implements the transformation through
\be
{\cal P}  a_s(\gr{p}) {\cal P} = \ii a_{-s}(-\gr{p}) \,,\qquad {\cal
  P}  b_s(\gr{p}) {\cal P} = -\ii b_{-s}(-\gr{p})\,.
\ee
If we consider the parity transformed quantum system ${\cal P}|
\Psi \rangle$ then the distribution function $\widetilde
f_{rs}(\gr{p})$ built from it satisfies
\be
\widetilde{f}_{rs}(\gr{p}) = f_{-r\,-s}(-\gr{p})\,,
\ee
that is all helicities and momenta are reversed. If we now consider the spinor
valued operator $\widetilde{\gr{F}}$ which is built from $\widetilde{f}_{rs}$, and using the property
\be
u_{-s}(-\gr{p}) =-\ii \gamma^0 u_s(\gr{p})\,,\qquad v_{-s}(-\gr{p}) =
-\ii \gamma^0 v_s(\gr{p})\,,
\ee
we find that it is related to the original operator as
\be
\widetilde{\gr{F}}(-\gr{p}) = \gamma^0 \cdot \gr{F}(\gr{p})\cdot
\gamma^0 = \gr{F}^\dagger(\gr{p})\,.
\ee
which is nothing but Eq. (\ref{WonderfulTransformation}) applied for a discrete parity transformation with $D({\cal P}) = \gamma^0$.
Since
\bea
\gamma^0 \cdot \gamma^\mu \cdot \gamma^0 &=&(-1)^\mu\,\gamma^\mu \\
\gamma^0 \cdot \gamma^\mu \gamma^5\cdot \gamma^0 &=&-(-1)^\mu\,\gamma^\mu
\gamma^5\,,
\eea
where we used the compact notation~\cite{Peskin}
\be\label{CompactPeskin}
(-1)^\mu = \begin{cases}1\,\,\,\,\qquad\mu=0\\-1\qquad \mu=1,2,3,\end{cases}
\ee
then we deduce that
\be
\widetilde{Q}^0(-\gr{p}) = - Q^0(\gr{p})\,\qquad \widetilde{Q}^i(-\gr{p}) = Q^i(\gr{p})\,.
\ee
The spatial part of the polarisation is unchanged and transforms like an
axial vector. The time component is reverted just to ensure that the
transverse property $p^\mu Q_\mu$ still holds after a parity transformation. 

\subsubsection{Charge conjugation}\label{SecChargeConjugation}

Let us now consider charge conjugation transformations. They are implemented through the unitary operator ${\cal C}$ as
\be
{\cal C}  a_s(\gr{p}) {\cal C} = b_{s}(\gr{p}) \,,\qquad {\cal C}  b_s(\gr{p}) {\cal C} = a_{s}(\gr{p})\,.
\ee
Hence, the distribution function of a charged conjugated system ${\cal C}|\Psi \rangle$ is related to the distribution function of the original state $|\Psi\rangle$ by 
\be\label{frsfrs}
\widetilde{f}_{rs}(\gr{p}) = f_{rs}(\gr{p})\,.
\ee
The corresponding spinor valued operators can be related using
\be\label{usCvs}
u_{s,\mathfrak{a}}(\gr{p}) = C_{\mathfrak{a}\mathfrak{b}} \bar v^\mathfrak{b}_s(\gr{p})\,\qquad \bar
u_s^\mathfrak{a}(\gr{p}) =C^{-1 \mathfrak{a}\mathfrak{b}} v_{s,\mathfrak{b}}(\gr{p})\,,
\ee
where in the chiral representations the components of
$C_{\mathfrak{a}\mathfrak{b}}$ are antisymmetric and such that
\be
C = (\ii \gamma^0 \gamma^2) \qquad C^{-1} = -(\ii \gamma^0 \gamma^2)\,.
\ee
From (\ref{frsfrs}) and (\ref{usCvs}) and the definitions (\ref{Fspinor}) of the spinor valued operators we get
\be
{\widetilde{F}_{\mathfrak{a}}}^{\,\,\mathfrak{b}}(p) = -
C_{\mathfrak{a}\mathfrak{c}}
{F_\mathfrak{d}}^\mathfrak{c}(p)C^{-1\mathfrak{d}\mathfrak{b}} \qquad
\Rightarrow \qquad \widetilde{\gr{F}}(p) = -C \gr{F}^T(p) C^{-1}\,.
\ee
Since charge conjugation is not a special case of Lorentz transformations, it does not take the form (\ref{WonderfulTransformation}).
Using the properties (see G.1.24 of \cite{BibleSpinors})
\be\label{CgammaC}
C \gr{\Gamma}^T C^{-1} = \eta_\Gamma^C \gr{\Gamma}
\,,\qquad \eta_\Gamma^C= \begin{cases}+1,\qquad {\bm \Gamma} =
  \mathds{1},\gamma^5,\gamma^\mu\gamma^5\\-1,\qquad {\bm \Gamma}=\gamma^\mu,\Sigma^{\mu\nu}\end{cases}
\ee
we deduce that $\widetilde{\gr{F}}(p)$ is deduced from $\gr{F}(p)$ by
a global sign change and the replacement $p^\mu \to -p^\mu$, which is
also equivalent to $m \to -m$, as already noticed in \S~\ref{SecSpinorValued}.

\section{Kinetic theory for Fermions in curved spacetime}\label{SecLiouville}

\subsection{Liouville equation}

In order to derive a Liouville equation in curved space-time which describes the evolution of the covariant components, we must distinguish between the massive and the massless case.

\subsubsection{Massive fermions}

In the previous sections we have shown that $I$ and ${\cal Q}$ are observer independent. In addition, in the local Minkowski frame, they are also parallel transported in the absence of collisions. The helicity of particles does not change in free propagation and, considering that the momentum $p^\mu$ is conserved, the vectors $\epsilon_\pm^\mu$ and $S^\mu$ used to build the quantities $I$ and ${\cal Q}$ remain unchanged. Hence, in the local Minkowski space we obtain the equations of motion
\be\label{eq:Minkowskievo}
\frac{\dd I}{\dd t} = 0\,, \qquad \frac{\dd {\cal Q}^\mu}{\dd t} = 0\,.
\ee

From the point of view of general relativity, these equations are only valid locally and neglect entirely the impact of the relativistic space-time. The intensity $I$ describes the total number of particles. The conservation of $I$ in the absence of collisions in Eq.~(\ref{eq:Minkowskievo}) is equivalent to mass or particle number conservation. The geometrical impact of general relativity does not change the number of particles and we may generalise the equation of motion by requiring the conservation of $I$ along a full geodesic
\be\label{DIDlambda}
\frac{DI}{D\lambda} = 0 \,,
\ee
where $\frac{D}{D\lambda}$ is the derivative along the particle trajectory parameterized by $\lambda$.

The vector ${\cal Q}$ is parallel transported in the local space-time and describes the polarisation of particles in an observer-independent way. Again, the geometrical nature of general relativity does not change the polarisation of particles and we require that ${\cal Q}$ is parallel transported along the non-trivial trajectory of the particles. 
Note that the observer dependant linear and circular polarisation may change non-trivially during the transport and require a specification of the dynamics of the observer.

Using the observer-independence, we are able to uniquely define the vector ${\cal Q}$ on our full space-time by employing the tetrads
\be\label{QalphaQmu}
{\cal Q}^\alpha = {\cal Q}^\mu [e_\mu]^\alpha  \,,
\ee
where we remind that the index $\mu$ is a tetrad component index, but the index $\alpha$ is a general coordinate index. Assuming parallel transport, we obtain the equation of motion
\be\label{DQDlambda}
\frac{D{\cal Q}^\alpha}{D\lambda} = 0 \,.
\ee
Note that ${\cal Q}$ is by definition orthogonal to the momentum $P$. This property is automatically conserved in the relativistic evolution as both, $P$ and $\cal{Q}$ are parallel-transported along the geodesic of a free particle. 
  
The variation of coordinates along the trajectory is given by $\dd x^\alpha/ \dd \lambda = P^\alpha$. The derivatives along the trajectory can be expressed in terms of the time-coordinate $\tau = x^0$ using $\dd \tau/\dd \lambda = P^0$, and using the dependencies $I = I(\tau,\gr{x},\gr{p})$ we find
\be\label{DIDtau}
\frac{\partial I(\tau,\gr{x},\gr{p})}{\partial\tau} + \frac{P^i}{P^0}\frac{\partial I(\tau,\gr{x},\gr{p})}{\partial x^i } + \frac{\dd p^i}{\dd\tau}\frac{\partial I(\tau,\gr{x},\gr{p})}{\partial p^i } =  0 \,,
\ee
where we stress that the momentum appearing in the intensity $I(\tau,\gr{x},\gr{p})$ is the local 3-momentum $\gr{p}$ whose components are expressed in the local tetrad basis of the observer. For the polarisation vector we obtain
\be\label{DQDtau}
\frac{\partial {\cal Q}^\alpha(\tau,\gr{x},\gr{p})}{\partial\tau} + \frac{P^i}{P^0}\frac{\partial {\cal Q}^\alpha(\tau,\gr{x},\gr{p})}{\partial x^i } + \frac{\dd p^i}{\dd\tau}\frac{\partial {\cal Q}^\alpha(\tau,\gr{x},\gr{p})}{\partial p^i } + \Gamma^{\alpha}_{\gamma \beta} \frac{P^\gamma}{P^0} {\cal Q}^\beta(\tau,\gr{x},\gr{p})=  0 \,,
\ee
where the $\Gamma^{\alpha}_{\gamma\beta}$ are the Christoffel symbols associated with the metric. These are the usual relativistic Boltzmann equations, with the second term describing the variation of the distribution function due to particle propagation and the third term describing the impact of a change in the local momentum (e.g. due to redshifting related to the expansion of the Universe). For the propagation of the polarisation vector, we find an additional geometrical term from the non-trivial parallel transport of vectors in curved spaces.

Finally, the change of the local momentum $\dd p^i/\dd\tau$ is obtained from the geodesic equation
\be
\frac{D P^\alpha}{D \lambda}=\frac{\dd P^\alpha}{\dd \lambda} + \Gamma^{\alpha}_{\gamma\beta} P^\beta P^\gamma = 0 \,\quad\Rightarrow\quad\com{\frac{\dd (p^\mu [e_\mu]^\alpha)}{\dd \tau} = -  \Gamma^{\alpha}_{\gamma\beta}\frac{P^\beta P^\gamma}{P^0}}  \frac{\dd p^\mu}{\dd \tau}= P^\alpha\frac{\dd [e^\mu]_\alpha}{\dd \tau} - [e^\mu]_\alpha \Gamma^{\alpha}_{\gamma\beta}\frac{P^\beta P^\gamma}{P^0}\,.
\ee

\subsubsection{Massless fermions}

In the massless case, linear polarisation and circular polarisation must be considered separately. Circular polarisation $V$ is transported exactly like the intensity $I$ in Eqs. (\ref{DIDlambda}) and (\ref{DIDtau}) because the direction of the helicity vector is identical to the momentum and therefore parallel-transported. However the linear polarisation vector (considered in general coordinates with $Q^\alpha = Q^\mu [e_\mu]^\alpha$) cannot be parallel transported because it is transverse to both the momentum $P^\alpha$ and the observer velocity $u^\alpha$, and the latter is not (necessarily) parallel transported. In the process of free streaming, any variation of $Q^\alpha$ in the direction of $P^\alpha$ is not physical as explained at the end of \S~(\ref{SecCovariantF}). Hence this unphysical degree of freedom must be eliminated by an appropriate projection so as to obtain an unambiguous equation for parallel transport. To that purpose, we use the screen projector (\ref{DefScreenH}) in general coordinates. 
\be\label{LiouvilleQmassless}
{{\cal H}^\beta}_\alpha \frac{D Q^\alpha}{D \lambda} = 0\,.
\ee
The transport of linear polarisation in the massless case is the same as the transport of the full polarisation vector in the massive case [Eqs. (\ref{DQDlambda}) and (\ref{DQDtau})], up to an additional screen projection which ensures that the double transverse property holds. This is similar to the parallel transport of linear polarisation for photons~\cite{Challinor:1999zj,Challinor:2000as,Tsagas:2007yx,Pitrou2008} which in that case is described by a doubly projected tensor (see App. \ref{AppBosons} for more details about the statistical description of vector bosons).
We remark that the coset $[Q^\alpha]$ is paralell transported.
\subsection{Angular decomposition and multipoles}\label{SecYlm}

The intensity $I(\gr{p})$ is easily decomposed into spherical
harmonics. Indeed, once an observer choice is made, that is its
four-velocity $u^\alpha$ is identified with the time-like vector of the
tetrad $[e_0]^\alpha$, we can define the
spatial momentum $\gr{p}$ and its direction unit vector
$\hat{\gr{p}}$. We then perform the usual spherical harmonics
decomposition
\be\label{Ilm}
I(\gr{p}) = \sum_{\ell m} I_{\ell m}(|\gr{p}|) Y_{\ell m}(\hat{\gr{p}})\,.
\ee
Alternatively one could utilise a decomposition based on symmetric
trace-free tensors which is equivalent ~\cite{Thorne1980,Pitrou2008}.

For the polarisation vector ${\cal Q}^\mu$ we remind ourselves of its decomposition as 
\be
Q^\mu(\gr{p}) =Q_+(\gr{p}) { \epsilon}^\mu_+(\gr{p}) + Q_-(\gr{p}) { \epsilon}^\mu_-(\gr{p}) + V(\gr{p}) S^\mu(\gr{p})\,.
\ee
For the angular decomposition we have to pay attention to the transformation properties when performing a spatial rotation of the coordinate system around the direction of $\hat{\gr{p}}$. The ordinary spherical harmonics, when evaluated at $\hat{\gr{p}}$ do not transform under this rotation and are thus not suitable to decompose objects which have a non-trivial transformation under this rotation. 
The polarisation vector $\cal Q$ transforms as an ordinary 4-vector (we have shown that it is observer-independent). 
However, this is not the case for the observer-dependent vectors and distribution functions used to build $\cal Q$.
The vector in direction of the spatial momentum $S^\mu$ is invariant under this particular rotation as it points in the direction $\hat{\gr{p}}$. \com{and corresponds to the spin $0$ part of an ordinary vector.} Employing the observer-independence of $\cal Q$ we therefore conclude that $V$ must be invariant under this rotation and may be decomposed into ordinary spherical harmonics. The polarisation vectors ${\epsilon}_{\pm}(\hat{\gr{p}})$ however transform with an additional spin $\mp 1$ complex rotation. To generate an observer-independent $\cal Q$ the corresponding $Q_{\pm}$ must transform with the opposite spin and they are decomposed into  spin-weighted spherical harmonics $Y^s_{lm}$~\cite{Goldberg1967} as
\be
Q_+(\gr{p}) \equiv \sum_{\ell m} Q^+_{\ell m}(|\gr{p}|) Y_{\ell
  m}^+(\hat{\gr{p}}) \,,\quad  Q_-(\gr{p}) \equiv Q^-_{\ell m}(|\gr{p}|) Y_{\ell
  m}^-(\hat{\gr{p}})\,.
\ee
Note that this discussion only concerns the observer dependence under a specific spatial rotation and that due to the definition of helicity an additional dependence mixing $V$ and $Q_{\pm}$ exists for more general rotations and boosts. 

$E$ and $B$ modes multipoles \com{, that are invariant under the subclass of spatial rotations around $\hat{p}$,} can be defined from $Q^\pm_{\ell m}
\equiv \mp (E_{\ell m} \pm
\ii B_{\ell m})$. Equivalently since $\gr{Q}(\gr{p})$ is a vector field
on the unit sphere in momentum space, it can be decomposed as the
gradient and the curl of two scalar functions as 
\be\label{twoPotentials}
Q_i(\gr{p}) = D_i
{\cal E}(\gr{p}) + \hat{p}^j{\epsilon_{ji}}^k D_k {\cal B}(\gr{p})\,,
\ee
where $D_i$ is the covariant derivative on the unit sphere and
$\hat{p}^j{\epsilon_{ji}}^k$ is the Levi-Civita tensor on the unit sphere. Decomposing the scalar functions ${\cal E}$ and ${\cal B}$ in
multipoles ${\cal E}_{\ell m}$ and ${\cal B}_{\ell m}$  as in the expansion (\ref{Ilm}) and using~\cite{DurrerBook}
\be
D^i Y_{\ell m} = \sqrt{\frac{\ell(\ell+1)}{2}}\left(-Y^+_{\ell m}
\epsilon_+^i + Y^-_{\ell m} \epsilon_-^i \right)
\ee
the two possible definitions for the $E$ and $B$ modes multipoles are
related by $E_{\ell m} = \sqrt{\ell(\ell+1)/2} {\cal E}_{\ell m}$ and $B_{\ell m} = \sqrt{\ell(\ell+1)/2} {\cal B}_{\ell m}$.
Again a similar expansion can be obtained by using symmetric trace-free tensors to expand the scalar functions ${\cal E}$ and
${\cal B}$ directly in Eq. (\ref{twoPotentials}).

\section{The Boltzmann equation}\label{SecBoltzmann}

\subsection{Time evolution and collisions}

So far we have discussed the free propagation of fermions. When in addition considering collisions, we will employ a separation of scales. We assume that the relativistic evolution is dominant on macroscopic scales, while individual collisions act on microscopic scales. We therefore may compute the collision term in the local tanget space corresponding to special relativity. Then averaging over the local Minkowski space-time of the observer we will provide an effective collision term for the relativistic evolution of the distribution functions.

We therefore introduce three separate scales, the microscopic scale of individual interactions, typically the Compton timescale of interacting particles. Then a mesoscopic scale over which we average the individual collisions, define our local distribution functions and describe the impact of the collisions on the averaged fluid. Finally, the macroscopic scale on which particles free stream on general relativistic geodesics. 

We begin with the description of collisions in the local frame of our observer. The full Hamiltonian $H$ can be separated into a free part $H_0$ and an interaction part $ H_{\rm I}$. We employ the Heisenberg picture in which the states are time-independent. The time evolution of our distribution function is given by (omitting to specify the momentum dependence of $f_{rs}$ and $N_{rs}$ for simplicity)
\be\label{dfdtfromdNdt}
\deltarel(0)\frac{\dd}{\dd t}f_{rs} = \langle \Psi | \frac{\dd N_{rs}}{\dd t}  | \Psi \rangle = \ii \langle \Psi |  [H, N_{rs}] |\Psi\rangle = \ii \langle \Psi |  [H_{\rm I}, N_{rs}] |\Psi\rangle \,,
\ee
where in the last identity we have used that helicity is conserved in the absence of collisions since $[H_0,N_{rs}] = 0$. We find a differential equation for the operator $N_{rs}$ and are able to write an approximate solution as closed integration if we restrict ourselves to a given order in the interaction Hamiltonian. We define the operator $N_{rs}^{(0)}$ characterising the ingoing states prior to the collision. To first order in the interaction we obtain
\be\label{eq:firstorder}
N_{rs}(t) = N_{rs}^{(0)} + i\int \limits_{0}^{t} \dd t' [H_{\rm I}(t'),N_{rs}^{(0)}]\,.
\ee
The interpretation of this equation is that at the time $t=0$ the system is starting to interact, but as the background does not yet contain any correlations between the interacting species we can still evaluate the collisions using the zeroth order number operator. This first order solution describes forward scatterings and we need to go to second order to find the first non-forward interactions.

We insert the first order solution~(\ref{eq:firstorder}) into Eq.~(\ref{dfdtfromdNdt}) and find to second order
\be
\frac{\dd N_{rs}}{\dd t} = \ii [H_{\rm I}(t), N_{rs}(t)] \approx \ii [H_{\rm I}(t), N_{rs}^{(0)}] - \int \limits_{0}^{t} \dd t' [H_{\rm I}(t), [H_{\rm I}(t'),N_{rs}^{(0)}]]\,.
\ee
The second order contribution describes an interaction which is active between the time $t'$ and $t$. We identify this timescale with our microscopic timescale $t_{\rm mic} = t-t'$, quantifying the timescale of individual particle interactions. The averaged fluid however does not change significantly on this timescale and evolves on the much larger mesoscopic time-scale $t_{\rm mes} \gg t_{\rm mic}$. Since we compute the derivative of the number operator $N$ with respect to the time $t$ we may identify the mesoscopic time $t_{\rm mes} = t$. 
Expressed in these parameters we obtain
\be
\frac{\dd N_{rs}}{\dd t_{\rm mes}} = \ii [H_{\rm I}(t_{\rm mes}), N_{rs}^{(0)}] - \int \limits_{0}^{t_{\rm mes}} \dd t_{\rm mic} [H_{\rm I}(t_{\rm mes}), [H_{\rm I}(t_{\rm mes} - t_{\rm mic}),N_{rs}^{(0)}]]
\ee
Ideally we would like to evaluate this equation at the initial time and set $t_{\rm mes} =0$ to compute the change of our initial states under the considered interactions. 
This choice however is mathematically inconsistent as we are mixing mesoscopic and microscopic timescales in the integration. Instead we average the resulting time-derivative, considering the time-reversal symmetry, over a box that is centered on the initial time and has a length of $2\epsilon$ which is chosen to be small compared to the scale of macroscopic evolution.
\be
\frac{\dd N_{rs}(0)}{\dd t}\Big|_{\rm classical} \equiv \ii [H_{\rm I}(0), N_{rs}^{(0)}] - \int \limits_{-\epsilon}^{\epsilon} \frac{\dd t_{\rm mes}}{2\epsilon}{\rm sgn}(t_{\rm mes})\int \limits_{0}^{t_{\rm mes}} \dd t_{\rm mic} [H_{\rm I}(t_{\rm mes}), [H_{\rm I}(t_{\rm mes}-t_{\rm mic}),N_{rs}^{(0)}]]\,.
\ee
We may split this integration into three regions. First, the central region $\epsilon \approx t_{\rm Compton}$, where $t_{\rm Compton}$ the typical Compton time scale of particles, is highly non-trivial, but this region is negligible compared to our entire integration volume. In the remaining positive and negative regions the integrand is constant in time. The reason is that the integral over the microscopic time already has sufficient support and is converged. The remaining time-dependence based on the mesoscopic time $t_{\rm mes}$ is not relevant as we have chosen the box small compared to the mesoscopic evolution and we may now set $t_{\rm mes} = 0$ yielding
\be
\frac{\dd N_{rs}(0)}{\dd t}\Big|_{\rm classical} \equiv  \ii [H_{\rm I}(0), N_{rs}^{(0)}] - \frac{1}{2}\int \limits_{-\epsilon}^{\epsilon} \dd t_{\rm mic} [H_{\rm I}(0), [H_{\rm I}(t_{\rm mic}),N_{rs}^{(0)}]] \,.
\ee
Finally we may extend the integration limit $\epsilon$ to infinity compared to the microscopic evolution using a separation of scales.

We note that the interaction Hamiltonian appearing in this equation may always be evaluated based on the non-interacting field value as we only utilise times which are small compared to the mesoscopic time.
Our expression is equivalent to those used in Refs. (\cite{1993NuPhB.406..423S,Kosowsky:1994cy,Beneke2010}).

We finally deduce from Eq. (\ref{dfdtfromdNdt}) that the classical evolution of the distribution function is given by
\com{\be
\deltarel(0)\frac{\dd}{\dd t}f_{rs} = \frac{\dd}{\dd t}\langle \Psi | N_{rs} | \Psi \rangle = \ii \langle \Psi_i |  [H_I(t), N_{rs}] |\Psi_i\rangle  - \langle \Psi_i | \int_0^{t} \dd t' [H_I(t'),[H_I(t), N_{rs}]] | \Psi_i\rangle\,.
\ee}
\be\label{QuantumBoltzmann}
\deltarel(0){\frac{\dd f_{rs}(t)}{\dd t}} = \ii \langle \Psi(t) |
[H_I(t), N^{(0)}_{rs}] |\Psi(t)\rangle -\frac{1}{2} \langle \Psi(t) | \int_{-\infty}^{\infty} \dd t_{\rm mic} [H_I(t),[H_I(t+t_{\rm mic}), N^{(0)}_{rs}]] |\Psi(t)\rangle\,,
\ee
where the first term on the rhs is the forward scattering term. It is responsible for refractive effects or flavor oscillations in matter (see Refs. \cite{Lesgourgues:2006nd,Lesgourgues:2012uu} for neutrino oscillations in cosmology) such as the MSW effect~\cite{1978PhRvD..17.2369W,1986NCimC...9...17M,Marciano:2003eq,1993NuPhB.406..423S}). The second term is the collision term and we define
\be\label{eq:defcoll}
\deltarel(0) C[f_{rs}(t)]\equiv -\frac{1}{2} \langle \Psi(t) | \int_{-\infty}^{\infty} \dd t_{\rm mic} [H_I(t),[H_I(t+t_{\rm mic}), N^{(0)}_{rs}]] |\Psi(t)\rangle
\ee
such that the Boltzmann equation (\ref{QuantumBoltzmann}) (when neglecting forward scattering and restoring the notation of the momentum dependence) is written
\be\label{QuantumBoltzmann2}
\deltarel(0) \frac{\dd f_{ss'}(t,p)}{ \dd t } = \deltarel(0) C[f_{ss'}(t,p)]\,.
\ee
A spinor space operator associated with this collision term is obtained by contraction with $u_{s'}(p) \bar u_s(p)$ (or $v_{s}(p) \bar v_{s'}(p)$ for antiparticles) as in Eq. (\ref{Fspinor}), and we define
\be\label{DefCollisionOperator}
C[\gr{F}(t,p)] \equiv \begin{cases}\sum_{s\,s' }C[f_{ss'}(t,p)]u_{s'}(p) \bar u_s(p)\,,\qquad {\rm particles}\\\sum_{s\,s' }C[f_{ss'}(t,p)]v_{s}(p) \bar v_{s'}(p)\,,\qquad\, {\rm antiparticles}\,.\end{cases}
\ee
The covariant parts of this spinor space collision operator, $I_C(p)$ and ${\cal Q}_C^\mu(p)$ are obtained exactly like in Eq. (\ref{BigDecompositionGood}). In the massless case the covariant parts are $I_C(p)$, $V_C(p)$ and ${Q}_C^\mu(p)$ and are obtained as in Eq.  (\ref{BigDecompositionGoodMassless}).

The classical Boltzmann equation is obtained when considering that this derivation, which has been made for a homogeneous system (see \S~\ref{SecNrs}), is in fact valid locally. That is in the derivation we assumed that the distribution function depends on time and momentum only $f_{rs}(t,p)$, but we now assume that it also depends on the position and employ $f_{rs}(t,\gr{x},p)$. This amounts to considering that there is a mesoscopic length scale under which the system can be considered as homogeneous such that the volume integral in the Hamiltonian $H_I=\int \dd^3 \gr{x} {\cal H}_I$ can be extended to infinity in the computation of the local collision term $C[f_{ss'}(t,\gr{x},p)]$. Expressed in terms of spinor valued operators the classical Boltzmann equation reads 
\be\label{dgrFdt}
\frac{\dd \gr{F}(t,\gr{x},p)}{ \dd t } =C[\gr{F}(t,\gr{x},p)]\,.
\ee

\subsection{General relativity and the classical Boltzmann equation in curved space-time} \label{sec:grcoll}

We connect the collision term derived in the mesoscopic Minkowski space-time to the macroscopic relativistic evolution. We assume that collisions are well described in special relativity and general relativistic corrections may be neglected. Under our assumptions, the covariant parts of the local spinor space collision operator (\ref{DefCollisionOperator}), do also depend on the position of space [e.g. we should write $I_C(t,\gr{x},p)$]. Using $\dd t /\dd \lambda = p^0=E$, the derivative $E \dd I/\dd t $ is identified with $D I / D \lambda$ of the parallel transport in equation (\ref{DIDlambda}), but now the effect of collisions is taken into account by the intensity part of Eq. (\ref{dgrFdt}), $\dd I/\dd t=I_C$. This means that we connect the time in the local inertial frame, in which we compute the collisions, with the time along our general relativistic geodesic. 

Similarly, expressing the polarisation part of the collision term in general coordinates as in Eq. (\ref{QalphaQmu}), that is with
\be
{\cal Q}^\alpha_C = {\cal Q}^\mu_C [e_\mu]^\alpha\,,
\ee
we identify $E \dd {\cal Q}^\alpha/\dd t $ with $D {\cal Q}^\alpha / D \lambda$ of the parallel transport equation (\ref{DQDlambda}) and we find
\be\label{BoltzmannFermions}
\frac{D I}{D \lambda} = E \,I_C\,,\qquad \frac{D Q^\alpha}{D \lambda} = E \,Q^\alpha_C\,.
\ee
This is the general relativistic Boltzmann equation for fermions, needed to compute the effect of both free streaming and collisions on the distribution function of fermions. When computing the collision term for particular examples in \S~\ref{SecWeak}, it is convenient to present the results with a prefactor $1/E$ so that the rhs of Eq. (\ref{BoltzmannFermions}) can be readily obtained.

When expressed in functions of the general time coordinate $\tau$, we just use $\dd \tau/\dd \lambda = P^0$ to deduce that the rhs of Eqs. (\ref{DIDtau}) and (\ref{DQDtau}) are respectively $E/P^0 I_C$ and $E/P^0 Q^\alpha_C$ when collisions are to be considered.

\subsection{Flavor description and flavor oscillations}

We will now consider the case of flavoured particles, following Ref. \cite{1993NuPhB.406..423S}. We add flavour indices ($i,j = 1 , \dots , n_{\rm flavour}$) to our creation and annihilation operators $a_{s}^{i}$ and the distribution function $f_{ri\,sj}$ is obtained from $a^{\dagger i}_{r} a^{j}_{s}$ following the same procedure as in \S \ref{SecNrs}. For simplicity of notation, we omit the spin indices $r,s$ in this section. These flavour-states are chosen to be diagonal in the interactions, but the corresponding mass matrix may not be diagonal and thus the flavour states are no longer conserved in free propagation. 

In our approach these oscillations are represented by forward scatterings. The free Hamiltonian takes the form
\be
H_0 = \int [\dd p] \sum_{ij}\sum_s a^{i \dagger}_s(p) \Omega_{ij} a^j_s(p) + b^{i \dagger}_s(p) \Omega_{ij} b^j_s(p) \,,
\ee
with $\Omega_{ij} = {\sqrt{{\bm p}^2 + M^2}}_{ij}$ a symmetric matrix and $M_{ij}$ the neutrino mass matrix. 
In the absence of interactions we find\footnote{Note that in Eq. (2.5) of Ref. \cite{1993NuPhB.406..423S} there is an extra minus sign for particles compared to antiparticles because the definition of the distribution function is the transposed of our definition $f_{rs}$.}
\be\label{eq:firstorderFlavors}
\frac{\dd N^{(0)}_{ij}(t)}{\dd t} =\ii  [H_{\rm 0}(t),N^{(0)}_{ij}]\,.
\ee
where the commutator between the free Hamiltonian and the number operator does not vanish. We explictly obtain 
\be
\frac{\dd f_{ij}}{\dd t} = \ii \sum_k\left(\Omega_{ik} f_{kj} - f_{ik}\Omega_{kj} \right) \,.
\ee
When considering both interactions and flavour oscillations, we assume a separation of scales between the flavour oscillations, relevant on the mesoscopic scale, and the collisions on the microscopic scale. This means that we allow for flavour oscillations in-between two individual interactions, but not during one single collision.
This assumption allows us to write up to second order in collisions
\begin{eqnarray} \nonumber
\deltarel(0){\frac{\dd f_{ij}(t)}{\dd t}} &=& 
\ii \langle \Psi(t) |
[H_0(t), N_{ij}] |\Psi(t)\rangle
+\ii \langle \Psi(t) |
[H_I(t), N^{(0)}_{ij}] |\Psi(t)\rangle 
\\&&
-\frac{1}{2} \langle \Psi(t) | \int_{-\infty}^{\infty} \dd t_{\rm mic} [H_I(t),[H_I(t+t_{\rm mic}), N^{(0)}_{ij}]] |\Psi(t)\rangle\,.
\end{eqnarray}
Defining the collision term as in Eq.~(\ref{eq:defcoll}) and neglecting forward scattering induced by the interaction Hamiltonian, we obtain the classical Boltzmann equation 
\be\label{dfijdtflavor}
\frac{\dd f_{ij}(t,p)}{ \dd t } =\ii \sum_k\left(\Omega_{ik}(p) f_{kj}(t,p) - f_{ik}(t,p)\Omega_{kj}(p) \right) + C[f_{ij}(t,p)]\,.
\ee
Note that only the spatial momentum is conserved (and not energy of
particles, because they change from one mass shell to another one) by the flavor changing term but we must not forget that when we consider a distribution function $f_{ij}$ we must extract its energy with $\sum_{ij}\Omega_{ij} f_{ij}$ and it is easily seen that this is conserved by the flavor changing term. To promote the Boltzmann equation to general relativity we follow the recipe presented in the previous sections. We construct the observer-independent quantities $I_{ij}$ and ${\cal Q}_{ij}$ from the distribution function $f_{ij}$ and connect them to general relativistic manifold by multiplying them with the tetrad. The flavour oscillations can be treated in the same way as the collision term. They are well described on the mesoscopic scale, independent from the geometrical corrections of general relativity relevant on the macroscopic scale.  As particles are propagating on the free geodesics of general relativity, the flavour of the particles is mixed. We therefore connect the time measured by the local observer describing the flavour oscillations to our relativistic time coordinate and, similarly to section~\ref{sec:grcoll} we find that the factor $E/P_{0}$ must be added in front of the flavour oscillations term [the first term on the rhs of Eq. (\ref{dfijdtflavor})] before incorporating it to the rhs of the Boltzmann equations (\ref{DIDtau}) and (\ref{DQDtau}). 

\section{Collisions mediated by weak interactions}\label{SecWeak}

In order to illustrate how the previous formalism should be
implemented in practice, we will focus on weak-interactions, and more
precisely on their low-energy limit. Since this has the advantage of involving
only fermions it is very well suited for the formalism introduced in
this article.

\subsection{General form of weak-interactions}

All weak interaction take the form of current-current interactions~\cite{Nachtmann1991} at low energy (low compared to the $W^\pm$ and $Z$ masses), that is they are given by
\be\label{StructureHI}
{\cal H}_I = -{\cal L}_I = \sgnslash\frac{G_F}{\sqrt{2}}\left[J^{\rm NC}_\mu
  J_{\rm NC}^\mu + J^{{\rm CC} \dagger}_\mu J_{\rm CC}^\mu
\right]\,,
\ee
where $G_F\simeq 1.17 \,\times\, 10^{-5}\,{\rm GeV}^{-2}$ is the Fermi constant of weak interactions.

\subsubsection{Neutral currents}

Neutral currents describe the exchange of $Z$ bosons and as these are not charged they mediate elastic scatterings that do not alter the involved types of particles and only transfer momentum, spin and energy. 

The neutral current is simply the sum of the neutral currents of all
particles undergoing weak interactions
\be\label{StructureJNC}
J_{NC}^\mu = J^\mu_{ee} + J^\mu_{\nu\nu} +\dots\,.
\ee
For neutrinos, the neutral current couples only the left chiralities and, noting ${\bm \nu}$ the neutrino quantum field, it is simply
\be\label{JNCneutrinos}
 J^\mu_{\nu\nu} \equiv e^\nu_- \,\bar{\gr{\nu}} \gamma^\mu(\mathds{1}-\gamma_5)\gr{\nu}\,,\qquad e^\nu_- \equiv \frac{1}{2}\,.
\ee
\com{\CF{we have not defined $\gr{\nu}$, $e^\nu_-$ etc.. do you think it is standard enough so that everyone knows what we mean? }}
with similar expressions for other flavors. However for electrons and
(similarly pions and taus) the neutral currents must be further
decomposed into left and right chiral interactions as
\be\label{JNCelectrons}
J^\mu_{ee} = \epsilon^e_-  J^{-\,\mu}_{ee}+ \epsilon^e_+
J^{+\,\mu}_{ee}\,,\qquad J^{-\,\mu}_{ee} \equiv \bar{\gr{e}}
\gamma^\mu(\mathds{1}-\gamma_5)\gr{e} \qquad  J^{+\,\mu}_{ee} \equiv \bar{\gr{e}} \gamma^\mu(\mathds{1}+\gamma_5)\gr{e} \,,
\ee
where we noted ${\bm e}$ the electronic quantum field. The chiral coupling constants are for electrons
\be\label{ChiralCouplings}
\epsilon^e_- \equiv -\frac{1}{2} + \sin^2\theta_W \,,\qquad \epsilon^e_+ \equiv \sin^2\theta_W \,,
\ee
with $\theta_W$ the Weinberg angle ($\sin^2 \theta_W \simeq 0.23$).

\subsubsection{Charged currents}

Opposed to the neutral currents, the charged currents describe the exchange of charged $W$-bosons and therefore are inelastic.  
The structure of the charged current is more complex since it couples eigenmass states of different flavors, thanks to the
Cabbibo-Kobayashi-Maskawa (CKM) matrix for quarks or the
Pontecorvo-Maki-Nakagawa-Sakata (PMNS) matrix for massive neutrinos. We ignore these complications for the examples that we shall consider and employ effective charged currents for the neutron/proton pair
which is involved in beta decays and related processes, and the
charged currents of the first two lepton flavors, that is of the
electron/neutrino and muon/muon neutrino pairs.
\bea
J_{CC}^\sigma &=& \cos \theta_C  J^\sigma_{pn} +  J^\sigma_{e \nu} +
J^\sigma_{\mu \nu_\mu }\,,\\
\eea
where $\cos \theta_C = 0.98$ is a Cabbibo-Kobayashi-Maskawa (CKM) angle.

The charged currents for electron/neutrino and muon/muon neutrino pairs
are coupling only the left chiralities
\be\label{JCCdetail}
J^\sigma_{e \nu} \equiv  \bar{\bm \nu}\gamma^\sigma
(\mathds{1}-\gamma^5) {\gr{e}}\,,\qquad  J^\sigma_{\mu \nu_\mu} \equiv
\bar{\bm \nu}_\mu\gamma^\sigma (\mathds{1}-\gamma^5) {\gr{\mu}}\,.
\ee
However, due to internal QCD effects, the coupling in the proton/pair
is not purely left chiral. The deviation from left chirality of the
coupling is parameterized by the $g_A$ parameter whose measured value
is approximately $1.25$~\cite{Nachtmann1991} and the corresponding charged current reads
\be\label{JCCpn}
J^\mu_{pn} \equiv \bar{\gr{p}} \gamma^\mu (\mathds{1}- g_A \gamma^5) \gr{n}\,.
\ee

When considering the cumulative effect of neutral currents and
charged currents, we can use the Fierz identities which for anticommuting fields give~\cite{Sarantakos:1982bp,1993NuPhB.406..423S}
\be\label{CrossingSymmetry}
 J^{\dagger\mu}_{e\nu} {(J_{e\nu})}_\mu\ = J^{-\,\mu}_{ee} {(J_{\nu\nu}^{-})}_\mu \,.
\ee
This means that the effect multiple charged currents can be replaced by equivalent neutral currents. In the collision term we may therefore replace the charged currents by modifying the neutral chiral coupling factors  (\ref{ChiralCouplings}), yielding
\be\label{epsilonmoinsplusun}
\epsilon^e_- \to \epsilon^e_- +1\,. 
\ee

\subsection{General current/current interaction}

For any reaction involving weak-interactions, a visual inspection of the interaction Hamiltonian is
sufficient to deduce which currents are involved in the process. In
all cases this amounts to considering the current-current coupling between four species $a,b,c,d$ given by
\be\label{HIabcd}
{\cal H}_I = -{\cal L}_I =\sgnslash g\left( J^{ac}_\mu  J^\mu_{bd}+ {\rm cc} \right)\,,
\ee
where, depending on the interaction, the same species may be represented by multiple indices. The chiral contributions of
these currents are parameterized by $\epsilon^{ac}_\pm$ and
$\epsilon^{bd}_\pm$ as
\be\label{GeneralCurrents}
 J^\mu_{ac} =\overline{\gr{\psi}}_c {\cal \chi}^\mu_{(ac)}\gr{\psi}_a \,, \qquad\qquad 
J^\mu_{bd} =\overline{\gr{\psi}}_d {\cal \chi}^\mu_{(bd)} \gr{\psi}_b\,,
\ee
with the notation
\bea\label{Omu}
{\cal \chi}^\mu_{(ac)} &\equiv& 2 \sum_{s=\pm 1}\epsilon^{ac}_s \gamma^\mu P_s = \epsilon^{ac}_+ \gamma^\mu(\mathds{1}+\gamma^5) + \epsilon^{ac}_- \gamma^\mu(\mathds{1}-\gamma^5)\\
{\cal \chi}^\mu_{(bd)} &\equiv& 2 \sum_{s=\pm 1}\epsilon^{bd}_s \gamma^\mu P_s = \epsilon^{bd}_+ \gamma^\mu(\mathds{1}+\gamma^5) + \epsilon^{bd}_- \gamma^\mu(\mathds{1}-\gamma^5)\,.
\eea

We now study the structure of the collision term due the general
interaction Hamiltonian (\ref{HIabcd}) with the currents (\ref{GeneralCurrents}), and we apply it further to specific cases, which all correspond to a choice of the species
$a,b,c,d$ together with the couplings  $\epsilon^{ac}_\pm$ and
$\epsilon^{bd}_\pm$ and the coupling constant $g$.

Let us investigate the total collision term for the particle species $a$
due to this interactions, which is represented by a sum of individual collision terms. First there is the decay of particle $a$ corresponding to the reaction
\be
a \to \bar b + c + d
\ee where we recall that $\bar b$ is the antiparticle species related to the particle
species $b$. We may typically neglect the three body reactions that would revert this decay, as they are only relevant in very high density environments. In addition to the decay, we have to consider the
two-body reactions
\bea
a+b &\leftrightarrow& c+d\\
a+\bar c &\leftrightarrow& \bar b+d\\
a+\bar d &\leftrightarrow& c+\bar b\,.
\eea
However all the related collision terms can be deduced from the one of $a+b \leftrightarrow c+d$ through crossing symmetry and we discuss this
procedure in the next section. Here we focus only on this specific reaction.

\subsection{General  method for the generic process $a+b \leftrightarrow c+d$} \label{SecGeneralabcd}

Let us introduce some compact notation with the multi-indices
\be
\alpha \equiv (s_\alpha, p_\alpha) \qquad \alpha' \equiv (s'_\alpha, p'_\alpha)\,.
\ee
These multi-indices contain all information characterising one single particle (its momentum and helicity). We will typically label ingoing states as unprimed and outgoing states with primed indices. For species $a$ we employ the multi-index $\alpha$ and 
similarly for species $b$ (resp. $c$ and $d$) we use the multi-indices $\beta$ (resp. $\gamma$ and $\delta$). The plane wave solutions are written in a compact form in this
notation. For instance for the species $a$ we write $u_\alpha \equiv u_{s_\alpha}(p_\alpha)$ and $v_\alpha \equiv v_{s_\alpha}(p_\alpha)$.
Furthermore this allows to write a compact relativistic Dirac delta function
which acts both on helicities and momenta as
\be
\deltarel_{\alpha \alpha'} \equiv \delta^{\rm K}_{s_\alpha s'_\alpha} \deltarel(p_\alpha-p'_\alpha)\,.
\ee
We denote the number operator associated with species $a$ as
\be
A_{\alpha \alpha'} \equiv N_{s_\alpha s'_\alpha}(p_\alpha,p'_\alpha) = a^\dagger_{s_\alpha}(p_\alpha) a_{s'_\alpha}(p'_\alpha)\,. 
\ee
We also define the Pauli blocking operator
\be
\widehat{A}_{\alpha \alpha'} \equiv \deltarel_{\alpha \alpha'}-{A}_{\alpha \alpha'}\,.
\ee
The expectation value of these operators is denoted as
\be
\langle A_{\alpha \alpha'}\rangle = \deltarel(p_\alpha-p'_\alpha)
  {\cal A}_{\alpha \alpha'}(p_\alpha)\,,\qquad \langle \widehat{A}_{\alpha \alpha'}\rangle = \deltarel(p_\alpha-p'_\alpha) \widehat{\cal A}_{\alpha \alpha'}(p_\alpha)
\ee
where we introduce the short-hand notation ${\cal A}_{\alpha
  \alpha'}(p_\alpha) = {\cal A}_{s_\alpha
  s'_\alpha}(p_\alpha)$. We recall that this quantity is exactly the one-particle distribution
function associated with species $a$. Note that for the Pauli blocking factor, $\widehat{\cal A}_{\alpha
  \alpha'}(p_\alpha)$ is a shorthand notation for $\delta^{\rm K}_{s_\alpha
  s'_\alpha} - {\cal A}_{s_\alpha  s'_\alpha}(p_\alpha)$. We
associate to the one-particle distribution function (resp. the Pauli blocking function) a spinor valued operator following the
procedure (\ref{Fspinor}) that we note ${A_\mathfrak{a}}^\mathfrak{b}$ (resp. ${\widehat{A}_\mathfrak{a}}^{\,\,\mathfrak{b}}$) in component notation or simply $\gr{A}$ (resp. $\widehat{\gr{A}}$) in operator notation. Having defined for species $a$ the number operator $A_{\alpha \alpha'}$, the distribution function ${\cal A}_{\alpha \alpha'}$ and the spinor-valued (observer-independant) operator $\gr{A}$, identically for species $b$ (resp. $c$ , $d$) we use $B_{\beta \beta'}$, ${\cal B}_{\beta \beta'}$ and $\gr{B}$ (resp. $C_{\gamma  \gamma'}$, ${\cal
  C}_{\gamma \gamma}'$ and $\gr{C}$, $D_{\delta \delta'}$, ${\cal
  D}_{\delta \delta'}$ and $\gr{D}$), and associated hatted notations for Pauli blocking factors.
Furthermore, for the antiparticles species $\bar a,\bar b,\bar c,\bar
d$ related to the species $a,b,c,d$, we use barred notation for number
operators (e.g. $\overline{A}_{\alpha \alpha'}$), distribution function
(e.g. $\overline{\cal A}_{\alpha \alpha'}$) and spinor valued operators (e.g. $\overline{\gr{A}}$), along with their hatted versions for Pauli blocking terms. Finally we define the collision term as in Eq. (\ref{QuantumBoltzmann2}), that is
\be\label{DefCollisionTerm}
\deltarel(0) C[{\cal A}_{ss'}(p)]\equiv -\frac{1}{2} \langle \int_{-\infty}^{\infty} \dd t' [H_I(0),[H_I(t'), A_{ss'}(p)]]
\rangle 
\ee
such that the quantum Boltzmann equation (\ref{QuantumBoltzmann}) for species $a$ is written as (when neglecting forward scattering)
\be\label{QuantumBoltzmann3}
\deltarel(0) \frac{\dd {\cal A}_{ss'}(p)}{ \dd t } = \deltarel(0) C[{\cal
A}_{ss'}(p)]\,.
\ee

Following the previous discussion, our goal is to compute the collision term $C[{\cal
A}_{ss'}(p)]$ corresponding to the reaction $a+b \leftrightarrow c+d$,
when considering an interaction Hamiltonian of the form (\ref{HIabcd}).
This interaction Hamiltonian contains the term
\be\label{HIofM}
{H}_I^{a+b\leftrightarrow c+d} \equiv \int [\dd p_\alpha] [\dd p_\beta] [\dd p_\gamma] [\dd p_\delta]
(2\pi)^3 \delta(\gr{p}_\alpha+\gr{p}_\beta-\gr{p}_\gamma-\gr{p}_\delta){\rm e}^{-\ii
  (p_\alpha^0+p_\beta^0-p_\gamma^0-p_\delta^0) t} {\cal M}^{a+b\leftrightarrow c+d} 
\ee
which is decribing the reaction $a+b \leftrightarrow
c+d$, and where we used the scattering operator for this reaction
\be
{\cal M}^{a+b\leftrightarrow c+d}  \equiv \sum_{\rm spins}\left(d^\dagger_\delta c^\dagger_\gamma b_\beta a_\alpha M_{\alpha\,\beta\to \gamma\,\delta}
  +b^\dagger_\beta a^\dagger_\alpha d_\delta c_\gamma M_{\gamma\,\delta\to \alpha\,\beta} \right)\,.
\ee
The $M$ matrices are 
\bea
M_{\alpha\,\beta\to \gamma\,\delta}  &\equiv&  M[(s_\alpha,p_\alpha)\,(s_\beta,p_\beta)\to
  (s_\gamma,p_\gamma)\,(s_\delta,p_\delta)]  \equiv g  [\bar u_\gamma{\cal \chi}^\mu_{(ac)} u_\alpha ][ \bar u_\delta {\cal \chi}^{(bd)}_\mu
u_\beta] \\
M_{\gamma\,\delta\to \alpha\,\beta}  &=&M^\star_{\alpha\,\beta\to
  \gamma\,\delta}  =g  [\bar u_\alpha {\cal \chi}^\mu_{(ac)} u_\gamma ][ \bar u_\beta {\cal \chi}_\mu^{(bd)} u_\delta]\,.
\eea

To compute the collision term we first need to compute the operator $ [{\cal M},[{\cal M},A_{ss'}]] $. Using the commutation rules given in Appendix~\ref{AppCommute}, we get
\beanosub\label{HHD}
 [{\cal M},[{\cal M},A_{ss'}]]   &\equiv& M_{\gamma'\delta' \to \alpha\beta}M_{\alpha'\beta'\to \gamma\delta}\left\{[d_{\delta'} c_{\gamma'} b_\beta^\dagger a_\alpha^\dagger, d_\delta^\dagger c_\gamma^\dagger
b_{\beta'} [ a_{\alpha'}, a^\dagger_{s} a_{s'} ] ]+ [d^\dagger_{\delta}
c^\dagger_{\gamma} b_{\beta'} a_{\alpha'}, d_{\delta'} c_{\gamma'} b_{\beta}^\dagger [ a_{\alpha}^\dagger, a^\dagger_{s} a_{s'} ] ] \right\}\nonumber\\
&=& M_{\gamma' \delta' \to \alpha\beta}M_{\alpha'\beta'\to \gamma\delta}\left\{ [d_{\delta'} c_{\gamma'} b_\beta^\dagger a_\alpha^\dagger, d_\delta^\dagger c_\gamma^\dagger
b_{\beta'} a_{s'}  ] \deltarel_{s\alpha'} - [d^\dagger_{\delta} c^\dagger_{\gamma} b_{\beta'}
a_{\alpha'}, d_{\delta'} c_{\gamma'} b_{\beta}^\dagger a^\dagger_{s}  ] \deltarel_{s'\alpha}
\right\}\nonumber\\
&=& M_{\gamma' \delta' \to \alpha\beta}M_{\alpha'\beta'\to \gamma\delta}\left\{ [d_{\delta'} c_{\gamma'} b_\beta^\dagger a_\alpha^\dagger, d_\delta^\dagger c_\gamma^\dagger
b_{\beta'} a_{s'}  ] \deltarel_{s\alpha'} + [d_{\delta'} c_{\gamma'} b_{\beta}^\dagger
a^\dagger_{s} , d^\dagger_{\delta} c^\dagger_{\gamma} b_{\beta'}
a_{\alpha'} ] \deltarel_{s'\alpha} \right\}\,.
\eeanosub
The commutators of this last expression are expressible simply in terms of
the number operators of the species. For instance using the
commutation rules of Appendix~\ref{AppCommute} we get
\be
[d_{\delta'} c_{\gamma'} b_\beta^\dagger a_\alpha^\dagger, d_\delta^\dagger c_\gamma^\dagger
b_{\beta'} a_{s'}  ] = A_{\alpha s' }B_{\beta \beta'}
\widehat{C}_{\gamma \gamma'} \widehat{D}_{\delta \delta'} -
\widehat{A}_{\alpha s' } \widehat{B}_{\beta \beta'} {C}_{\gamma \gamma'} {D}_{\delta \delta'}\,.
\ee
Hence we obtain
\beanosub\label{MMAresult}
&&
 [{\cal M},[{\cal M},A_{ss'}]]  = M^\star_{\alpha\,\beta\to
   \gamma'\,\delta'}M_{\alpha'\,\beta'\to \gamma\,\delta}\nonumber\\
 &&\left\{\deltarel_{s \alpha'} \left[B_{\beta \beta'}A_{\alpha s'} 
\widehat D_{\delta \delta'} \widehat C_{\gamma \gamma'}  -D_{\delta
  \delta'} C_{\gamma \gamma'}
\widehat B_{\beta \beta'} \widehat A_{\alpha
  s'}\right]+\deltarel_{\alpha s'} \left[B_{\beta \beta'}A_{s \alpha'} 
\widehat D_{\delta \delta'} \widehat C_{\gamma \gamma'}  -D_{\delta
  \delta'} C_{\gamma \gamma'}
\widehat B_{\beta \beta'} \widehat A_{s\alpha'}\right]\right\}\,.
\eeanosub
We now employ this result in Eqs. (\ref{HIofM}) and (\ref{DefCollisionTerm}).
We integrate a total of five momentum integrals (each one being itself three-dimensional in
momentum space) using the Dirac distributions. Of these, four Dirac functions are contained in the expectation values
of the number operators associated to the four species, and there is
an extra Dirac function $\deltarel_{s \alpha'}$ from the collision term ensuring local energy and momentum conservation. Eventually, taking the expectation in the quantum state, we get
\beanosub\label{ColAss}
 \deltarel(0) C[{\cal A}_{ss'}(p)]&=& \vol {\cal K} M^\star[(s_\alpha,p)\,(s_\beta,p_\beta)\to (s'_\gamma,p_\gamma)\,(s'_\delta,p_\delta)]M[(s'_\alpha,p)\,(s'_\beta,p_\beta)\to (s_\gamma,p_\gamma)\,(s_\delta,p_\delta)]\nonumber\\
&&\left\{\delta^{\rm K}_{s\alpha'}\left[-{\cal B}_{\beta \beta'}(p_\beta){\cal A}_{\alpha s'}(p) 
\widehat{\cal D}_{\delta \delta'}(p_\delta) \widehat{\cal
  C}_{\gamma\gamma'}(p_\gamma) +{\cal D}_{\delta \delta'}(p_\delta)
{\cal C}_{\gamma \gamma'}(p_\gamma)
\widehat{\cal B}_{\beta \beta'}(p_\beta) \widehat{\cal A}_{\alpha s'}(p)\right]\right.\nonumber\\
&&\left. +\delta^{\rm K}_{\alpha s'}
\left[-{\cal B}_{\beta \beta'}(p_\beta){\cal A}_{s \alpha'}(p) 
\widehat{\cal D}_{\delta \delta'}(p_\delta) \widehat{\cal C}_{\gamma
  \gamma'}(p_\gamma)  +{\cal D}_{\delta \delta'}(p_\delta) {\cal
  C}_{\gamma \gamma'}(p_\gamma)
\widehat{\cal B}_{\beta \beta'}(p_\beta) \widehat{\cal A}_{s\alpha'}(p)\right]\right\}
\eeanosub
with the integration on momenta
\be\label{calK}
{\cal K} = \frac{1}{2} \int [\dd p_\delta][\dd p_\gamma][\dd
p_\beta](2\pi)^4\delta^{(4)}(p_\delta+p_\gamma-p_\beta-p) \,.
\ee
We note that:
\begin{itemize}
\item The collision term is made of two types of terms. The first terms on
the second and the third line of Eq. (\ref{ColAss}) correspond to
\emph{scattering out} processes, that is collisions which due to the minus sign deplete the
distribution function associated with species $a$ and they correspond
to $a+b \to c+d $. The second term on the second and third line
correspond conversely to \emph{scattering in} processes, which increase the
distribution function of species $a$, and they represent interactions $c+d \to
a+b$. 

\item For scattering out processes, the collision term is proportional
to the distribution function of the initial states (species $a$ and
$b$), but also to the Pauli blocking function of the final states
(species $c$ and $d$), and the reverse is true for the scattering in processes.

\item The distribution functions are Hermitian, that is ${\cal  A}^\star_{s s'}(p) = {\cal A}_{s' s'}(p)$ as in
  Eq.~(\ref{HermitianProperty}). Let us now consider
  $C[{\cal A}_{ss'}(p)]^\star$. Given the Hermiticity of the distribution functions and thus
  of the Pauli blocking functions, with a simple renaming of all
  primed indices as unprimed indices (and also of unprimed
  indices as primed indices), it is straightforward to show that this is equal to $C[{\cal A}_{s's}(p)]$, hence the collision term is
  also Hermitian as expected.

\item In the previous computation when checking the Hermiticity, the second and third line of
  Eq. (\ref{ColAss}) are interchanged. Terms of the second line are
  proportional to $\delta^{\rm K}_{s\alpha'}$ and correspond
  physically to the scattering of the helicity index $s'$, and
  conversely in the third line the terms are proportional to
  $\delta^{\rm K}_{\alpha s'}$ and it corresponds to the scattering of
  the helicity index $s$. Hence we see that the collision term
  possesses four terms corresponding to the \emph{in/out} contributions and
  the $s/s'$ contributions.

\item Finally even though we computed the collision term for a
  homogenous system in a Minkowski space-time, the total volume, which appears as
  $\delta^{(3)}(0)$, drops out from both the left and the right hand
  side of Eq. (\ref{ColAss}) and thus of
  Eq. (\ref{QuantumBoltzmann3}). Hence, as argued before Eq. (\ref{dgrFdt}), we can consider that this collision
  term is valid locally, allowing us to consider in a classical
  macroscopic description that all distribution functions should be
  considered with a dependence on the point of space-time. We started
  a computation with total number of particles in a quantum system,
  but we end up using it with \emph{number densities} of particles,
  considering that the collisions are point-like.

\end{itemize}

The procedure to follow is now transparent. The helicity indices of
the distribution functions (or the related Pauli blocking functions)
are contracted with the plane waves solutions contained in the $M$
matrices. From Eqs. (\ref{Fspinor}) this is exactly what is needed to build the
spinor space operators related to each species. Since only the indices
$s$ and $s'$ remain uncontracted in Eq. (\ref{ColAss}), we contract them
with $u_{s'}(p) \bar u_s(p)$ (or $v_{s}(p) \bar v_{s'}(p)$ for antiparticles) so as to form a spinor space collision operator $C[\gr{A}(p)] $ as specified in the definition (\ref{DefCollisionOperator}). 
\com{\be
C[\gr{A}(p)] \equiv \sum_{s\,s' }C[{\cal A}_{ss'}(p)]u_{s'}(p) \bar u_s(p)\,.
\ee}
Note that the contraction of $\delta^{\rm
  K}_{\alpha s'}$ with $u_{s'}(p) \bar u_s(p)$ or $v_{s}(p) \bar v_{s'}(p)$ gives simply $\sgnslash\slashed{p}+ M $ as can be seen from Eqs. (\ref{Closure}).

We finally obtain
\beanosub\label{CAoperator}
C[\gr{A}(p)] &=&  -\frac{g^2}{2 E_a} {\cal K}\left\{{\rm Tr}[\gr{B} \cdot {\cal \chi}_\mu^{(bd)}  \cdot \hat{\gr{D}} \cdot
  {\cal \chi}_\nu^{(bd)} ] \times \left[\gr{A} \cdot {\cal \chi}^\mu_{(ac)}  \cdot \widehat{\gr{C}} \cdot {\cal
    \chi}^\nu_{(ac)}  \cdot (\sgnslash \slashed{p}+M )+(\sgnslash\slashed{p}+M)\cdot {\cal \chi}^\mu_{(ac)}  \cdot \widehat{\gr{C}} \cdot  {\cal \chi}^\nu_{(ac)} \cdot \gr{A}\right]\right.\nonumber\\
&&\left.\qquad \qquad-(\gr{A} \leftrightarrow \widehat{\gr{A}},\gr{B} \leftrightarrow \widehat{\gr{B}},\widehat{\gr{C}} \leftrightarrow {\gr{C}},\widehat{\gr{D}} \leftrightarrow {\gr{D}})\right\}
\eeanosub
where the momentum dependence $\gr{A}(p)$ and $\gr{B}(p_\beta)$,
$\gr{C}(p_\gamma)$,  $\gr{D}(p_\delta)$ (and similarly for Pauli
blocking operators) are omitted for a more compact notation.
The structure of the collision term is again manifest. The first line
corresponds to scattering out processes. It is made of two terms which
thanks to the property (\ref{HermitianSpinor}) of all operators
appearing ($\{\gr{A}, \gr{B}, \widehat{\gr{C}},\widehat{\gr{D}}, {\cal  \chi}^\mu_{(ac)},{\cal \chi}^\mu_{(bd)}\}$), ensure that the property (\ref{HermitianSpinor})  is satisfied for the collision operator. As for the second line, it
corresponds to the scattering in processes, and differs only by an
overal sign and the exchange of the distribution and Pauli blocking functions.

This collision term $C[\gr{A}]$, being itself an operator in spinor
space, can be decomposed into its covariant parts $I_{C[A]}$and ${\cal Q}_{C[A]}^\mu$ as in the decomposition (\ref{BigDecompositionGood}).  These components can be found by
multiplying by the appropriate $X \in {\cal O}$ and taking the trace, that is using the extractions (\ref{ExtractCovParts1}) or (\ref{ExtractCovParts2}) in the massless case.  Since all
operators involved in the collision term are made of $\gamma^\mu$ or $\gamma^5$ matrices,
the problem is reduced to taking traces of products of these
operators. The corresponding expressions are gathered in Appendix~\ref{TraceTechnologie}.  This procedure is conceptually simple and standard, but it
can become rather involved when there is a large number of operators
in the products. Indeed, letting aside the $\gamma^5$ operators, the largest products involve two
$\gamma^\mu$ matrices for each distribution function or Pauli blocking
function from the $\widetilde{\Sigma}_{\mu\nu}$. Then considering the
$\gamma^\mu$ of the operators (\ref{Omu}), and the one contained in
$\sgnslash\slashed{p}+M$, we see that when extracting the components
on $\gamma^\mu$ (resp. $\Sigma_{\mu\nu}$) of the collision term, we
end up with a total of eight (resp. nine) $\gamma^\mu$
operators. However, this systematic computation can be handled by a computer algebra
package such as {\it xAct}~\cite{xAct} and this is particularly powerful
since it also takes care of all simplifications involving space-time indices.

In particular, when using Eqs. (\ref{DecompositionFormula}) to extract the intensity part of the collision term (\ref{CAoperator}), we find using the decomposition (\ref{BigDecompositionGood2}) for $\gr{A}(p)$ and the property $(\slashed{p}+M)(\slashed{p}+M) = 2M (\slashed{p}+M)$ (deduced from $\slashed{p} \slashed{p}= \sgnslash m^2$) its general expression
\be\label{ICA}
\frac{1}{2}I_{C[A]}(p) =  -\frac{g^2}{2 E_a} {\cal K}\left\{{\rm Tr}[\gr{B} \cdot {\cal \chi}_\mu^{bd}  \cdot \hat{\gr{D}} \cdot
  {\cal \chi}_\nu^{bd} ] {\rm Tr}\left[\gr{A} \cdot {\cal \chi}^\mu_{ac}  \cdot \widehat{\gr{C}} \cdot {\cal
    \chi}^\nu_{ac}\right]-(\gr{A} \leftrightarrow \widehat{\gr{A}},\gr{B} \leftrightarrow \widehat{\gr{B}},\widehat{\gr{C}} \leftrightarrow {\gr{C}},\widehat{\gr{D}} \leftrightarrow {\gr{D}})\right\}\,.
\ee
This general expression is also valid in the massless case. To show this we first need that for any two four-vectors $A^\mu$ and $B^\mu$, $\{\slashed{A},\slashed{B}\}=\sgnslash 2 A^\mu B_\mu \mathds{1}$ thanks to Eq. (\ref{DefClifford}), so that $\{\slashed{p},\slashed{k}\}=2 \mathds{1}$ and $\{\slashed{p}, \slashed{Q}\}=0$. Using the extraction expression (\ref{ExtractIMassless}) applied on the collision term (\ref{CAoperator}), and commuting the operators with the previous properties to force the appearance of $\slashed{p}\slashed{p}=0$, leads also to Eq. (\ref{ICA}).

The collision operator (\ref{CAoperator}) has the general form
\be\label{GenCA}
C[\gr{A}(p)] = -[(\sgnslash\slashed{p}+M)\cdot{\bm K} \cdot \gr{A}(p)+\gr{A}(p)\cdot{\bm K} \cdot (\sgnslash\slashed{p}+M)] + [(\sgnslash\slashed{p}+M)\cdot\widehat{\bm K} \cdot \widehat{\gr{A}}(p)+\widehat{\gr{A}}(p)\cdot\widehat{\bm K} \cdot (\sgnslash\slashed{p}+M)]\,,
\ee
where ${\bm K}={\bm K}[\gr{B},\widehat{\gr{C}},\widehat{\gr{D}}]$ is an operator [satisfying property (\ref{HermitianSpinor})] depending on other species distribution functions integrated over momenta, and $\widehat{\bm K}={\bm K}[\widehat{\gr{B}},{\gr{C}},{\gr{D}}]$ is its hatted version. For other types of interactions than the current-current weak interactions that we considered (e.g. considering the effect of the finite mass of the vector boson exchanged), the collision operator for the species $a$ would still have the general form (\ref{GenCA}) but with a different ${\bm K}$. If the interactions are with bosons, the hatted expressions in ${\bm K}$ have to be understood in the sense of stimulated emission instead of Pauli blocking as discussed in Appendix \ref{AppBosons}. 
From the general form (\ref{GenCA}), the intensity part is in general obtained from
\be
\frac{1}{2}I_{C[A]}(p) = - {\rm Tr}[{\bm K}.\gr{A}(p)]+{\rm Tr}[\widehat{\bm K}.\widehat{\gr{A}}(p)]\,.
\ee

\subsection{Explicit form of the collision term for $a+b
  \leftrightarrow c+d$}\label{SecExplicitCabcd}

\subsubsection{Notation}
 
Let us introduce some notation which allows to give explicitly the collision term
for the species $a$ corresponding to the reaction $a+b \leftrightarrow
c+d$, as obtained from the method exposed in the previous
section. From the covariant components of the distribution function
which are $I(p)$ and ${\cal Q}^\mu(p)$, we can build scalar, vectors and
tensors with which the collision term is better expressed. Let us first introduce the chiral vectors ${\cal I}^{\pm}_\mu(p)$ (see appendix (\ref{SecInterpretationImu}) for physical interpretation) and the achiral vector ${\cal I}^{0}_\mu(p)$
\be
2 {\cal I}^\pm_\mu \equiv I p_\mu \pm M {\cal Q}_\mu = I p_\mu 
\pm M Q_\mu \pm M V S_\mu\,, \qquad 2 {\cal I}^0_\mu \equiv I p_\mu\,,
\ee
where we remind the notation $M=m$ (resp. $M=-m$) for a particle
(resp. for an antiparticle) and where the dependence of all quantities
on the momentum $p^\mu$ is omitted. Note that the left chiral vector for a particle ${\cal I}^-_\mu$ is equal to the right chiral vector ${\cal I}^+_\mu$ of the corresponding
antiparticle and vice versa.  We also define the Pauli blocking vectors
\be
\widehat{\cal I}^\pm_\mu \equiv p_\mu-{\cal I}^\pm_\mu\,,\qquad \widehat{\cal I}^0_\mu \equiv p_\mu- {\cal I}^0_\mu\,,\qquad \widehat{\cal Q}^\mu = - {\cal Q}^\mu\,.
\ee
In the massless case, using that $m S^\mu \to p^\mu$ we must take the definitions 
\be
2 {\cal I}^\pm_\mu \equiv(I \pm \hatM V)p_\mu  \,.
\ee
where we remind that $\hatM$ is $1$ for particles and $-1$ for antiparticles.
Note that the purely linear polarisation $Q^\mu$ does not contribute
in the definitions of the chiral vectors in the massless case.
We then define scalar quantities 
\be
{\cal I}\equiv \frac{I}{2}\,,\qquad \widehat{\cal I} = 1- {\cal
  I}\,,\qquad \qquad {\cal J} \equiv M {\cal I} = M \frac{I}{2}
\,,\qquad \widehat{\cal J} \equiv M \widehat{\cal I} = M \left(1-\frac{I}{2}\right)\,,
\ee
and tensor quantities
\bea
{\cal S}_{\mu\nu}&\equiv& p_{[\mu} {\cal
  Q}_{\nu]}\qquad \widehat{\cal S}_{\mu\nu} = - {\cal S}_{\mu\nu}\\
{\cal S}^\pm_{\mu\nu} &=& {\cal S}_{\mu\nu} \sgnzi {\cal J} g_{\mu\nu}={\cal
  S}_{\mu\nu} \sgnzi \frac{MI}{2}  g_{\mu\nu}\,,\label{defScal}
\eea
where again the hatted notation refers to Pauli blocking forms.
In the massless limit ${\cal J}=\widehat{\cal J}= 0$ and ${\cal
  S}^\pm_{\mu\nu} = {\cal S}_{\mu\nu}$. Furthermore ${\cal S}_{\mu\nu} =
p_{[\mu} Q_{\nu]}$, and it is non-vanishing only if there
is linear polarisation.

The collision term is expressed by means of various contractions of
these scalars, vectors and tensors associated with the various species
involved in the collision. In order to obtain a compact result, we introduce the following contractions (we remind that
indices $a,b,c,d$ refer to the species considered):
\be
[{\cal S}_a \cdot {\cal S}_b] \equiv {{\cal S}_a^\mu}_\nu {{\cal
    S}_b^\nu}_\mu\,,\qquad[{\cal S}_a\cdot {\cal S}_b \cdot {\cal S}_c \cdot {\cal S}_d]
\equiv {{\cal S}_a^\sigma}_\lambda {{\cal S}_b^\lambda}_\mu {{\cal
    S}_c^\mu}_\nu {{\cal S}_d^\nu}_\sigma\,,
\ee

\be
{\cal
  I}_a\cdot {\cal S}_b \cdot {\cal I}_c \equiv {\cal I}^\mu_a{\cal S}^b_{\mu\nu} {\cal I}_c^\nu \,,\qquad{\cal I}_a\cdot {\cal S}_b \cdot  {\cal S}_c
\cdot {\cal I}_d \equiv {\cal I}^a_\mu {{\cal S}_b^{\mu}}_\nu {{\cal S}_c^{\nu}}_\lambda {\cal
  I}_d^\lambda \,,
\ee

\be
({\cal S}_a \cdot {\cal I}_b)^\mu = {{\cal S}_a^\mu}_\nu {\cal
  I}_b^\nu\,,\qquad({\cal S}_a \cdot {\cal S}_b \cdot  {\cal I}_c)^\mu = {{\cal
    S}_a^\mu}_\nu {{\cal S}_b^\nu}_\sigma {\cal
  I}_c^\sigma\,,\qquad({\cal S}_a \cdot {\cal S}_b \cdot {\cal S}_c \cdot {\cal I}_d)^\mu = {{\cal
    S}_a^\mu}_\nu {{\cal S}_b^\nu}_\sigma  {{\cal S}_c^\sigma}_\lambda {\cal
  I}_d^\lambda\,.
\ee
 
\subsubsection{Intensity of the collision term}

The intensity part of the collision term takes the general form 
\be
I_{C[A]} = \frac{2^{9}}{2 E_a} g^2 {\cal K}
\left[{\cal T}_I(\widehat{\gr{A}},\widehat{\gr{B}},\gr{C},\gr{D}) - {\cal T}_I(\gr{A},\gr{B},\widehat{\gr{C}},\widehat{\gr{D}})\right]\,,
\ee
where the first contribution corresponds to scattering out processes
and the second one to scattering in processes. The Kernel of the
collision is common to both types of processes and reads
\beanosub\label{CIKernel}
{\cal T}_I(\gr{A},\gr{B},\gr{C},\gr{D})&\equiv&\Big\{\sum_{r=\pm}(\epsilon^{ac}_r)^2 (\epsilon^{bd}_r)^2({\cal I}^r_a \cdot
  {\cal I}^r_b) ({\cal I}^r_c \cdot {\cal I}^r_d)+\sum_{r=\pm}(\epsilon^{ac}_r)^2 (\epsilon^{bd}_{-r})^2({\cal I}^r_a \cdot
  {\cal I}^{-r}_d) ({\cal I}^r_c \cdot {\cal I}^{-r}_b)\\
&+&\sum_{r=\pm}(\epsilon^{ac}_r)^2 \epsilon^{bd}_+ \epsilon^{bd}_{-}\Big(\sgnii \frac{1}{2}({\cal I}^r_a \cdot
  {\cal I}^r_c) [{\cal S}_b \cdot {\cal S}_d] \sgnzz ({\cal I}^r_a \cdot {\cal S}_b \cdot
  {\cal S}_d  \cdot {\cal
   I}^r_c) \sgnzz ({\cal I}^r_c \cdot {\cal S}_b \cdot
  {\cal S}_d  \cdot {\cal
   I}^r_a )\nonumber\\
&&\qquad \qquad\qquad+({\cal I}^r_a  \cdot {\cal S}_d \cdot {\cal I}^r_c) {\cal J}_b +  ({\cal
   I}^r_c  \cdot {\cal S}_b \cdot  {\cal I}^r_a ) {\cal J}_d \sgnii {\cal J}_b {\cal J}_d ({\cal I}^r_a \cdot
  {\cal I}^r_c)\Big)\nonumber\\
&+&\sum_{r=\pm}(\epsilon^{bd}_r)^2 \epsilon^{ac}_+ \epsilon^{ac}_{-}\Big(\sgnii \frac{1}{2}({\cal I}^r_b \cdot
  {\cal I}^r_d) [{\cal S}_a \cdot {\cal S}_c] \sgnzz ({\cal I}^r_b \cdot {\cal S}_a \cdot
  {\cal S}_c  \cdot {\cal
   I}^r_d) \sgnzz ({\cal I}^r_d \cdot {\cal S}_a \cdot
  {\cal S}_c  \cdot {\cal
   I}^r_b )\nonumber\\
&&\qquad \qquad\qquad+({\cal I}^r_b  \cdot {\cal S}_c \cdot {\cal I}^r_d) {\cal J}_a +  ({\cal
   I}^r_d  \cdot {\cal S}_a \cdot  {\cal I}^r_b ) {\cal J}_c \sgnii {\cal J}_a {\cal J}_c ({\cal I}^r_b \cdot
  {\cal I}^r_d)\Big)\nonumber\\
&+&\epsilon^{ac}_- \epsilon_+^{ac} \epsilon^{bd}_-
  \epsilon^{bd}_+[-[{\cal S}_a \cdot {\cal S}_c][{\cal S}_b \cdot {\cal S}_d] + 2  [{\cal S}_a
\cdot {\cal S}_b \cdot {\cal S}_d \cdot {\cal S}_c]+2 [{\cal S}_a
\cdot {\cal S}_d \cdot {\cal S}_b \cdot {\cal S}_c]\nonumber\\
&&\qquad \qquad\quad\left. - [{\cal S}_a \cdot {\cal S}_b] {\cal J}_c {\cal J}_d - [{\cal S}_c \cdot
{\cal S}_d] {\cal J}_a {\cal J}_b + [{\cal S}_a \cdot {\cal S}_d] {\cal J}_b {\cal J}_c + [{\cal S}_b \cdot {\cal S}_c] {\cal J}_a {\cal J}_d + 4
{\cal J}_a {\cal J}_b {\cal J}_c {\cal J}_d\right]\Big\}\,.\nonumber
\eeanosub
It is rather involved since we have considered general currents
(\ref{GeneralCurrents}) which include both left and right chiral
coupling and we have allowed all species to be polarized. However we see immediately that if there are only left
chiral couplings, only the first term in this expression
survives. 

\subsubsection{Polarisation of the collision term}

The polarisation part of the collision term must be transverse. Let us
introduce the projector operator
\be
{H^\mu}_\nu \equiv \delta_\nu^\mu \sgnii m^{-2}p^\mu p_\nu= \sgnminusslash S^\mu S_\nu \sgnii \epsilon_+^\mu \epsilon_{-\,\nu} \sgnii \epsilon_-^\mu \epsilon_{+\,\nu}\,.
\ee
The polarisation part of the collision term takes the general form
\be
{\cal Q}_{C[A]}^\mu = \frac{2^{9}}{2 E_a} g^2 {\cal K} {{(H_a)}^\mu}_\nu \left[{\cal
    T}_{\cal Q}^\nu(\widehat{\gr{A}},\widehat{\gr{B}},\gr{C},\gr{D}) - {\cal
    T}_{\cal Q}^\nu(\gr{A},\gr{B},\widehat{\gr{C}},\widehat{\gr{D}})\right]\,,
\ee
where ${{(H_a)}^\mu}_\nu $ is the projector for species $a$, that is orthogonally to $p_a$, and where the Kernel for polarisation is
\beanosub\label{CQKernel}
{\cal
    T}_{\cal Q}^\nu(\gr{A},\gr{B},\gr{C},\gr{D}) &\equiv&
\Big\{
\sum_{r=\pm}(\epsilon^{ac}_r)^2 (\epsilon^{bd}_{r})^2\left[\sgnii({\cal S}^r_a \cdot {\cal I}^{r}_b)^\nu  ({\cal I}^r_c \cdot  {\cal I}^{r}_d)\right]+\sum_{r=\pm}(\epsilon^{ac}_r)^2 (\epsilon^{bd}_{-r})^2\left[\sgnii({\cal S}^r_a \cdot {\cal I}^{-r}_d)^\nu  ({\cal I}^r_c \cdot  {\cal I}^{-r}_b)\right]\\
&+&\sum_{r=\pm}\epsilon^{ac}_{-} \epsilon^{ac}_{+} (\epsilon^{bd}_{r})^2\Big[({\cal I}^0_a \cdot {\cal S}_c \cdot {\cal
  I}^r_b){\cal I}^{r\,\nu}_d+({\cal I}^0_a \cdot {\cal S}_c \cdot {\cal
  I}^r_d){\cal I}^{r\,\nu}_b + ( {\cal  I}^r_b \cdot  {\cal
  I}^r_d)({\cal S}_c \cdot {\cal I}^0_a )^\nu -  ( {\cal  I}^r_b \cdot  {\cal
  I}^0_a)({\cal S}_c \cdot {\cal  I}^r_d )^\nu \nonumber\\
&&\qquad \qquad \qquad-  ( {\cal  I}^r_d \cdot  {\cal I}^0_a)({\cal S}_c
\cdot {\cal  I}^r_b )^\nu \sgnzz ( {\cal  I}^r_b \cdot  {\cal I}^0_a) {\cal J}_c {\cal
  I}_d^{r\,\nu} \sgnii  ( {\cal  I}^r_d \cdot  {\cal I}^0_a) {\cal J}_c {\cal
  I}_b^{r\,\nu} + ({\cal I}^r_b \cdot {\cal S}^r_c \cdot {\cal
  I}^r_d) {\cal I}_a^{-r\,\nu} \Big]\nonumber\\
&+&\sum_{r=\pm}\epsilon^{bd}_{-} \epsilon^{bd}_{+}
(\epsilon^{ac}_{r})^2\Big[\sgnii({\cal S}^r_a \cdot {\cal S}^{-r}_b \cdot {\cal S}^r_d \cdot
{\cal I}^r_c )^\nu  \sgnii ({\cal S}^r_a \cdot {\cal S}_d \cdot {\cal S}_b \cdot
{\cal I}^r_c )^\nu \sgnzz \frac{1}{2}[{\cal S}_b \cdot {\cal S}_d]({\cal S}^r_a \cdot {\cal I}^r_c)^\nu\Big]\nonumber\\
&+&\epsilon^{ac}_- \epsilon_+^{ac}
\epsilon^{bd}_-\epsilon^{bd}_+\Big[\sgnii 2 ({\cal S}_d \cdot {\cal S}_b \cdot {\cal S}_c \cdot
  {\cal I}^0_a)^\nu \sgnii 2 ({\cal S}_c \cdot {\cal S}_d \cdot {\cal S}_b \cdot
  {\cal I}^0_a)^\nu \sgnii 2 ({\cal S}_c \cdot {\cal S}_b \cdot {\cal S}_d \cdot
  {\cal I}^0_a)^\nu \nonumber\\
&&\qquad \qquad\qquad \sgnii 2 ({\cal S}_b \cdot {\cal S}_d \cdot {\cal S}_c \cdot
  {\cal I}^0_a)^\nu \sgnzz 2 [{\cal S}_b \cdot {\cal S}_d]({\cal S}_c \cdot {\cal I}^0_a)^\nu  +M_a {\cal Q}_a^\nu
{\cal J}_b [{\cal S}_c \cdot {\cal S}_d]-M_a {\cal Q}_a^\nu {\cal J}_d
  [{\cal S}_b \cdot {\cal S}_c] \nonumber\\
&&\qquad \qquad\qquad +4 M_a {\cal Q}_a^\nu {\cal J}_b {\cal
    J}_c {\cal J}_d \sgnzz  2 {\cal J}_c {\cal J}_d ({\cal S}_b \cdot
  {\cal I}_a^0)^\nu \sgnii 2 {\cal J}_c {\cal J}_b ({\cal S}_d \cdot {\cal I}_a^0)^\nu\Big]\Big\}\,.\nonumber
\eeanosub

\subsubsection{Particular case of a massless species $a$}\label{SecmasslessGeneral}

Expression (\ref{CQKernel}) is valid also for massless case but in that case it describes only the linear polarisation part. Hence, the projector ${{(H_a)}^\mu}_\nu$ must be replaced by the screen projector ${{({\cal H}_a)}^\mu}_\nu$ [see Eq. (\ref{DefScreenH})] since linear polarisation is orthogonal to both $p^\mu$ and $u^\mu$. That is the polarisation part of the collision term in the case the species $a$ is massless takes the form
\be
{\cal
    T}_{\cal Q}^\nu(\gr{A},\gr{B},\gr{C},\gr{D}) ={\cal  T}_{Q}^\nu(\gr{A},\gr{B},\gr{C},\gr{D}) +{\cal T}_{V}(\gr{A},\gr{B},\gr{C},\gr{D}) S^\mu\,,
\ee
and ${\cal  T}_{Q}^\nu(\gr{A},\gr{B},\gr{C},\gr{D})$ is formally
equal to the rhs of Eq. (\ref{CQKernel}) up to this change of projector. In the massless case, the circular polarisation part of the collision term has the structure
\be
 V_{C[A]} = \frac{2^{9}}{2 E_a} g^2 {\cal K} \left[{\cal
    T}_{V} (\widehat{\gr{A}},\widehat{\gr{B}},\gr{C},\gr{D}) - {\cal
    T}_{V} (\gr{A},\gr{B},\widehat{\gr{C}},\widehat{\gr{D}})\right]\,,
\ee
and the Kernel for circular polarisation is 
\beanosub\label{CVKernel}
\hatM_a {\cal
    T}_{V} (\gr{A},\gr{B},\gr{C},\gr{D})&\equiv&\Big\{\sum_{r=\pm}\,r\,(\epsilon^{ac}_r)^2 (\epsilon^{bd}_r)^2({\cal I}^r_a \cdot
  {\cal I}^r_b) ({\cal I}^r_c \cdot {\cal I}^r_d)+\sum_{r=\pm}\,r\,(\epsilon^{ac}_r)^2 (\epsilon^{bd}_{-r})^2({\cal I}^r_a \cdot
  {\cal I}^{-r}_d) ({\cal I}^r_c \cdot {\cal I}^{-r}_b)\nonumber\\
&&+\sum_{r=\pm}\,r\,(\epsilon^{ac}_r)^2 \epsilon^{bd}_+ \epsilon^{bd}_{-}\Big[\sgnii \frac{1}{2}({\cal I}^r_a \cdot
  {\cal I}^r_c) [{\cal S}_b \cdot {\cal S}_d]  \sgnzz ({\cal I}^r_a \cdot {\cal S}_b \cdot
  {\cal S}_d  \cdot {\cal
   I}^r_c) \sgnzz ({\cal I}^r_c \cdot {\cal S}_b \cdot
  {\cal S}_d  \cdot {\cal
   I}^r_a )\nonumber\\
&&\qquad \qquad\qquad\qquad\qquad+({\cal I}^r_a  \cdot {\cal S}_d \cdot {\cal I}^r_c) {\cal J}_b +  ({\cal
   I}^r_c  \cdot {\cal S}_b \cdot  {\cal I}^r_a ) {\cal J}_d \sgnii {\cal J}_b {\cal J}_d ({\cal I}^r_a \cdot
  {\cal I}^r_c)\Big]\Big\}\,.\nonumber
\eeanosub
It is possible to check that the purely circular polarisation part in
the massive case, obtained from the projection along the helicity
vector~(\ref{defSmu}) as $\sgnminusslash S_\mu
{\cal Q}_{C[A]}^\mu$ tends to $V_{C[A]}$ in the massless limit,
showing the consistency of the expressions obtained.

\subsection{Crossing symmetries and charge conjugation}\label{SecCrossingSymmetry}

The effect of parity and charge conjugation has been investigated in \S~\ref{SecDiscrete}. When applied on the currents which enter the interaction Hamiltonian (\ref{HIabcd}), the transformation properties are found to be
\bea
{\cal C}J^\mu_{ac}(\epsilon_-^{ac},\epsilon_+^{ac}) {\cal C} &=&J^{\mu\star}_{ac}(\epsilon_+^{ac},\epsilon_-^{ac}) \slabel{CJC}\\
{\cal P}J^\mu_{ac}(\epsilon_-^{ac},\epsilon_+^{ac}) {\cal P} &=&(-1)^\mu \,J^{\mu}_{ac}(\epsilon_+^{ac},\epsilon_-^{ac})\slabel{PJP}\,, 
\eea
where we used the compact notation (\ref{CompactPeskin}). To be explicit both transformations interchange the role of $\epsilon_-^{ac}$ and $\epsilon_+^{ac}$. The additional effect of parity is not relevant because in the general Hamiltonian currents are coupled together and so $((-1)^\mu)^2=1$. The additional effect of charge conjugation is to add a complex conjugation. When there are no CP violating coupling matrices, such as CKM or PMNS matrices with complex phases, then this has also no effect. Combining both P and C transformations leaves thus the Hamiltonian (\ref{HIabcd}) invariant. CP invariance means that the collision term for the reaction $\bar a + \bar b \leftrightarrow \bar
c + \bar d$ can be either computed by applying the charge conjugation operator, or by applying the parity operation which amounts to a simple interchange of $\epsilon_-^{ac}$ and $\epsilon_+^{ac}$.

When applying the charge conjugation operation on the collision term (\ref{CAoperator}), it amounts simply to considering that all quantities built from the distribution functions (such as ${\cal I}^\pm_\mu$, $S_{\mu\nu}$ etc...) now should refer to the antiparticle species. This means that the collision Kernels for the reaction $\bar a + \bar b \leftrightarrow \bar
c + \bar d$ are exactly the same except that the covariant components
of the antiparticles now enter its definition. That is for the
collision term of species $\bar a$ due to the process $\bar a + \bar b \leftrightarrow \bar
c + \bar d$, the Kernel for the intensity part of the collision is
given by
\be
{\cal   T}_{I} (\overline{\gr{A}},\overline{\gr{B}},\overline{\gr{C}},\overline{\gr{D}})\,.
\ee
And given the CP symmetry, it can be checked explicitly that this is
exactly equivalent to the interchange $\epsilon_-^{ac} \leftrightarrow
\epsilon_+^{ac}$ and $\epsilon_-^{bd} \leftrightarrow \epsilon_+^{bd}$
in ${\cal T}_I(\gr{A},\gr{B},\gr{C},\gr{D})$. For instance a contribution of the form 
\be
\sum_r (\epsilon^{ac}_r)^2 (\epsilon^{bd}_r)^2({\cal I}^r_a \cdot
  {\cal I}^r_b) ({\cal I}^r_c \cdot {\cal I}^r_d)
\ee
when considered for antiparticles is equivalent to the operation $\epsilon_-^{ac} \leftrightarrow \epsilon_+^{ac}$ and $\epsilon_-^{bd} \leftrightarrow \epsilon_+^{bd}$ because ${\cal I}^\pm_\mu$ for a particle is equal to ${\cal I}^{\mp}_\mu$ of the corresponding antiparticle.

Now that we have computed the collision term for $a+ b \leftrightarrow c+d$ and deduced in the most straightforward manner the one associated to $\bar a + \bar b \leftrightarrow \bar
c + \bar d$, we can obtain all other related reactions by crossing
symmetry. 
Let us for instance consider the processes $a \leftrightarrow \bar b +c
+d$. The part of the interaction Hamiltonian responsible for this
process, $H_I^{a\leftrightarrow \bar b + c +d}$ can be deduced from
$H_I^{a+b \leftrightarrow c+d}$ given in Eq. (\ref{HIofM}) by the
replacement
\be
b_s \to \bar b^\dagger_s\,,\quad b^\dagger_s
\to \bar b_s\,,\qquad u_s \to v_s\,, \quad \bar u_s \to \bar v_s\,.
\ee
From Eqs.~(\ref{Magic4}), we see that for the species $b$, the Pauli
blocking operator of the particle species $b$, $\widehat B_{rs}$, (resp
the number operator $B_{rs}$) is replaced by
number operator of the antiparticle species $\bar b$,
$\overline{B}_{rs}$, (resp. the Pauli blocking operator $\widehat{\overline{B}}_{rs}$), which was expected since we have changed an initial
state for a final state. The change from a particle to an antiparticle
is also expected and consistent with the contraction of the operators
which is now made with $v_s$ and $\bar v_s$ instead of $u_s$ and $\bar u_s$ when defining the
associated spinor valued operators as in Eqs. (\ref{Fspinor}). For instance the intensity part due to the
process $a + \bar c \leftrightarrow \bar b +d$ on the species $a$ is
\be
I_{C[A]} = \frac{2^{9}}{2 E_a} g^2 {\cal K}
\left[{\cal T}_I(\widehat{\gr{A}},\overline{\gr{B}},\widehat{\overline{\gr{C}}},\gr{D}) - {\cal T}_I(\gr{A},\widehat{\overline{\gr{B}}},\overline{\gr{C}},\widehat{\gr{D}})\right]\,,
\ee
where we made clear with the barred notation that the covariant quantities related to the species $\bar b$ and
$\bar c$ are those of antiparticles. These rules for crossing symmetry
apply as well for the polarisation part of the collision term.

\subsection{Structure of the collision Kernels}\label{SecStructure}

Before applying these general results to particular examples corresponding to physical situations, let us comment on the general structure of the collision terms.
\begin{itemize}
\item The intensity Kernel satisfies the property
\be
{\cal T}_I(\gr{A},\gr{B},\gr{C},\gr{D}) = {\cal T}_I(\gr{C},\gr{D},\gr{A},\gr{B})\,.
\ee
Since the simple interaction (\ref{HIabcd}) has a CP symmetry, then it
must also have a T (time-reversal) symmetry to ensure CPT symmetry and
this is reflected by this relation. This property could also have been found from Eq. (\ref{ICA}).

\item In the case where the species $a$ whose collision term is considered is massless, the Kernel for the circular polarisation (\ref{CVKernel}) does not involve couplings of the left and right chiral coupling constants $\epsilon_-^{ac}$ and $\epsilon_+^{ac}$, since there are no terms proportional to $\epsilon_-^{ac}\epsilon_+^{ac}$. Furthermore, if no species is initially polarized (circularly or linearly), then if the chiral coupling constants are equal ($\epsilon_-^{ac}=\epsilon_+^{ac}$ and $\epsilon_-^{bd}=\epsilon_+^{bd}$), the Kernel for circular polarisation vanishes. A departure from a purely unpolarized state must be due to a difference between the left and right chiral couplings in one of the currents.

\item In the massless limit it is also instructive to consider the
particular case when one of the chiral couplings constants
$\epsilon_-^{ac}$ or $\epsilon_+^{ac}$ vanishes, corresponding to
either a purely left or purely right chiral coupling in the current of
the $ac$ pair. In that case, we can focus on the purely left or purely
right helicity part of the collision term by considering
\beanosub\label{TIpmV}
{\cal   T}_{I} (\gr{A},\gr{B},\gr{C},\gr{D})&\pm& \hatM_a {\cal
  T}_{V} (\gr{A},\gr{B},\gr{C},\gr{D})=(\epsilon^{ac}_\pm)^2 (\epsilon^{bd}_\pm)^2({\cal I}^\pm_a \cdot
  {\cal I}^\pm_b) ({\cal I}^\pm_c \cdot {\cal I}^\pm_d)+(\epsilon^{ac}_\pm)^2 (\epsilon^{bd}_{\mp})^2({\cal I}^\pm_a \cdot
  {\cal I}^{\mp}_d) ({\cal I}^\pm_c \cdot {\cal I}^{\mp}_b)\nonumber\\
&+&(\epsilon^{ac}_\pm)^2 \epsilon^{bd}_+ \epsilon^{bd}_{-}\Big[\sgnii\frac{1}{2}({\cal I}^\pm_a \cdot
  {\cal I}^\pm_c) ({\cal S}_b \cdot {\cal S}_d) \sgnzz ({\cal I}^\pm_a \cdot {\cal S}_b \cdot
  {\cal S}_d  \cdot {\cal   I}^\pm_c) \sgnzz ({\cal I}^\pm_c \cdot {\cal S}_b \cdot  {\cal S}_d  \cdot {\cal  I}^\pm_a )\nonumber\\
&&\qquad \qquad \qquad +({\cal I}^\pm_a  \cdot {\cal S}_d \cdot {\cal I}^\pm_c) {\cal J}_b +  ({\cal
   I}^\pm_c  \cdot {\cal S}_b \cdot  {\cal I}^\pm_a ) {\cal J}_d \sgnii {\cal J}_b {\cal J}_d ({\cal I}^\pm_a \cdot
  {\cal I}^\pm_c)\Big]\,.
\eeanosub
If the species $a$ is a particle $\hatM_a=1$ (resp. an
antiparticle $\hatM_a=-1$), and if $\epsilon_+^{ac}=0$ (resp. $\epsilon_-^{ac}=0$), then this expression means that only the left
helicities $I_a-V_a$ are affected by the collision term (resp. only the
right helicities $I_a + V_a$ ) since the collision term for the opposite helicity cancels exactly. If the species $a$ is also not
polarized linearly, then the linear polarisation part of the collision
term vanishes as well under these assumptions. Hence we recover that in the massless case, for
interactions which have a definite chirality for a given species and if the distribution of particles is maximally circularly polarized
($I=V$ or $I=-V$), then it
remains so even after interactions. The other helicity states can only be
sourced if the interactions are not purely chiral (both $\epsilon_\pm^{ac}\neq 0$) or if the particles are massive. These results are explained by the direct identification between helicity and chirality in the massless limit.

For instance at very high energies when we can consider that both
neutrinos and electrons are massless, neutrinos are only left chiral
and antineutrinos are only right chiral and they remain so because the currents with which they couple (\ref{JCCdetail}) and (\ref{JNCneutrinos})  are always purely left chiral, whereas for electrons both helicities are
populated since the neutral currents (\ref{JNCelectrons}) involve left
and right chiral coupling constants.

\item This feature is a particular case of a more general feature of
  the collision term. Indeed, in \S~\ref{SecGeneralabcd}, we have commented on the fact that
  the collision term only affects one of the two spin indices in the
  distribution function ${\cal A}_{rs}$. This means that a state with
  only positive helicities ${\cal A}_{++}$ (resp. ${\cal A}_{--}$) either evolves under one
  collision to ${\cal A}_{++}$ (resp. ${\cal A}_{--}$), ${\cal A}_{-+}$ or ${\cal
    A}_{+-}$. And the two latter states can evolve toward ${\cal
    A}_{--}$ (resp. ${\cal A}_{++}$) after a second collision. A
  distribution function which is fully circularly polarized (such that
  $I=\pm V$), that is which is made only of helicity states of one kind, must first interact to
  generate linear polarisation and then interact again to generate
  helicity states of the other kind. In the massless case and for the
  case $\epsilon_+^{ac}=0$, given that chirality is the same thing as
  helicity, then there is no transition possible from ${\cal A}_{--}$
  to ${\cal A}_{-+}$ or ${\cal A}_{+-}$ and the distribution with
  purely left helicities remains so even after collisions.

\end{itemize}

\section{Application to standard reactions}\label{SecStandard}

In practice the kernels for the collision term obtained in the
previous section are far too general. Indeed, we shall consider
cases in which only one species is polarized or could be polarized by
the collisions. For instance when considering electrons, apart in certain
circumstances, it is not useful to describe their polarisation state
since electromagnetic interactions will contribute to erase any
polarisation. Furthermore at high energies compared to the masses of
neutrinos, it is reasonable to approximate the neutrinos as being
massless. Under these two types of simplifications (absence of
polarisation or of mass for some species) the collision Kernels take
simpler forms. Finally for nearly all applications of interest, some
chiral coupling constants vanish.

\subsection{Processes involving $4$ bodies}

\subsubsection{Muon decay}

The muon decay is due to the interaction between the muon ($\mu^-$)/muon
neutrino ($\nu_\mu$) charged current and the electron ($e^-$)/neutrino ($\nu$) charged current. Furthermore it involves only left-chiral
couplings. It thus corresponds to the case
\be\label{abcd1}
a=\mu^-,\quad c= \nu_\mu,\quad b = \nu,\quad d = e^-\,,\qquad
\epsilon_{+}^{ac}=\epsilon_{+}^{bd}= 0\,,\quad \epsilon_{-}^{ac}=\epsilon_{-}^{bd}= 1\,,\quad g = \frac{G_F}{\sqrt{2}}.
\ee

We remind that for the decay reaction $a \leftrightarrow \bar b +c +d$, the collision term is deduced from the reaction $a+b \leftrightarrow c+d$ by crossing symmetry. For instance from Eqs. (\ref{CIKernel}) and (\ref{CQKernel}), we deduce that the covariant parts of the collision term are of the form
\bea\label{CollisionTermMuon}
I_{C[A]} &=& \frac{2^{8}}{2 E_a} G_F^2 {\cal K}
\left[{\cal T}_I(\widehat{\gr{A}},\overline{\gr{B}},\gr{C},\gr{D}) - {\cal T}_I(\gr{A},\widehat{\overline{\gr{B}}},\widehat{\gr{C}},\widehat{\gr{D}})\right]\slabel{TImuon}\\
{\cal Q}_{C[A]}^\mu &=& \frac{2^{8}}{2 E_a} G_F^2 {\cal K} {{(H_a)}^\mu}_\nu \left[{\cal
    T}_{\cal Q}^\nu(\widehat{\gr{A}},\overline{\gr{B}},\gr{C},\gr{D}) - {\cal
    T}_{\cal Q}^\nu(\gr{A},\widehat{\overline{\gr{B}}},\widehat{\gr{C}},\widehat{\gr{D}})\right]\slabel{TQmuon}
\eea
and where it is stressed by a barred notation that the covariant quantities related to the species $a$, $c$ and $d$ are those of particles, and those for the species $\bar b$ are those of antiparticles. The Kernels are the most simple ones and read
\bea\label{TIQMuon}
{\cal T}_I(\gr{A},\gr{B},\gr{C},\gr{D}) &=& ({\cal I}^-_a \cdot  {\cal I}^-_b) ({\cal I}^-_c \cdot {\cal I}^-_d)\\
{\cal
    T}_{\cal Q}^\nu(\gr{A},\gr{B},\gr{C},\gr{D}) &=&\sgnminusslash({\cal S}^-_a \cdot {\cal I}^{-}_b)^\nu  ({\cal I}^-_c \cdot  {\cal I}^{-}_d)\,.\slabel{TQmuon2}
\eea
If we neglect the three body reactions, we need only to consider the
scattering out processes which correspond to the muon
decay. Furthermore if all species are unpolarized, and if the density
of species $b,c,d$ is low enough such that we can neglect their Pauli
blocking effects, then the intensity part of the collision term becomes 
\be
I_{C[A]}(p) =-\frac{32}{E_a} G_F^2 I_a(p)\int [\dd p_b][\dd
p_c][\dd
p_d](2\pi)^4\delta^{(4)}(p_d+p_c-p_b-p) (p
\cdot p_b) (p_c \cdot p_d)\,.
\ee
As a check, the muon lifetime is recovered from this collision term
evaluated at null spatial momentum of  ($\gr{p}=0$) thanks to the definition
$I_{C[A]}(\gr{p}=0) = \dd I_a/\dd t (\gr{p}=0) \equiv -\Gamma_a I_a(\gr{p}=0)$. Using $p \cdot p_b = \sgnslash
m_a E_b$, the muon lifetime is obtained as
\be\label{Gammamuon}
\Gamma_a=32 G_F^2 \int [\dd p_b][\dd
p_c][\dd
p_d](2\pi)^4\delta^{(3)}(\gr{p}_d+\gr{p}_c-\gr{p}_b)\delta^{(1)}(E_d+E_c-E_b-m_a)\,E_b\, (\sgnslash p_c \cdot p_d)\,,
\ee
and this is exactly the expression that would be obtained from the Fermi golden rule.

\subsubsection{Muon annihilation on neutrinos ($\mu^- +\nu \leftrightarrow \nu_\mu + e^-$)}

This reaction corresponds to the general case $a+b \leftrightarrow c+d$ for which the parameters are those of Eqs. (\ref{abcd1}). Again, since the currents coupled are purely left chiral, the structure of the collision term is much more transparent than the general case and illuminating.
The covariant components of the collision term are the same as Eqs.~(\ref{CollisionTermMuon}) up to a crossing symmetry. That is the intensity and polarisation parts are
\bea\label{CollisionTermMuonNeutrinos}
I_{C[A]} &=& \frac{2^{8}}{2 E_a} G_F^2 {\cal K}
\left[{\cal T}_I(\widehat{\gr{A}},\widehat{\gr{B}},\gr{C},\gr{D}) - {\cal T}_I(\gr{A},\gr{B},\widehat{\gr{C}},\widehat{\gr{D}})\right]\slabel{TImuonNeutrinos}\\
{\cal Q}_{C[A]}^\mu &=& \frac{2^{8}}{2 E_a} G_F^2 {\cal K} {{(H_a)}^\mu}_\nu \left[{\cal
    T}_{\cal Q}^\nu(\widehat{\gr{A}},\widehat{\gr{B}},\gr{C},\gr{D}) - {\cal
    T}_{\cal Q}^\nu(\gr{A},\gr{B},\widehat{\gr{C}},\widehat{\gr{D}})\right]\slabel{TQmuonNeutrinos}
\eea
with the Kernels (\ref{TIQMuon}). Let us comment on its structure.

\begin{itemize}
\item The circular polarisation part is extracted by evaluating $\sgnminusslash S_\mu {\cal Q}_{C[A]}^\mu$. In the case where we can consider the muon massless ($m_a =0$) the circular polarisation part of the collision is such that
\bea\label{MagicTIV}
{\cal T}_I(\gr{A},\gr{B},\gr{C},\gr{D}) + {\cal T}_V(\gr{A},\gr{B},\gr{C},\gr{D}) &=&0\\
{\cal T}_I(\gr{A},\gr{B},\gr{C},\gr{D}) - {\cal T}_V(\gr{A},\gr{B},\gr{C},\gr{D}) &=& ({\cal I}^-_a \cdot  {\cal I}^-_b) ({\cal I}^-_c \cdot {\cal I}^-_d)
\eea
and this can also be seen from Eq. (\ref{TIpmV}) applied to the special case (\ref{abcd1}). This means that if there are only left helical muons (the distribution is maximally circularly polarised with $I=-V$ so that ${\cal I}^+_\mu=0$), then it remains so in the massless limit. Furthermore, if linear polarisation is not present, then the linear polarisation part of (\ref{TQmuon2}) also vanishes, meaning that no linear polarisation can be generated. To summarize, with only left chiral interactions in the massless case, a distribution which has only left helicities remains entirely left chiral even after interactions take place, as expected. This property has already been commented at the end of \S~\ref{SecStructure}.

\com{The Kernel (\ref{TQmuon2}) has to be projected in Eq. (\ref{TQmuon}) to
form the polarisation part of the collision. From the definition
(\ref{defScal}) for ${\cal S}^-_{\mu\nu}$ we see that the term
involving ${\cal S}_{\mu\nu}$ cannot contribute if muons are initially
not polarized. Even if they are linearly polarized, this contribution does not
enhance polarisation since either the projection acts on $p_a$ and thus it cancels, or it acts on ${\cal
  Q}_a^\mu$, but if it is a scattering out process this reduces
polarisation, and if it is a scattering in process, then this is a
Pauli blocking form and also reduces polarisation. Hence one can
replace $({\cal S}^-_a \cdot {\cal I}^{-}_b)^\nu$ by $\sgnii M_a  {\cal I}_a
{\cal I}^{-\,\nu}_b$. The first conclusion we draw is that
polarisation can be generated for muons only because they have a
mass.} 
\item Since muons are not massless, we must still consider that interactions may generate linear polarisation. For instance, in the case where all species $a,b,c,d$ are unpolarized, we find $({\cal S}^-_a \cdot {\cal I}^{-}_b)^\nu = \sgnii M_a  {\cal I}_a {\cal I}^{-\,\nu}_b $ and collisions generate polarisation  with the source term proportional to the muon mass. Furthermore, since ${\cal I}^{-\,\nu}_b= {\cal I}_b p_b^\nu$ and the polarisation part of the collision term is proportional to
${{(H_a)}^\mu}_\nu   p_b^\nu$, the induced polarisation is transverse to $p_a^\mu$, the momentum of species $b$ . Polarisation is generated in these collisions since the pure left-chiral interactions are no longer entirely left-helical since the muons have a mass, and thus purely left chiral interactions are allowed to change the helicity of states.

\item Let us focus on the linear polarisation part of the collision term, still assuming that all species are initially unpolarised. It is sufficient to consider the spatial components of (\ref{TQmuonNeutrinos}). Neglecting the
  Pauli blocking effects, the part due to scattering in processes for
  the purely linear polarisation part of the collision term in the
  reaction $a+b \leftrightarrow c+d$ has the structure
\be\label{Qidipole}
Q_{C[A]}^i(p) \propto m_a \int \frac{[\dd p_c] [\dd p_d]}{E_b}\delta(E_c+E_d-E-E_b) (\delta^i_j \sgnzz \hat p^i \hat p_j) (p_c^j+p_d^j) I_c(\gr{p}_c) I_d(\gr{p}_d)(p_c \cdot p_d)\,.
\ee
The integrals on momenta $[\dd p_c][\dd p_d]$ can be separated into direction and
magnitude. One integral on magnitude (either $|\gr{p}_c|$ or
$|\gr{p}_d|$)  can be used to remove the Dirac delta function on energies. Let us for simplicity ignore the directional dependence in the factors $(p_c \cdot p_d)$ and $1/E_b$. The integral on the direction $\hat{\gr{p}}_c$ of the factor $p_c^j
I_c(\gr{p}_c) $ selects by definition the dipolar contribution
($\ell=1$ when expanded in spherical harmonics) of species $c$, and
similarly  the integral on $\hat{\gr{p}}_d$ of the factor $p_d^j
I_d(\gr{p}_d) $ selects the dipolar part of species $d$.  Furthermore the projection operator $(\delta^i_j \sgnzz \hat p^i \hat p_j)$ ensures that the linear polarisation generated
has also exactly a dipolar structure (we remind from
\S~\ref{SecYlm} that the angular expansion is performed with spin-$\pm1$ spherical harmonics). The reasoning is more complex  when correctly taking into account the directional dependence in $(p_c \cdot p_d)$ and $1/E_b$ but the conclusion remains the same, and it is that we need a dipolar structure in the initial species $c$ or $d$ to create polarisation in the massive particle $a$. This could also have been guessed from the angular decompositions of \S~\ref{SecYlm} since linear polarisation is decomposed in spin-$\pm1$ spherical harmonics, and these cannot have a monopole ($\ell=m=0$). 

\item However, when the distributions have no dipolar structure (for instance as those of homogeneous cosmology), nothing prevents circular polarisation to develop, since circular polarisation is expanded in standard spherical harmonics. A different observer, that is an observer boosted with
respect to the one which sees only purely monopolar distributions,
would see distributions with a dipolar structure, and thus would see that the
collision term produces linear polarisation. There is no contradiction
because what appears as purely circular polarisation for an observer,
appears as both circular and linear polarisation for the boosted
one. Only the total polarisation ${\cal Q}^\mu$ is observer independent. In the massless case, both circular and linear polarisation are observer independent separately, but as we have seen, linear polarisation is not generated, and for all observers only intensity and circular polarisation are affected by collisions with the Kernels (\ref{MagicTIV}).
\end{itemize}

\subsubsection{Neutron decay}

The neutron decay is due to the coupling of the neutron ($n$)/proton
($p$) charged current and the neutrino/electron charged current. While
the latter is purely left chiral, the former has both chiral couplings
due to the effective constant $g_A$ defined in
Eq. (\ref{JCCpn}). Hence we must consider the case
\be\label{abcd2}
a=n,\quad c=p,\quad b = \nu,\quad d = e^-\,,\quad
\epsilon_{+}^{ac}=\frac{1-g_A}{2}\,,\quad \epsilon_{+}^{bd}= 0\,,\quad \epsilon_{-}^{ac}=\frac{1+g_A}{2}\,,\quad \epsilon_{-}^{bd}= 1\,,\quad g = \frac{G_F\cos \theta_C}{\sqrt{2}}\,.
\ee
The neutron decay is deduced by crossing symmetry from the reaction
$n+\nu \leftrightarrow p + e^-$ (and so is the reaction $n+
e^+ \leftrightarrow p + \bar \nu$). As in the muon decay, the covariant parts of the collision term are of
the form (\ref{CollisionTermMuon}), but now the Kernels are slightly
more complicated due to the non-vanishing $\epsilon_{+}^{ac}$. For instance the intensity part of the collision term for the neutron
decay takes the same form as (\ref{TImuon}), with an extra CKM angle factor $(\cos \theta_C)^2$, but with the Kernel
\beanosub
{\cal T}_I(\gr{A},\gr{B},\gr{C},\gr{D})&=&\left(\frac{1+g_A}{2}\right)^2 ({\cal I}^-_a \cdot  {\cal I}^-_b) ({\cal I}^-_c \cdot {\cal I}^-_d)+\left(\frac{1-g_A}{2}\right)^2 ({\cal I}^+_a \cdot
  {\cal I}^{-}_d) ({\cal I}^+_c \cdot {\cal I}^{-}_b)\\
&+&\left(\frac{1-g_A^2}{4}\right)\Big(\sgnii \frac{1}{2}({\cal I}^-_b \cdot
  {\cal I}^-_d) [{\cal S}_a \cdot {\cal S}_c] \sgnzz ({\cal I}^-_b \cdot {\cal S}_a \cdot
  {\cal S}_c  \cdot {\cal
   I}^-_d) \sgnzz ({\cal I}^-_d \cdot {\cal S}_a \cdot
  {\cal S}_c  \cdot {\cal
   I}^-_b )\nonumber\\
&&\qquad \qquad\qquad+({\cal I}^-_b  \cdot {\cal S}_c \cdot {\cal I}^-_d) {\cal J}_a +  ({\cal
   I}^-_d  \cdot {\cal S}_a \cdot  {\cal I}^-_b ) {\cal J}_c \sgnii {\cal J}_a {\cal J}_c ({\cal I}^-_b \cdot
  {\cal I}^-_d)\Big)\,.\nonumber
\eeanosub
If no species is polarized, the Kernel takes the simpler form
\beanosub
{\cal
  T}_I(\gr{A},\gr{B},\gr{C},\gr{D}) &=& {\cal I}_a {\cal I}_b {\cal
  I}_c {\cal I}_d \left[\left(\frac{1+g_A}{2}\right)^2 (p_a \cdot p_b)
  (p_c \cdot p_d)+\left(\frac{1-g_A}{2}\right)^2 (p_a \cdot
p_d) (p_b \cdot p_c)\right.\\
&&\qquad \qquad \quad\left.\sgnzz \left(\frac{g_A^2-1}{4}\right)M_a M_c (p_b \cdot p_d)\right]\,.\nonumber
\eeanosub
This form of the Kernel matches Eq. (3.25) of
Ref.~\cite{Blaschke:2016xxt}. The life-time of the neutron at zero temperature (that is when we can
neglect the Pauli blocking factors), is obtained from the same method
as for the lifetime of the muon (\ref{Gammamuon}) and we get
\beanosub
\Gamma_a &=& \sgnslash 32 (G_F \cos \theta_C)^2 \int [\dd p_b][\dd
p_c][\dd
p_d](2\pi)^4\delta^{(3)}(\gr{p}_d+\gr{p}_c-\gr{p}_b)\delta^{(1)}(E_d+E_c-E_b-m_a)\nonumber\\
&&\times\left[ \left(\frac{1+g_A}{2}\right)^2  E_b\, (p_c \cdot
  p_d)+\left(\frac{1-g_A}{2}\right)^2 E_d (p_b \cdot p_c) + \left(\frac{g_A^2-1}{4}\right)m_c (p_b \cdot p_d)\right]\,.
\eeanosub
It is customary to evaluate this decay rate in the infinite mass limit
for the protons and neutrons, that is using the rule
\be
E_c=m_c\,,\qquad p_c \cdot p_d = \sgnslash m_c E_d\,,\qquad p_b \cdot p_c =\sgnslash m_c E_b\,,\qquad p_b
\cdot p_d = \sgnslash E_b E_d \sgnii \gr{p}_b \cdot \gr{p}_d\,.
\ee 
Integrating over the spatial momentum of the protons we obtain
\bea
\Gamma_a &\simeq & \frac{(G_F \cos \theta_C)^2}{(2\pi)^5} \int \dd^3 \gr{
p}_b \dd^3\gr{p}_d\delta^{(1)}(E_d+m_c-E_b-m_a)\left[(1+3 g_A^2)
+(1-g_A^2) \frac{\gr{p}_b \cdot \gr{p}_d}{E_b E_d}\right]\\
&\simeq &(1+3 g_A^2) \frac{(G_F \cos \theta_C)^2}{2 \pi^3} \int |\gr{p}_d| E_d  (E_d+m_c-m_a )^2 \dd E_d\,,
\eea
where in the last step we have used that an integral on spatial
momenta directions of $\gr{p}_p \cdot \gr{p}_d$ vanishes since it is odd, and we have considered the neutrinos as being totally massless. Hence we recover the standard result for the neutron decay in the $V-A$ theory (see e.g. Refs.~\cite{Lopez:1997ki,Lopez:1998vk,Esposito:1999sz,Serpico:2004gx,RevModPhys.83.1173}).

\subsubsection{Big-bang nucleosynthesis reactions}

We have already mentioned in \S~\ref{SecCrossingSymmetry} that from the collision term for neutrons
due to the reaction $n+\nu \leftrightarrow p + e^-$ we can deduce the
neutron decay and $n+ e^+ \leftrightarrow p + \bar \nu$ thanks to
crossing symmetry. To be complete, let us illustrate the procedure to get
all the collision terms for these four species, focusing on a few examples.

\begin{itemize}
\item In order to get the collision term for neutrinos, we must consider the reaction
$\nu+n \leftrightarrow e^-+p$, that is we must consider
\be\label{abcd3}
a=\nu,\quad c=e^-,\quad b = n,\quad d = p\,,\quad
\epsilon_{+}^{bd}=\frac{1-g_A}{2}\,,\quad \epsilon_{+}^{ac}= 0\,,\quad \epsilon_{-}^{bd}=\frac{1+g_A}{2}\,,\quad \epsilon_{-}^{ac}= 1\,,\quad g = \frac{G_F \cos \theta_C}{\sqrt{2}}\,.
\ee
\item In order to get the collision term for electrons we must consider the
reaction $e^-+p \leftrightarrow \nu+n $, that is the case
\be\label{abcd4}
a=e^-,\quad c=\nu,\quad b = p,\quad d = n\,,\quad
\epsilon_{+}^{bd}=\frac{1-g_A}{2}\,,\quad \epsilon_{+}^{ac}= 0\,,\quad \epsilon_{-}^{bd}=\frac{1+g_A}{2}\,,\quad \epsilon_{-}^{ac}= 1\,,\quad g = \frac{G_F \cos \theta_C}{\sqrt{2}}\,.
\ee
\item In order to get the collision term for antineutrinos which is
  due to the reaction $\bar \nu +p \leftrightarrow e^+ +n $, we must
  consider the reaction $\bar \nu +\bar n \leftrightarrow e^+ +\bar p$
  and deduce the former by crossing symmetry as explained in \S~\ref{SecCrossingSymmetry}. This means that we consider the case 
\be\label{abcd5}
a=\bar \nu,\quad c=e^+,\quad b = n,\quad d = p\,,\quad
\epsilon_{+}^{bd}=\frac{1-g_A}{2}\,,\quad \epsilon_{+}^{ac}= 0\,,\quad \epsilon_{-}^{bd}=\frac{1+g_A}{2}\,,\quad \epsilon_{-}^{ac}= 1\,,\quad g = \frac{G_F \cos \theta_C}{\sqrt{2}}\,.
\ee
\end{itemize}
In all these case, there is a polarisation component in the collision
term. Its detailed expression is more complex than Eq. (\ref{TQmuon2})
because the couplings are not purely left chiral but the general
structure remains the same. Circular polarisation is generated in all
cases, and linear polarisation is generated if the distribution functions of the interacting species have some dipolar distribution.
Since the electron/neutrino current is still purely left chiral, then
if these were the only interactions, the electrons and neutrinos would
develop circular polarisation. If this is certainly
the case for massless neutrinos, which develop maximal circular
polarisation, this would not be the case for electrons which have both right and chiral couplings thanks to neutral
current interactions, and also more importantly due to electromagnetic
interactions which are far stronger in general and would erase any
type of polarisation.

All the four-body reactions related to the neutron beta decay are gathered in Table \ref{Table1}.

{\renewcommand{\arraystretch}{2}%
\begin{figure}[!htb]
\begin{tabular}{|c|c|c|c|}
\hline
{\rm Reaction} & {\rm Particles names} & {\rm Chiral couplings} & $2^{-7}\,G_F^{-2} \,E \,I_{C[A]} = {\cal K} \times $\\[0.1cm]
\hline
   $n+\nu \leftrightarrow p+e^-$& $a+b \leftrightarrow c+d$ &  $\epsilon_-^{ac}=\frac{(1+g_A)}{2}\quad \epsilon_+^{ac}=\frac{(1-g_A)}{2}\quad \epsilon_-^{bd}=1\quad\epsilon_+^{bd}=0$ & ${\cal T}_I(\widehat{\gr{A}},\widehat{\gr{B}},\gr{C},\gr{D}) - {\cal T}_I(\gr{A},{\gr{B}},\widehat{\gr{C}},\widehat{\gr{D}})$\\
   \hline
   $p+e^-\leftrightarrow n+\nu $& $a+b \leftrightarrow c+d$ &  $\epsilon_-^{ac}=\frac{(1+g_A)}{2}\quad \epsilon_+^{ac}=\frac{(1-g_A)}{2}\quad \epsilon_-^{bd}=1\quad\epsilon_+^{bd}=0$ & ${\cal T}_I(\widehat{\gr{A}},\widehat{\gr{B}},\gr{C},\gr{D}) - {\cal T}_I(\gr{A},{\gr{B}},\widehat{\gr{C}},\widehat{\gr{D}})$\\
   \hline
   $\nu+n \leftrightarrow e^-+p$& $a+b \leftrightarrow c+d$ &  $\epsilon_-^{bd}=\frac{(1+g_A)}{2}\quad \epsilon_+^{bd}=\frac{(1-g_A)}{2}\quad \epsilon_-^{ac}=1\quad\epsilon_+^{ac}=0$ & ${\cal T}_I(\widehat{\gr{A}},\widehat{\gr{B}},\gr{C},\gr{D}) - {\cal T}_I(\gr{A},{\gr{B}},\widehat{\gr{C}},\widehat{\gr{D}})$\\
   \hline
   $e^-+p\leftrightarrow \nu+n $& $a+b \leftrightarrow c+d$ &  $\epsilon_-^{bd}=\frac{(1+g_A)}{2}\quad \epsilon_+^{bd}=\frac{(1-g_A)}{2}\quad \epsilon_-^{ac}=1\quad\epsilon_+^{ac}=0$ & ${\cal T}_I(\widehat{\gr{A}},\widehat{\gr{B}},\gr{C},\gr{D}) - {\cal T}_I(\gr{A},{\gr{B}},\widehat{\gr{C}},\widehat{\gr{D}})$\\
   \hline
   $n+e^+\leftrightarrow p+\bar \nu$ & $a+\bar d \leftrightarrow c+\bar b$ &  $\epsilon_-^{ac}=\frac{(1+g_A)}{2}\quad \epsilon_+^{ac}=\frac{(1-g_A)}{2}\quad \epsilon_-^{bd}=1\quad\epsilon_+^{bd}=0$ & ${\cal T}_I(\widehat{\gr{A}},\overline{\gr{B}},\gr{C},\widehat{\overline{\gr{D}}}) - {\cal T}_I(\gr{A},\widehat{\overline{\gr{B}}},\widehat{\gr{C}},\overline{\gr{D}})$\\
   \hline
   $p+\bar \nu\leftrightarrow n+e^+$ & $a+\bar d \leftrightarrow c+\bar b$ &  $\epsilon_-^{ac}=\frac{(1+g_A)}{2}\quad \epsilon_+^{ac}=\frac{(1-g_A)}{2}\quad \epsilon_-^{bd}=1\quad\epsilon_+^{bd}=0$ & ${\cal T}_I(\widehat{\gr{A}},\overline{\gr{B}},\gr{C},\widehat{\overline{\gr{D}}}) - {\cal T}_I(\gr{A},\widehat{\overline{\gr{B}}},\widehat{\gr{C}},\overline{\gr{D}})$\\
   \hline
   $e^++n\leftrightarrow \bar \nu+p$ & $a+\bar d \leftrightarrow c+\bar b$ &  $\epsilon_-^{bd}=\frac{(1+g_A)}{2}\quad \epsilon_+^{bd}=\frac{(1-g_A)}{2}\quad \epsilon_-^{ac}=1\quad\epsilon_+^{ac}=0$ & ${\cal T}_I(\widehat{\gr{A}},\overline{\gr{B}},\gr{C},\widehat{\overline{\gr{D}}}) - {\cal T}_I(\gr{A},\widehat{\overline{\gr{B}}},\widehat{\gr{C}},\overline{\gr{D}})$\\
   \hline
   $\bar \nu+p\leftrightarrow e^++n$ & $a+\bar d \leftrightarrow c+\bar b$ &  $\epsilon_-^{bd}=\frac{(1+g_A)}{2}\quad \epsilon_+^{bd}=\frac{(1-g_A)}{2}\quad \epsilon_-^{ac}=1\quad\epsilon_+^{ac}=0$ & ${\cal T}_I(\widehat{\gr{A}},\overline{\gr{B}},\gr{C},\widehat{\overline{\gr{D}}}) - {\cal T}_I(\gr{A},\widehat{\overline{\gr{B}}},\widehat{\gr{C}},\overline{\gr{D}})$\\
   \hline
   $n\leftrightarrow \bar \nu+ p+e^-$ & $a\leftrightarrow \bar b+ c+d$ &  $\epsilon_-^{ac}=\frac{(1+g_A)}{2}\quad \epsilon_+^{ac}=\frac{(1-g_A)}{2}\quad \epsilon_-^{bd}=1\quad\epsilon_+^{bd}=0$ & ${\cal T}_I(\widehat{\gr{A}},\overline{\gr{B}},\gr{C},\gr{D}) - {\cal T}_I(\gr{A},\widehat{\overline{\gr{B}}},\widehat{\gr{C}},\widehat{\gr{D}})$\\
   \hline
\end{tabular}
\caption{Main four-body reactions related to the neutron beta decay. The general intensity Kernel ${\cal T}_I$ is given in \S~\ref{SecExplicitCabcd} by Eq. (\ref{CIKernel}). The polarisation part of the collision term ${\cal Q}^\mu_{C[A]}$ is obtained by using instead ${{H_a}^\mu}_\nu {\cal T}_{\cal Q}^\nu$, where the Kernel for polarisation is given in the general case by Eq. (\ref{CQKernel}). If species $a$ is massless, one must distinguish between the purely linear polarisation part and the circular polarisation part of the collision term as explained in \S~\ref{SecmasslessGeneral}. In all cases, the collision term is the one of particle $a$ or antiparticle $\bar a$. In the arguments of the Kernels, we have either particles or antiparticles operators, and in the latter cases we use a barred notation to stress that the associated covariant components (${\cal I}^\pm_\mu$, ${\cal S}_{\mu\nu}$ etc...) which enter the expression of the Kernel are those of antiparticles. Furthermore for final states, operators are indicated with a hatted notation, since we must take the Pauli blocking forms, that is the hatted forms, of the covariant components defined in \S~\ref{SecExplicitCabcd}.}
\label{Table1}
\end{figure}}

\subsection{Processes involving $2$ bodies}

There are in general two types of $2$-bodies interactions. Either reactions of thermalisation of the type $a+b \leftrightarrow a+b$ or annihilation reactions which enforce chemical equilibrium such as $a+ \bar a \leftrightarrow
b+\bar b$~\cite{Dicus1982}. As analyzed in \S~\ref{SecCrossingSymmetry}, using
crossing symmetry one can deduce the latter from the
former. Additionally all cases can be reduced to the interaction of
neutral currents with a simple Fierz reordering as in
Eq. (\ref{CrossingSymmetry}). The two-body process can thus be considered as special cases of the
$4$-body processes for which 
\be\label{Conditiontwobody}
c=a\,,\qquad d=b\,,
\ee
and the interaction Hamiltonian due to neutral current takes the form
\be
{\cal H}_I = -{\cal L}_I =\sgnslash g J^{aa}_\mu  J^\mu_{bb}\,, \qquad g = 2\frac{G_F}{\sqrt{2}}\,.
\ee
This extra factor $2$ in the coupling constant $g$ can be viewed as the contribution of the complex conjugated terms in Eq. (\ref{HIabcd}).
The general derivation of the collision term follows the same general
steps as in \S~\ref{SecGeneralabcd}. However Eq. (\ref{HHD}) is
modified since operators of species $c$ and $a$ do not trivially
commute as they are the same species and similarly for species $b$ and
$d$. We must instead use
\be
\langle [{\cal M},[{\cal M},A_{ss'}]] \rangle \equiv M^\star_{\alpha\,\beta\to
   \gamma'\,\delta'}M_{\alpha'\,\beta'\to \gamma\,\delta} \langle [b_\beta^\dagger b_{\delta'}  a_\alpha^\dagger a_{\gamma'} ,
b_\delta^\dagger b_{\beta'} [ a_\gamma^\dagger a_{\alpha'},
a^\dagger_{s} a_{s'} ] ] \rangle\,,
\ee
\beanosub\label{HHD2}
&&\langle [b_\beta^\dagger b_{\delta'}  a_\alpha^\dagger a_{\gamma'} ,
b_\delta^\dagger b_{\beta'} [ a_\gamma^\dagger a_{\alpha'},
a^\dagger_{s} a_{s'} ] ] \rangle =  \langle[B_{\beta \delta'}  A_{\alpha \gamma'} , B_{\delta \beta'} (
 A_{\gamma s'} \deltarel_{s \alpha'}-A_{s \alpha'}
  \deltarel_{\gamma s'}) \rangle\\
&&\qquad \qquad= \langle\left\{({B}_{\beta \beta'}\widehat{B}_{\delta
     \delta'}+{B}_{\beta \delta'}{B}_{\delta \beta'})
  \left[({A}_{\alpha s'}\widehat{A}_{\gamma \gamma'}+{A}_{\alpha
      \gamma'}{A}_{\gamma s'})\deltarel_{s \alpha'}- ({A}_{\alpha
      \alpha'}\widehat{A}_{s \gamma'}+{A}_{\alpha
      \gamma'}{A}_{s
      \alpha'})\deltarel_{\gamma s'}\right]\right.
  \nonumber\\
&&\qquad \qquad \quad\left.-({B}_{\delta \delta'}\widehat{B}_{\beta
    \beta'}+{B}_{\beta
    \delta'}{B}_{\delta \beta'})\left[ ({A}_{\gamma
      \gamma'}\widehat{A}_{\alpha s'}+{A}_{\alpha
      \gamma'}{A}_{\gamma
      s'})\deltarel_{s \alpha'}-({A}_{s\gamma'}\widehat{A}_{\alpha
      \alpha'}+{A}_{\alpha \gamma'}{A}_{s\alpha'})\deltarel_{\gamma s'}\right]\right\}\rangle\nonumber\,,
\eeanosub
where for the first equality we have use property (\ref{CorrectMasterCommute}) and for the second line we have used the property (\ref{MagicChaos}). The difference lies in the fact that there are additional terms with distribution functions
such as ${\cal B}_{\beta \delta'}$, ${\cal B}_{\delta \beta'}$ or
${\cal A}_{\alpha   \gamma'}$, that is such that indices are not pairs $\alpha s'$, $s \alpha'$, $\beta \beta'$,  $\gamma \gamma'$ or
$\delta \delta'$ as in Eq. (\ref{MMAresult}). However, when taking the expectation in the quantum state this introduces Dirac delta
functions [from Eq. (\ref{AverageNrs})], and for these types of terms it corresponds to a process where
the momenta of species $a$ and $b$ between the
initial and final state are unchanged, that is to processes where there is no
momentum exchange at all. It is expected that these processes should
be irrelevant and that scattering out and
scattering in processes due to these terms cancel identically. We
conclude that the analysis of \S~\ref{SecGeneralabcd}, and the general
features of the collision term we have subsequently highlighted, can be
completely transposed for the two-body processes, provided we use
(\ref{Conditiontwobody}).
In order to give concrete example we apply it to some two-body interactions mediated by weak interactions.

\subsubsection{neutrino/muon neutrino scattering}\label{Secnunumuon}

The interactions between neutrinos of different types (e.g. electronic
neutrinos and muonic neutrinos) are only due to neutral currents
with a pure left chiral coupling. The effect of the reaction $\nu +
\nu_\mu \leftrightarrow \nu +
\nu_\mu$ thus corresponds to the case
\be\label{abab1}
a=\nu\,,\quad b = \nu_\mu\,,\quad c=\nu\,,\quad d = \nu_\mu\,,\qquad
\epsilon_-^{ac} = \epsilon_-^{bd} = e^\nu_-=\frac{1}{2}\,,\quad \epsilon_+^{ac} = \epsilon_+^{bd} = 0\,,\quad g = 2 \frac{G_F}{\sqrt{2}}.
\ee
If we cannot neglect the masses of neutrinos, then the covariant parts
of the collision term take the form
\bea\label{CollisionTermNuNumu}
I_{C[A]} &=& \frac{2^{10}}{2 E_a} G_F^2 {\cal K}
\left[{\cal T}_I(\widehat{\gr{A}},\widehat{\gr{B}},\gr{C},\gr{D}) - {\cal T}_I(\gr{A},\gr{B},\widehat{\gr{C}},\widehat{\gr{D}})\right]\\
{\cal Q}_{C[A]}^\mu &=& \frac{2^{10}}{2 E_a} G_F^2 {\cal K} {{(H_a)}^\mu}_\nu \left[{\cal
    T}_{\cal Q}^\nu(\widehat{\gr{A}},\widehat{\gr{B}},\gr{C},\gr{D}) - {\cal
    T}_{\cal Q}^\nu(\gr{A},\gr{B},\widehat{\gr{C}},\widehat{\gr{D}})\right]\slabel{TQNuNuMuon}
\eea
where the Kernels are $1/2^4$ times those of the muon decay (\ref{TIQMuon}), that is
\bea
{\cal T}_I(\gr{A},\gr{B},\gr{C},\gr{D}) &=& \frac{1}{2^4}({\cal I}^-_a \cdot  {\cal I}^-_b) ({\cal I}^-_c \cdot {\cal I}^-_d)\\
{\cal
    T}_{\cal Q}^\nu(\gr{A},\gr{B},\gr{C},\gr{D}) &=&\sgnminusslash \frac{1}{2^4}({\cal S}^-_a \cdot {\cal I}^{-}_b)^\nu  ({\cal I}^-_c \cdot  {\cal I}^{-}_d)\,.
\eea
If we can consider that the electronic neutrinos are
massless, then we must also consider the collision term for circular
polarisation since Eq. ({\ref{TQNuNuMuon}}) only applies to the linear
polarisation part when the projector ${{(H_a)}^\mu}_\nu$ is replaced by the screen projector ${{({\cal H}_a)}^\mu}_\nu$. The circular polarisation part of the collision
term reads simply
\be\label{CollisionTermNuNumuV}
\hatM_a V_{C[A]} = \frac{2^{10}}{2 E_a} G_F^2 {\cal K}
\left[{\cal T}_V(\widehat{\gr{A}},\widehat{\gr{B}},\gr{C},\gr{D}) - {\cal T}_V(\gr{A},\gr{B},\widehat{\gr{C}},\widehat{\gr{D}})\right]
\ee
\be
{\cal T}_V({\gr{A}},{\gr{B}},\gr{C},\gr{D})  = - \frac{1}{2^4}({\cal I}^-_a \cdot  {\cal I}^-_b) ({\cal I}^-_c \cdot {\cal I}^-_d)\,.
\ee

The effect of the reaction $\nu +
\bar \nu_\mu \leftrightarrow \nu + \bar \nu_\mu$, which in our general
notation is $a+ \bar d \leftrightarrow c + \bar b$, is obtained by a simple crossing symmetry. For instance the intensity part of the
Kernel would be for that process
\be
I_{C[A]} = \frac{2^{10}}{2 E_a} G_F^2 {\cal K} \left[{\cal T}_I(\widehat{\gr{A}},\overline{\gr{B}},\gr{C},\widehat{\overline{\gr{D}}}) - {\cal T}_I(\gr{A},\widehat{\overline{\gr{B}}},\widehat{\gr{C}},\overline{\gr{D}})\right]\,,
\ee
where we emphasized again with a barred notation that the covariant
components of antiparticles are those of antiparticles.

For completeness, we must stress again that the effect of
antineutrino-muonic antineutrino reactions ($\bar \nu +
\bar \nu_\mu \leftrightarrow \bar \nu +\bar \nu_\mu$) on
antineutrinos is obtained by charge conjugation, that is by
considering the case
\be\label{abab1bar}
a=\bar\nu\,,\quad b = \bar \nu_\mu\,,\quad c=\bar \nu\,,\quad d = \bar
\nu_\mu\,,\qquad
\epsilon_-^{ac} = \epsilon_-^{bd} = e^\nu_-=\frac{1}{2}\,,\quad \epsilon_+^{ac} = \epsilon_+^{bd} = 0\,,\quad g = 2 \frac{G_F}{\sqrt{2}}\,.
\ee 
This means that the collision term takes the same form as Eqs. (\ref{CollisionTermNuNumu}) but where
all covariant components should now refer to antiparticle species. For instance the intensity part takes the form
\be
I_{C[A]} = \frac{2^{10}}{2 E_a} G_F^2 {\cal K}
\left[{\cal T}_I(\widehat{\overline{\gr{A}}},\widehat{\overline{\gr{B}}},\overline{\gr{C}},\overline{\gr{D}}) - {\cal T}_I(\overline{\gr{A}},\overline{\gr{B}},\widehat{\overline{\gr{C}}},\widehat{\overline{\gr{D}}})\right]\,.
\ee
The effect of the process $\bar \nu + \nu_\mu \leftrightarrow \bar
\nu +\nu_\mu$ is then obtained by a crossing symmetry as in \S~\ref{SecCrossingSymmetry}.

\subsubsection{neutrino/neutrino scattering}\label{Secnunu}

Neutrino-neutrino scattering ($\nu +\nu \leftrightarrow \nu+\nu$) and
neutrino-antineutrino scattering  ($\nu +\bar \nu \leftrightarrow
\nu+\bar \nu$) are special cases of the previous electronic
neutrino-muonic neutrino scattering but there are a few crucial differences in the derivation of the collision term. 

Let us focus first on neutrino autointeractions. It could be viewed as a special case of Eq. (\ref{abab1}) in which $a=b=\nu=c=d$. but with two differences. First when examining Eqs. (\ref{StructureHI}) and (\ref{StructureJNC}) we realize that this would require to consider $g=G_F/\sqrt{2}$ which is half of the valued used for the interaction between two different species of neutrinos. Since $H_I$ enters as a square, this is a reduction by factor $4$. Furthermore, when species $a$ and $b$ are the same species, in the the first line of Eq. (\ref{HHD2}) we must now use
\beanosub\label{HHD2bis}
&&\
[a_\delta^\dagger a_{\beta'} a_\gamma^\dagger a_{\alpha'},
a^\dagger_{s} a_{s'} ] =   A_{\delta \beta'}A_{\gamma s'}\deltarel_{s\alpha'} + A_{\delta s'}A_{\gamma \alpha'}\deltarel_{s\beta'}
- A_{\delta \beta'}A_{s \alpha'}  \deltarel_{\gamma s'}-A_{s \beta'}A_{\gamma\alpha'}  \deltarel_{\delta s'}\,,
\eeanosub
that is at this stage of the computation we have twice as many terms and this can be seen as a consequence from the fact that the initial particles are the same~\cite{Dolgov1997}. 
Computing the equivalent of the second line of Eq. (\ref{HHD2}), using property (\ref{MagicChaos}) we find four times more physical terms, which correspond to the product of two allowed pairings in initial states and two in final states. For instance, eliminating the non-physical processes, we find the contribution
\be
\langle A_{\beta \delta'}  A_{\alpha \gamma'}A_{\delta \beta'}A_{\gamma s'}\rangle \deltarel_{s\alpha'}= (\langle A_{\beta \beta'}\rangle \langle A_{\alpha s'}\rangle + \langle A_{\beta s'}\rangle \langle A_{\alpha \beta'}\rangle )(\langle \widehat{A}_{\alpha \delta'}\rangle \langle \widehat{A}_{\gamma\beta'}\rangle+\langle \widehat{A}_{\alpha \beta'}\rangle \langle \widehat{A}_{\gamma\delta'}\rangle)\deltarel_{s\alpha'}\,.
\ee
This factor $4$ can also be viewed as the fact that we now have two Feynmann diagrams which add coherently~\cite{Flowers:1976kb}.
Taking all factors into account at the level of the collision term [$1/4$ because the coupling constant $g$ is reduced by half, an extra factor $2$ in the first line of Eq. (\ref{HHD2}) and an extra $4$ in the second line of (\ref{HHD2})], we end up with an enhancement by a factor $2$ compared to neutrino scattering between two different flavors of neutrinos.

Let us focus now on the case of neutrinos scattering on antineutrinos of the same flavor. Again since it comes from the autointeraction of a neutrino flavor neutral current, the coupling constant is also $g=G_F/\sqrt{2}$, that is reduced by a factor $2$ compared to scattering between neutrinos and antineutrinos of different flavors. However, when computing in details the terms which contribute to the Hamiltonian for such process we find $4$ terms instead of one. Indeed we recover a double product, so we recover the factor $2$ which was initially lost. But we also get twice more terms because there are two Feynmann diagrams for the process (see graphs (a) and (c) of \cite{Flowers:1976kb}). And with a simple Fierz reordering, which is possible because neutrinos interact with a purely left chiral coupling, they add up coherently. Taking all these factors into account, the interaction Hamiltonian is enhanced by a factor $2$ compared to interactions between neutrinos and antineutrinos of different flavors, resulting in a factor $4$ enhancement of the collision term.

To summarize, when considering interactions between neutrinos ($\nu+\nu \leftrightarrow \nu+\nu$) one must consider the two-body case (\ref{abab1}) in the particular case $a=b=c=d=\nu$ and multiply the result by a factor $2$ (this point was omitted in Ref. \cite{Hannestad:1995rs}). And when considering interactions between neutrinos and antineutrinos of the same flavor ($\nu+\bar \nu \leftrightarrow \nu+\bar \nu$) one must consider the two-body interaction in the particular case $a=c=\nu$, $b=d=\bar \nu$ and multiply the result by a factor $4$ in agreement with Ref. \cite{Dolgov1997}. In particular, a simple crossing symmetry allows to get the former reactions $\nu+\nu \leftrightarrow \nu+\nu$ from the $\nu+\bar \nu \leftrightarrow \nu+\bar \nu$ only up to a factor $1/2$. We can interpret this reduction by a factor two using the fact that outgoing particles are identical and one must not double count the outgoing states.

\subsubsection{neutrino/electron scattering}

Contrary to neutrino-neutrino scattering, electron-neutrino
scattering is due to both charged and neutral currents. However the
Fierz reordering reduces the problem to an interaction of neutral currents with modified chiral couplings. Using Eqs. (\ref{ChiralCouplings}) and (\ref{epsilonmoinsplusun}), the effect of $\nu+e^-
\leftrightarrow \nu+e^-$ on neutrinos corresponds to the case
\be\label{abab3}
a= c= \nu\,,\quad b = d = e^-\,,\qquad \epsilon_{-}^{bd} =
\epsilon^e_- +1\,,\quad \epsilon_{+}^{bd} = \epsilon^e_+\,,\quad
\epsilon_-^{ac} = e^\nu_-=\frac{1}{2}\,,\quad \epsilon_+^{ac} = 0\,,\quad g=2 \frac{G_F}{\sqrt{2}}\,.
\ee
The effect of $\nu+e^+ \leftrightarrow \nu+e^+$ is obtained by a crossing symmetry.
The effect of $\bar \nu+e^+ \leftrightarrow \bar
\nu+e^+$ on antineutrinos is obtained from charge conjugation of
(\ref{abab3}), that is it corresponds to the case
\be\label{abab4}
a= c= \bar \nu\,,\quad b = d = e^+\,,\qquad \epsilon_{-}^{bd} =
\epsilon^e_- +1\,,\quad \epsilon_{+}^{bd} = \epsilon^e_+\,,\quad
\epsilon_-^{ac} = e^\nu_-=\frac{1}{2}\,,\quad \epsilon_+^{ac} = 0\,,\quad g=2 \frac{G_F}{\sqrt{2}}\,,
\ee
and the effect of $\bar \nu+e^- \leftrightarrow \bar
\nu+e^-$ is obtained from symmetry crossing.

From the electrons perspective, the effect of the  $e^-+\nu \leftrightarrow e^-+\nu$  on electrons corresponds to the case
\be
b= d= \nu\,,\quad a = c = e^-\,,\qquad \epsilon_{-}^{ac} =
\epsilon^e_- +1\,,\quad \epsilon_{+}^{ac} = \epsilon^e_+\,,\quad
\epsilon_-^{bd} = e^\nu_-=\frac{1}{2}\,,\quad \epsilon_+^{bd} = 0\,,\quad g= 2\frac{G_F}{\sqrt{2}}\,.
\ee
and again all other reactions can be obtained from crossing symmetry or charge conjugation as in \S~\ref{SecCrossingSymmetry}.

In all these cases, the generation of linear polarisation has the
general form (\ref{Qidipole}). Hence the general conditions to
generated polarisation for a given species through two-body
interactions are the same. The species under consideration must be
massive, and a dipolar structure must be present in either of the two
interacting species. For neutrinos, linear polarisation is
poorly generated because of their tiny mass and the ratio of the linear polarisation part to the intensity part of the collision term is typically of order $m_\nu/E_\nu$. However circular polarisation develops to favor left helicities. For electrons, a linear polarisation can develop, but it is generically erased by the scattering out contributions of the more efficient electromagnetic interactions of the type $e^-+e^- \leftrightarrow e^-+e^-$. Only in some special situations it is necessary to describe correctly the polarisation state of fermions.
\com{A typical
configuration to generate linear polarisation in electrons would be to consider very diluted electrons interacting with a very dense beam of
neutrinos, since a beam presents a dipolar structure.}

Finally, we can check that in the unpolarized case, these results for neutrino/electrons interaction and those for neutrino/neutrinos interactions obtained in \S~\ref{Secnunumuon} and \ref{Secnunu} are exactly the results of Ref. \cite{Grohs:2015tfy}. However note that as mentionned in this reference, there is a typo in the annihilation of neutrino and antineutrinos into electrons and positron in tables $1$ and $2$ of \cite{Dolgov1997}, and thus Tables 1.5 and 1.6 of \cite{NeutrinoBook}. The process described in these tables should be of the form $\nu+\bar \nu \leftrightarrow e^-+e^+$ and not $\nu+\bar \nu \leftrightarrow e^++e^-$. Up to this typographical correction our results agree also in the unpolarized case with Refs. \cite{Dolgov1997,NeutrinoBook} and we gather all reactions in Table~\ref{Table2}.

{\renewcommand{\arraystretch}{2}%
\begin{figure}[!htb]
\begin{tabular}{|c|c|c|c|}
\hline
{\rm Reaction} & {\rm Particles names} & {\rm Chiral couplings} & $2^{-9}\,G_F^{-2} \,E \,I_{C[A]} = {\cal K} \times $\\[0.1cm]
\hline
   $\nu+\nu_\mu \leftrightarrow \nu+\nu_\mu$& $a+b \leftrightarrow c+d$ &  $\epsilon_-^{ac}=\tfrac{1}{2}\quad \epsilon_+^{ac}=0\quad \epsilon_-^{bd}=\tfrac{1}{2}\quad\epsilon_+^{bd}=0$ & ${\cal T}_I(\widehat{\gr{A}},\widehat{\gr{B}},\gr{C},\gr{D}) - {\cal T}_I(\gr{A},{\gr{B}},\widehat{\gr{C}},\widehat{\gr{D}})$\\
   \hline
   $\nu+\bar\nu_\mu \leftrightarrow \nu+\bar \nu_\mu$& $a+\bar d \leftrightarrow c+\bar b$ &  $\epsilon_-^{ac}=\tfrac{1}{2}\quad \epsilon_+^{ac}=0\quad \epsilon_-^{bd}=\tfrac{1}{2}\quad\epsilon_+^{bd}=0$ & ${\cal T}_I(\widehat{\gr{A}},\overline{\gr{B}},\gr{C},\widehat{\overline{\gr{D}}}) - {\cal T}_I(\gr{A},\widehat{\overline{\gr{B}}},\widehat{\gr{C}},\overline{\gr{D}})$\\
   \hline
   $\nu+\bar \nu \leftrightarrow \bar \nu_\mu+\nu_\mu$& $a+\bar c \leftrightarrow \bar b+d$ &  $\epsilon_-^{ac}=\tfrac{1}{2}\quad \epsilon_+^{ac}=0\quad \epsilon_-^{bd}=\tfrac{1}{2}\quad\epsilon_+^{bd}=0$ & ${\cal T}_I(\widehat{\gr{A}},\overline{\gr{B}},\widehat{\overline{\gr{C}}},\gr{D}) - {\cal T}_I(\gr{A},\widehat{\overline{\gr{B}}},\overline{\gr{C}},\widehat{\gr{D}})$\\
   \hline
   $\nu+\nu \leftrightarrow \nu+\nu$& $a+b \leftrightarrow c+d$ &  $\epsilon_-^{ac}=\tfrac{1}{2}\quad \epsilon_+^{ac}=0\quad \epsilon_-^{bd}=\tfrac{1}{2}\quad\epsilon_+^{bd}=0$ & $2\left[{\cal T}_I(\widehat{\gr{A}},\widehat{\gr{B}},\gr{C},\gr{D}) - {\cal T}_I(\gr{A},{\gr{B}},\widehat{\gr{C}},\widehat{\gr{D}})\right]$\\
   \hline
   $\nu+\bar\nu\leftrightarrow \nu+\bar \nu$& $a+\bar d \leftrightarrow c+\bar b$ &  $\epsilon_-^{ac}=\tfrac{1}{2}\quad \epsilon_+^{ac}=0\quad \epsilon_-^{bd}=\tfrac{1}{2}\quad\epsilon_+^{bd}=0$ & $4\left[{\cal T}_I(\widehat{\gr{A}},\overline{\gr{B}},\gr{C},\widehat{\overline{\gr{D}}}) - {\cal T}_I(\gr{A},\widehat{\overline{\gr{B}}},\widehat{\gr{C}},\overline{\gr{D}})\right]$\\
   \hline
   $\nu+e^- \leftrightarrow \nu+e^-$& $a+b \leftrightarrow c+d$ &  $\epsilon_-^{ac}=\tfrac{1}{2}\quad \epsilon_+^{ac}=0\quad \epsilon_-^{bd}=\epsilon^e_-+1\quad\epsilon_+^{bd}=\epsilon_+^e$ & ${\cal T}_I(\widehat{\gr{A}},\widehat{\gr{B}},\gr{C},\gr{D}) - {\cal T}_I(\gr{A},{\gr{B}},\widehat{\gr{C}},\widehat{\gr{D}})$\\
   \hline
   $\nu+e^+ \leftrightarrow \nu+e^+$& $a+\bar d \leftrightarrow c+\bar b$ &  $\epsilon_-^{ac}=\tfrac{1}{2}\quad \epsilon_+^{ac}=0\quad \epsilon_-^{bd}=\epsilon^e_-+1\quad\epsilon_+^{bd}=\epsilon_+^e$ & ${\cal T}_I(\widehat{\gr{A}},\overline{\gr{B}},\gr{C},\widehat{\overline{\gr{D}}}) - {\cal T}_I(\gr{A},\widehat{\overline{\gr{B}}},\widehat{\gr{C}},\overline{\gr{D}})$\\
   \hline
   $\nu+\bar \nu \leftrightarrow e^+ +e^-$& $a+\bar c \leftrightarrow \bar b +d$ &  $\epsilon_-^{ac}=\tfrac{1}{2}\quad \epsilon_+^{ac}=0\quad \epsilon_-^{bd}=\epsilon^e_-+1\quad\epsilon_+^{bd}=\epsilon_+^e$ & ${\cal T}_I(\widehat{\gr{A}},\overline{\gr{B}},\widehat{\overline{\gr{C}}},\gr{D}) - {\cal T}_I(\gr{A},\widehat{\overline{\gr{B}}},\overline{\gr{C}},\widehat{\gr{D}})$\\
   \hline
\end{tabular}
\caption{Main two-body reactions for the collision term  of neutrinos. See Table \ref{Table1} for more details about the notation. The electronic neutrino is noted $\nu$ and the muonic neutrino is noted $\nu_\mu$. Similar reactions for antineutrinos can be deduced with a global charge conjugation on all these reactions, and thus on all the collision Kernels.}
\label{Table2}
\end{figure}}

\subsection{Majorana neutrinos}

In the case where we cannot neglect the mass of neutrinos, there are
two possibilities. 
\begin{itemize}
\item
Either the neutrinos are Dirac particles, as we
have implicitly assumed so far, and right helicity states are usually
underpopulated because the linear polarisation collision term is
reduced a factor $m_\nu/E_\nu$ compared to the intensity part of the
collision term. Furthermore to populate the right helicity states ($f_{++}$) one
needs another interaction, according to the structure described at the
end of \S~\ref{SecStructure} and this is also reduced by a factor
$m_\nu/E_\nu$. To a good approximation the quantum neutrino field is
excited only in left helicity particles and right helicity
antiparticles and the distribution can be considered as being
maximally circularly polarized ($I=-V$ for the neutrinos and $I=V$ for
the antineutrinos).
\item
Another possibility is that neutrinos are Majorana particles. In that
case they are their own antiparticles. The quantum field (\ref{DefQuantumField}) is modified
by replacing $\bar a^\dagger_s$ with $a^\dagger_s$~\cite{Pal:2010ih,BibleSpinors} and consequently in
the interaction Hamiltonian we also replace $\bar a_s$ by $a_s$. From
Eq.~(\ref{CJC}) and using that for a neutral current $J_{aa}^{\mu\star}=
J_{aa}^\mu$ we see that for a Majorana particle, the effect is that
both left and right chiral couplings must be equal in the associated neutral current. That is we must consider that for Majorana neutrinos
\be\label{epsilonsMajorana}
\epsilon^{aa}_-=\epsilon^{aa}_+=e^\nu_-=\frac{1}{2}\,.
\ee
This can be seen more directly by noting from Eqs. (\ref{usCvs}) and (\ref{CgammaC}) that
\be
\bar v_r \gamma^\mu(\mathds{1}-\gamma^5) v_s = \bar u_s \gamma^\mu(\mathds{1}+\gamma^5) u_r\,.
\ee
Hence, the collision term for Majorana neutrinos interacting with electrons or
positrons are not just the sum of the collision terms of Dirac
neutrinos and Dirac antineutrinos. Thanks to $CP$ symmetry, we know
already (see \S~\ref{SecCrossingSymmetry}) that the collision term of Dirac antineutrinos can be viewed as a collision term for neutrinos with
$\epsilon_+^{ac}=1/2$ and $\epsilon_{-}^{ac}=0$. The
interaction of Majorana neutrinos with electrons corresponds to
\be
a=c=\nu\,\qquad b=d=e^-\,,\qquad \epsilon_+^{ac}=\epsilon_-^{ac}=e^\nu_-=\frac{1}{2}
\,,\quad \epsilon_{-}^{bd} = \epsilon^e_- +1\,,\quad \epsilon_{+}^{bd} =
\epsilon^e_+\,,\quad g=2\frac{G_F}{\sqrt{2}}\,.
\ee 
Terms proportional to $(\epsilon_-^{ac})^2$
(resp. $(\epsilon_+^{ac})^2$) are those encountered for Dirac neutrinos
(resp. Dirac antineutrinos), and terms proportional to
$\epsilon_-^{ac} \epsilon_+^{ac}$ are characteristic of the Majorana
nature.

In the case where neutrinos are of Majorana type but massless ($m_a=m_c=0$), then if they are not linearly polarized, that is if there is no mixing between
their Dirac neutrino and Dirac antineutrino nature (${\cal
  S}_a^{\mu\nu}={\cal S}_c^{\mu\nu}=0$), then from Eq. (\ref{CQKernel})
we check that this remains the case. Furthermore in the intensity part
of the collision term (\ref{CIKernel}), the interference terms
proportional to $\epsilon_-^{ac}\epsilon_+^{ac}$ also vanish and their
Majorana nature has no physical effect. Finally, still assuming massless Majorana neutrinos interacting with electrons, the circular polarisation part of the collision term (\ref{CVKernel}) remains null
if the distribution is initially not polarized, that is if there are
as many Dirac neutrinos as Dirac antineutrinos. These conclusions are
of course similar when considering interactions with positrons or even
autointeractions between neutrinos. To conclude, we recover that in the massless case there is no experiment which can distinguish between Dirac and Majorana neutrinos.
\end{itemize}

\section*{Conclusion}

We have shown that the statistical description of fermions is based on a scalar function, the intensity $I$, and a polarisation vector ${\cal Q}^\mu$ which are both observer independent. The latter encompasses both linear and circular polarisation, but these concepts are observer dependent except in the massless case. 

The proper identification of the covariant components $I$  and ${\cal Q}^\mu$ allows to extend the statistical description of fermions to curved space-times, and they should be considered as functions of the position in space-time and of particle momentum. Hence this approach is more general than the more usual description based solely on the distribution functions $f_{rs}$ which is restricted to a Minkowski spacetime, and which requires to consider a given observer to define the helicity states. The connection between these two possible descriptions is made using the spinor valued operator (\ref{Fspinor}) and its decomposition (\ref{BigDecompositionGood}) which is a central result of this article.

We then derive a classical Boltzmann equation for the covariant components and we show in the case of weak interactions that all collision terms are expressed in terms of $I$ and ${\cal Q}^\mu$. We compute the general collision terms for neutron beta decay and related processes (results gathered in Table \ref{Table1}), and the collision terms for neutrino two-body interactions with other neutrinos or with electrons (results gathered in Table \ref{Table2}), for which we recover the standard results of literature when polarisation is ignored.

The formalism used for the statistical description of fermions is reminiscent of the double projected tensor used to describe photons. However, since linear polarisation for fermions is expanded in spin-$1$ spherical harmonics, it is generically sourced by distributions with at least a dipolar structure, whereas for photons, linear polarisation is sourced by distribution with at least a quadrupolar dependence as it is expanded in spin-$2$ spherical harmonics. Furthermore, a second difference  is that linear polarisation of fermions is not sourced by the charged current and neutral current couplings if the particle considered is massless.

This difference between fermions and photons can also be understood independently by analyzing the underlying spin structure of the number operators. Since we have combined two spin $1/2$ operators in the number operator of fermions and in their distribution function, it is expected that a spin $1$ quantity is involved in the statistical description, and this role is played by the linear polarisation $Q^\mu$. If fermions are massless, the parallel transport of linear polarisation described by the Liouville equation (\ref{LiouvilleQmassless}) is in all ways similar to the transport of photon polarisation in the eikonal approximation of a light beam. Similarly, when performing the statistical description of photons, we combine two spin-$1$ quantities in the number operator, hence a spin-$2$ quantity arises as the linear polarisation tensor $\bar T_{\mu\nu}$ [see Appendix. \ref{AppBosons}]. It is no surprise that the description of linear polarisation for a gas of photons is similar to the description of gravitational waves in the eikonal approximation \cite{Blanchet2014}. If in a quantum theory of gravitation, we were to describe statistically the gravitons which are spin-$2$ particles, we would surely find that their linear polarisation is described by a spin-$4$ quantity.

In the bosonic case (e.g. for photons) it is known that polarisation is related to the quadrupole moments, but these are typically suppressed in thermal equilibrium. For fermions, polarisation can be generated in thermal equilibrium as it is related to the dipoles and monopoles which cannot be suppressed. We find that both, circular polarisation and linear polarisation appear naturally. \com{In fact, chiral oscillations naturally induce circular polarisation and considering the observer-dependence of $V$ in the massive case, this automatically implies the existence of linear polarisation for a large class of observers. }

Finally, we stress that flavor and flavor oscillations can also be described in this formalism, hence it can be used to compute the dynamical evolution of one-particle distribution functions in various contexts, provided the molecular chaos assumption and the classical statistical description hold. However in most cases, linear polarisation is either erased by scattering out processes (e.g. for electrons) or poorly generated (for small mass neutrinos) and the usual approach based only on the intensity $I$ is sufficient.


\acknowledgments
C.F. is supported by the Wallonia-Brussels Federation grant ARC11/15-040 and the Belgian Federal Office for Science, Technical \& Cultural Affairs through the Interuniversity Attraction Pole P7/37.
C.F. thanks IAP for kind hospitality during the preparation of this article. C.P. is supported by French state funds managed by the ANR within the Investissements d'Avenir programme under reference ANR-11-IDEX-0004-02, and thanks J.-P. Uzan, S. Renaux-Petel and G. Faye for discussions on the kinetic theory.

\bibliography{Biblioweak}

\appendix

\section{Statistical description for vector bosons}\label{AppBosons}

The procedure to build distribution functions is very similar and much
simpler for vector bosons. Indeed one starts by considering the number operator (\ref{defNrs}) and the associated distribution function $f_{rs}(p)$ defined by Eq. (\ref{deffab}), but this time with creation and annihilation operators associated with bosons, and which satisfy commutation rules instead of anticommutation rules. 

In the Lorentz gauge, for a given momentum $p^\mu$ (and given an observer $u^\mu$ identified with the time-like vector $[e_0]^\mu$ of a tetrad), a polarisation basis is given by the three
vectors
\be
\epsilon_0^\mu(\gr{p}) \equiv S^\mu \,,\quad \epsilon_-^\mu(\gr{p})\,,\quad \epsilon_+^\mu(\gr{p})\,.
\ee
which are all unit vectors, mutually orthogonal and transverse to $p^\mu$. 
A quantum vector field is expanded as
\be\label{DefQuantumFieldA}
\psi(p) = \sum_{s=-1,0,1} \int [\dd p]\left[ {\rm e}^{\sgnslash \ii p \cdot x }
a^\dagger_s(\gr{p}) \epsilon_s^{\star \mu}(\gr{p})+ {\rm e}^{\sgnminusslash \ii p \cdot x }
a_s(\gr{p}) \epsilon_s^\mu(\gr{p})\right]\,
\ee
and the quantities which transform like (\ref{Trulea}), but with the
vector representation $D(\Lambda)=\Lambda$, are \be
\sum_{s=-1,0,1} a_s \epsilon_s^\mu\,,\qquad \sum_{s=-1,0,1} a^\dagger_s \epsilon_s^{\star\mu}\,.
\ee
A covariant distribution tensor is thus directly obtained by considering
\be\label{Deffmunu}
f^{\mu\nu}(p)\equiv\sum_{r,s=-1,0,1} f_{rs}(p) \epsilon_r^{\star \mu}(\gr{p})\epsilon_s^\nu(\gr{p})\,,
\ee
and by construction it is transverse
($f_{\mu\nu}p^\nu=f_{\mu\nu}p^\mu=0$). Since the spin indices refer to the representation of the rotation group $SU(2)$, for vector bosons the spin indices $-1,0,1$ correspond to the irreducible representation $\gr{3}$, and the distribution function lives in the  (reducible) representation $\gr{3} \otimes \gr{3}$. By decomposing $f_{\mu\nu}$ in antisymmetric, symmetric traceless, and trace part as
\be
f_{\mu\nu}(p) = \frac{1}{2}T_{\mu\nu}(p) -  \frac{\ii}{2m} p^\lambda\epsilon_{\lambda \mu \nu \sigma} V^{\sigma}(p) \sgnii \frac{1}{3}\perp_{\mu\nu} I(p)\,,\quad \perp_{\mu\nu} \equiv \eta_{\mu\nu} \sgnii \frac{p_\mu p_\nu}{m^2}\,,
\ee
where $T_{\mu\nu} = T_{(\mu\nu)}$, and with the transverse conditions $T_{\mu\nu}p^\mu=0$ and $V_{\mu}p^\mu=0$, we have decomposed it in irreducible components as $\gr{3} \otimes \gr{3} = \gr{5} \oplus \gr{3} \oplus \gr{1}$. We recall that the irreducible representations $2 \ell+1$ of the rotation group are spatial symmetric traceless tensor of rank $\ell$~\cite{Thorne1980}, which here are $T_{\mu\nu}$, $V_\mu$ and $I$.

In order to further decompose the angular dependence, that is the dependence in the direction of the spatial momentum $\hat p^i$, we decompose it according to the sub-representation of the rotation group $SO(2)$ around the helicity direction $S^\mu$ according to
\bea
T_{\mu\nu} = \bar T \left(S_\mu S_\nu \sgnzz \frac{1}{3}\perp_{\mu\nu}\right) +2 S_{(\mu} \bar T_{\nu)} + \bar T_{\mu\nu}\,,\qquad V_\mu = V S_\mu + \bar V_\mu \,.
\eea
The decomposition in spherical harmonics or spin-weighted spherical harmonics is then 
\begin{subequations}
\begin{align}
I&=f_{++}+f_{00}+f_{--}\,,\qquad &I=\sum_{\ell m}I_{\ell m}(|\gr{p}|) Y_{\ell m}(\hat{\gr{p}})\\
V&= f_{++}-f_{--}\,,\quad &V = \sum_{\ell m} V_{\ell m}(|\gr{p}|) Y_{\ell m}(\hat{\gr{p}}) \\
\bar V^\mu &= \bar V^+ \epsilon_+^\mu + \bar V^- \epsilon_-^\mu\,,\quad \bar V^+ =f_{- 0}-f_{0+}\,,\quad \bar V^- =f_{0-}-f_{+0}\,,\quad &\bar V^\pm = \sum_{\ell m} \bar V^\pm_{\ell m}(|\gr{p}|) Y^\pm_{\ell m}(\hat{\gr{p}}) \\
\bar T &= 2 f_{00}-(f_{++}+f_{--})\,,\qquad &\bar T=\sum_{\ell m}\bar T_{\ell m}(|\gr{p}|) Y_{\ell m}(\hat{\gr{p}})\\
\bar T^\mu &= \bar T^+ \epsilon_+^\mu + \bar T^- \epsilon_-^\mu\,,\quad \bar T^\pm = f_{0\pm}+f_{\mp 0}\,,\quad &\bar T^\pm = \sum_{\ell m} \bar T^\pm_{\ell m}(|\gr{p}|) Y^\pm_{\ell m}(\hat{\gr{p}}) \\
\bar T^{\mu\nu} &= \bar T^{++} \epsilon_+^\mu \epsilon_+^\nu+ \bar T^{--} \epsilon_-^\mu\epsilon_-^\nu\,,\quad \bar T^{\pm\pm} = 2 f_{\mp\pm}\,,\quad &\bar T^{\pm\pm} = \sum_{\ell m} \bar T^{\pm\pm}_{\ell m}(|\gr{p}|) Y^{\pm2}_{\ell m}(\hat{\gr{p}}) \,.
\end{align}
\end{subequations}
If the boson vector is massless, as it is the case for photons, then
the null-helicity states are not populated because they are not
physical so $f_{00}=f_{0\pm}=f_{\pm0}=0$, such that $\bar T = - I$,
$\bar T^\pm=\bar V^\pm=0$. For a given observer with four-velocity
$u^\mu$, and using the spatial momentum direction vector ($\hat p^0=0$, $\hat
p^i = p^i/|\gr{p}|$) seen by this observer, the decomposition of the distribution tensor has only four degrees of freedom and reads
\be\label{fmunuphotons}
f_{\mu\nu}(p) = \frac{1}{2}\left[\bar T_{\mu\nu}(p) \sgnii {\cal
    H}_{\mu\nu} I(p)- \ii (u^\lambda\epsilon_{\lambda \mu \nu \sigma}
 \hat p^{\sigma}) V(p) \right]\,.
\ee
$I=f_{++}+f_{--}$, $V=f_{++}-f_{--}$ and $\bar T^{\pm\pm}\equiv Q\pm \ii U = 2 f_{\mp\pm}$ are the usual Stokes parameters\footnote{It is sometimes customary in the cosmic microwave background context to define the distribution function as~\cite{DurrerBook} $f^{\rm CMB}_{r\,s} \equiv f_{-r\,s}$. Accordingly, the tensor valued function (\ref{Deffmunu}) is defined as $f^{\mu\nu} = \sum_{rs}f^{\rm CMB}_{rs}\epsilon_r^{\mu}\epsilon_s^\nu$. With this definition the Stokes parameters are $I\equiv f^{\rm CMB}_{-+}+f^{\rm CMB}_{+-}$, $V \equiv f^{\rm CMB}_{-+} - f^{\rm CMB}_{+-}$ and $Q\pm \ii U \equiv 2 f^{\rm CMB}_{\pm\pm}$.} corresponding to intensity, circular polarisation and linear polarisation. In that case, the distribution tensor is doubly transverse, that is transverse to the momentum $p^\mu$ and also  to the observer velocity $u^\mu$. In the $\gr{e}_\theta,\gr{e}_\phi$ basis its components are the Hermitian $2\times 2$ matrix~\cite{Hu:1997hp,Tsagas:2007yx,DurrerBook}
\be
\frac{1}{2}\left( \begin{array}{cc}
I+Q & U-\ii V  \\
U+\ii V & I-Q  \end{array} \right)\,.
\ee
This structure can also be understood from the fact that when considering the massless representation of the Poincar\'e group, the little group is no more $SO(3)$ but $ISO(2)$~\cite{Weinberg1} so the spin structure is classified according to $SO(2)\simeq U(1)$. Hence the number operator and thus the distribution function $f_{rs}$ correspond to the tensor product $\gr{2}_1  \otimes \gr{2}_1$ (in terms of representations of $SO(2)$) and the latter is decomposed in irreducible representations of $SO(2)$ as $\gr{2}_2 \oplus \gr{1} \oplus \gr{1}$ where $\gr{2}_1$ (resp. $\gr{2}_2$) is the spin-$1$ (resp. spin-$2$) representation of $SO(2)$. In practice and for the physical consequences, the main difference with massless fermions is that linear polarisation is expanded in spin-$\pm2$ spherical harmonics~\cite{Hu:1997hp} for massless vector bosons whereas we use spin-$\pm1$ spherical harmonics for fermions. For massless bosons, linear polarisation is thus sourced by quadrupolar distributions at least (spin-$\pm2$ spherical harmonics have no monopole nor dipole) whereas for fermions it is sourced by dipolar distributions at least (spin-$\pm1$ spherical harmonics have no monopole). Finally, just as the collision term for fermions is built out of the covariant parts $I$ and ${\cal Q}^\mu$, collision terms involving photons are in general built from the double transverse tensor (\ref{fmunuphotons}), see e.g. the case of Compton interactions in Refs.~\cite{Portsmouth:2004ee,Pitrou2008}.

Note also that in all derivations involving bosons, the Pauli blocking operators are everywhere replaced by stimulated emission operators. Adopting the notation $\widehat{N}_{rs}\equiv \deltarel_{rs}+N_{rs}$, for the stimulated emission operators, the properties (\ref{MagicChaos}) and (\ref{BasicCommutators}), which are central in the derivation of the collision term, remain true for bosons. Hence the collision term involving vector bosons, or fermions and bosons, are still formally given by an expression such as (\ref{ColAss}), see e.g. Eq. (39) of Ref. \cite{Beneke2010}.

\section{Commutation dictionary}\label{AppCommute}

We remind the compact notation with multi-indices which for a particle
species $a$ is
\be
A_{\alpha \alpha'} = N_{s_\alpha s'_\alpha}(p_\alpha,p'_\alpha) =
a^\dagger_{s_\alpha}(p_\alpha) a_{s'_\alpha}(p'_\alpha) \qquad
\deltarel_{\alpha \alpha'} = \delta^{\rm K}_{s_\alpha
  s'_\alpha} \deltarel(p_\alpha-p'_\alpha)\,.
\ee
From anticommuting rules (\ref{AntiCommuteRule}) we get
\be\label{Magic4}
a_{\alpha'}a^\dagger_{\alpha} = \left(\deltarel_{\alpha\alpha'}-
  A_{\alpha\alpha'}\right) \equiv \widehat{A}_{\alpha \alpha'}\qquad 
a^\dagger_{\alpha} a_{\alpha'}=A_{\alpha\alpha'}\,.
\ee
These basic rules are used to express the results in terms either of the number operator or of the Pauli blocking operator, and since different species commute we can use for instance
\be
a_{\alpha'}b_{\beta'} c_{\gamma'} d^\dagger_{\delta} a^\dagger_\alpha b_\beta^\dagger c^\dagger_\gamma
d_{\delta'} = \hat A_{\alpha\alpha'}  \hat B_{\beta\beta'}
\hat C_{\gamma\gamma'}  D_{\delta \delta'}\,.
\ee
It is then immediate to obtain commutators of the form
\bea\label{BasicCommutators}
&&[ a_{\alpha'}b_{\beta'},a^\dagger_\alpha b_\beta^\dagger] =\widehat A_{\alpha\alpha'}\widehat
B_{\beta\beta'}- A_{\alpha\alpha'}B_{\beta\beta'} \\
&&[ a_{\alpha'}b_{\beta'} c_{\gamma'} d_{\delta'},a^\dagger_\alpha b_\beta^\dagger c^\dagger_\gamma
d^\dagger_\delta]=\widehat{A}_{\alpha\alpha'}\widehat{B}_{\beta\beta'}\widehat{C}_{\gamma\gamma'}\widehat{D}_{\delta\delta'}-A_{\alpha\alpha'}B_{\beta\beta'}C_{\gamma\gamma'}D_{\delta\delta'}\,.
\eea
For commutators involving strictly more than two operators associated with a given species, we need the commutation rules
\be\label{CommuteaA}
[a_1 , A_{23}] = \deltarel_{12} a_3\qquad [a^\dagger_1 , A_{23}] = - \deltarel_{13} a^\dagger_2
\ee
\be\label{CorrectMasterCommute}
[A_{12} , A_{34}] = A_{14} \deltarel_{23} - A_{32} \deltarel_{14}\,,\qquad
[A_{12} A_{34} , A_{56}] = A_{12} (A_{36}\deltarel_{54} -A_{54}\deltarel_{36})+A_{34}( A_{16}\deltarel_{52} - A_{52}\deltarel_{16})  
\ee
where here $1,2,3,4,5,6$ represent multi-indices.
\com{We also find (still using $1,2,1',2'$ as multi-indices)
\be
a_{1'} a_{2'} a_1^\dagger a_2^\dagger = -\widehat A_{11'}\widehat
A_{22'}+\widehat A_{12'}\widehat A_{21'} \qquad
\ee
Using the molecular chaos assumption (\ref{Chaos}) 
\be
\langle a_1^\dagger
a_2^\dagger a_{1'} a_{2'}  \rangle = -\langle A_{11'}\rangle \langle
A_{22'} \rangle +\langle A_{12'} \rangle \langle A_{21'} \rangle
\ee
we deduce the expectation value of the commutator
\bea\label{Magic1}
\langle[ a_{1'} a_{2'} , a_1^\dagger a_2^\dagger ] \rangle&=&-\langle\widehat A_{11'}\rangle \langle\widehat
A_{22'}\rangle+\langle\widehat A_{12'}\rangle\langle\widehat A_{21'}\rangle+\langle A_{11'}\rangle \langle A_{22'}\rangle-\langle A_{12'}\rangle \langle A_{21'}\rangle\\
&=& \deltarel_{11'} \langle A_{22'} \rangle +\deltarel_{22'} \langle A_{11'}
\rangle-\deltarel_{11'}\deltarel_{22'}+\deltarel_{12'}\deltarel_{21'}-\deltarel_{12'}
\langle A_{21'} \rangle- \deltarel_{2 1'} \langle  A_{12'} \rangle\,.
\eea}

\section{Trace technology}\label{TraceTechnologie}

The traces of products of $\gamma^\mu$ operators can be found
recursively and are non-vanishing only when there is an even number of
these operators. Using the identity valid for even $n$
\be
\frac{1}{2}\{\gamma^{\mu_1},\gamma^{\mu_2}\dots \gamma^{\mu_n}\} = \sgnslash\sum_{i=2}^n (-1)^i g^{\mu_1 \mu_i}\gamma^{\mu_1}\dots \gamma^{\mu_{i-1}} \gamma^{\mu_{i+1}}\dots \gamma^{\mu_n},
\ee
and the cyclicity of the trace, the first few recursions are
\bea
T^{\mu\nu} &\equiv& {\rm Tr}[\gamma^\mu \gamma^\nu] = \sgnslash 4 g^{\mu\nu}\\
T^{\mu\nu\rho\sigma} &\equiv&{\rm Tr}[\gamma^\mu \gamma^\nu
\gamma^\rho\gamma^\sigma] = \sgnslash g^{\mu\nu}T^{\rho\sigma} \sgnii g^{\mu\rho}T^{\nu\sigma} \sgnzz g^{\mu\sigma}T^{\nu\rho}=
4\left(g^{\mu\nu}g^{\rho\sigma}+g^{\mu\sigma}g^{\nu\rho}-g^{\mu\rho}g^{\nu\sigma}\right)\\
T^{\mu\nu\rho\sigma\alpha\beta} &\equiv&{\rm Tr}[\gamma^\mu \gamma^\nu
\gamma^\rho\gamma^\sigma \gamma^\alpha \gamma^\beta]
=\sgnslash g^{\mu\nu}T^{\rho\sigma\alpha\beta} \sgnii g^{\mu\rho}T^{\nu\sigma\alpha\beta} \sgnzz g^{\mu\sigma}T^{\nu\rho\alpha\beta} \sgnii g^{\mu\alpha}T^{\nu\rho\sigma\beta} \sgnzz g^{\mu\beta}T^{\nu\rho\sigma\alpha}\,.
\eea
When $\gamma^5$ matrices are also present in the product, we can
simplify it first by gathering them thanks to the anticommutation rule
$\gamma^\mu \gamma^5 = - \gamma^5 \gamma^\mu$. If there is an even
number of $\gamma^5$, then using $(\gamma^5)^2= \mathds{1}$ the
problem reduces to a simple product of $\gamma^\mu$ already treated above. If there is an
odd number of $\gamma^5$ then the problem is reduced to a product of
one $\gamma^5$ with products of $\gamma^\mu$. Again the traces of these
types of products can be obtained by recursion. Also if there is an odd number of $\gamma^\mu$ or if it is strictly smaller
than $4$, this type of product vanishes. Using the identity
\be
\gamma^5 \gamma^\mu \gamma^\nu\gamma^\rho = \sgnslash\gamma^5\left(g^{\mu\nu} \gamma^\rho - g^{\mu\rho}\gamma^\nu + g^{\nu\rho}\gamma^\mu \right)\sgnzz \ii \epsilon^{\mu\nu\rho\sigma}\gamma_\sigma,
\ee
the first recursions are
\bea
T_5^{\mu\nu\rho\sigma} &=& {\rm Tr}[\gamma^5 \gamma^\mu \gamma^\nu
\gamma^\rho\gamma^\sigma] = 4 \ii \epsilon^{\mu\nu\rho\sigma}\\
T_5^{\mu\nu\rho\sigma\alpha\beta} &=& {\rm Tr}[\gamma^5 \gamma^\mu \gamma^\nu
\gamma^\rho\gamma^\sigma\gamma^\alpha \gamma^\beta] =
\sgnslash g^{\mu\nu}T_5^{\rho\sigma\alpha\beta} \sgnii g^{\mu\rho}T_5^{\nu\sigma\alpha\beta} \sgnzz g^{\nu\rho}T_5^{\mu\sigma\alpha\beta} \sgnzz \ii
{\epsilon^{\mu\nu\rho}}_\lambda T^{\lambda\sigma\alpha\beta}\com{\\
T_5^{\mu\nu\rho\sigma\alpha\beta\lambda\tau} &=& {\rm Tr}[\gamma^5 \gamma^\mu \gamma^\nu
\gamma^\rho\gamma^\sigma\gamma^\alpha \gamma^\beta \gamma^\lambda \gamma^\tau] =
g^{\mu\nu}T_5^{\rho\sigma\alpha\beta\lambda\tau}-g^{\mu\rho}T_5^{\nu\sigma\alpha\beta\lambda\tau}+g^{\nu\rho}T_5^{\mu\sigma\alpha\beta\lambda\tau}+\ii
{\epsilon^{\mu\nu\rho}}_\lambda T^{\lambda\sigma\alpha\beta\lambda\tau}}\,.
\eea
The automatic commutation of the $\gamma^5$ operators, and  the
recursive computation of the traces of large products together with
their simplifications are very well suited for a computer algebra
package and we used {\it xAct}~\cite{xAct} to deal with these large
but otherwise systematic computations.

\section{Two-components spinors}\label{AppTwoComponents}

\subsection{Spinors conventions}

The two-components spinor formalism is exposed in
Ref.~\cite{BibleSpinors}. We do not repeat all the definitions in this
section and we only gather the useful relations for this article. The two-component spinors indices
are raised and lowered with the antisymmetric tensors
\bea
&&\epsilon^{\dot{\alpha} \dot{\beta}},\qquad \epsilon^{\alpha \beta},\qquad \epsilon^{\dot{\alpha}
  \dot{\beta}} = (\epsilon^{\alpha \beta})^\star\,\qquad\epsilon^{12}
= 1\\
&&\epsilon_{\dot{\alpha} \dot{\beta}},\qquad \epsilon_{\alpha \beta},\qquad \epsilon_{\dot{\alpha}
  \dot{\beta}} = (\epsilon_{\alpha \beta})^\star\,\qquad\epsilon_{12}= -1\,.
\eea
Indices are conventionally always raised and lowered from the left,
for instance as
\be
\xi^\alpha = \epsilon^{\alpha \beta} \xi_\beta\quad \xi_\alpha = \epsilon_{\alpha \beta} \xi^\beta\,,
\ee
and the same convention applies for dotted indices. The consequence of
this convention is that
\be
{\epsilon^{\alpha}}_\beta = \epsilon^{\alpha\gamma}
\epsilon_{\gamma {\beta}}= \epsilon_{{\beta}\gamma}\epsilon^{\gamma
  \alpha}= {\epsilon_\beta}^\alpha = \delta^{\alpha}_{\beta}\,,\qquad \epsilon^{\alpha\beta} = - \epsilon^{\alpha\alpha'} \epsilon^{\beta \beta'} \epsilon_{\alpha' \beta'}\,.
\ee
In practice this means that
\be
\chi^\alpha \psi_\alpha = -\chi_\alpha \psi^\alpha\,.
\ee
Using the Pauli matrices $\vec{\sigma}$ we define the soldering forms
\bea
\sigma^\mu_{\alpha \dot{\beta}} &=& (1,\vec{\sigma}) \,,\qquad \sigma_{\mu
  \,\alpha \dot{\beta}} = \sgnslash (1,-\vec{\sigma}) \\
\bar\sigma^{\mu\,\dot{\alpha}\beta} &=& (1,-\vec{\sigma}) \,,\qquad
\bar\sigma_{\mu}^{\dot{\alpha} \beta} = \sgnslash (1,\vec{\sigma}) \,.
\eea
They are related since
\be
\bar\sigma^{\mu \dot\alpha \beta} = \sigma^{\mu \beta \dot\alpha} =
\epsilon^{\beta \beta'}\epsilon^{\dot \alpha \dot
  \alpha'}\sigma^\mu_{\beta' \dot \alpha'}\,.
\ee
The two-component spinors are particularly useful because there is a correspondence between vectors and bi-spinors $V_{\alpha \dot
  \beta}$ when using the soldering forms
\be\label{ConnectVectorandbispinor}
V^\mu = \frac{1}{2} \bar\sigma^{\mu\,\dot \beta \alpha}V_{\alpha \dot \beta} \,,\qquad
V_{\alpha \dot \beta} = \sgnslash V^\mu \sigma_{\mu\, \alpha \dot \beta}\,.
\ee
These relations can be proved by using the identities
\bea
\sigma^\mu_{\alpha \dot \alpha} \bar \sigma_\mu^{\dot \beta \beta} &=&
\sgnslash 2
\delta_\alpha^\beta \delta_{\dot \alpha}^{\dot \beta}\\
\sigma^\mu_{\alpha \dot \alpha} \sigma_{\mu\beta \dot \beta} &=& \sgnslash 2
\epsilon_{\alpha \beta} \epsilon_{\dot \alpha \dot \beta}\\
\bar \sigma^{\mu\dot \alpha \alpha} \bar \sigma_{\mu}^{\dot \beta
  \beta} &=& \sgnslash 2
\epsilon^{\alpha \beta} \epsilon^{\dot \alpha \dot \beta}\,.
\eea
In particular for a spinor $\xi_\alpha$ and its associated
$\xi^{\dagger \dot \alpha}$, we can define a vector
\be\label{vectorfromspin}
\Xi^\mu \equiv \xi^{\alpha} \sigma^{\mu}_{\alpha \dot \beta}
\xi^{\dagger {\dot \beta}} = \xi^\dagger_{\dot \beta}\bar\sigma^{\mu\,\dot \beta
  \alpha} \xi_\alpha = \xi \sigma^\mu \xi^\dagger = \xi^\dagger \bar \sigma^\mu \xi\,,
\ee
and it has the property to be a null vector $\xi_\mu \xi^\mu =0$.

\subsection{Clifford algebra with two-components spinors}\label{AppClifford}

A Dirac spinor is expressed in terms of two-components spinors by the
decomposition in a direct sum
\be\label{DecDiracinTCS}
\Psi_\mathfrak{a} = \left( \begin{array}{c}
\xi_\alpha\\
\eta^{\dagger \dot\alpha}  \end{array} \right)\,,
\ee
and this relates the Dirac spinor indices $\mathfrak{a}$ to the two-component spinor indices.
In the chiral representation, the Dirac $\gamma^\mu$ matrices are made
of $2\times2$ blocks and read
\be\label{Explicitgammamu}
{({\gamma^\mu})_\mathfrak{a}}^\mathfrak{b} = \left( \begin{array}{cc}
0 & \sigma^\mu_{\alpha \dot\beta}  \\
\bar \sigma^{\mu \dot \alpha \beta} & 0  \end{array} \right)\,.
\ee
From the properties
\be
{[\sigma^\mu \bar \sigma^\nu + \sigma^\nu \bar
  \sigma^\mu]_\alpha}^\beta= \sgnslash 2 g^{\mu\nu} {\delta_\alpha}^\beta \,,\qquad
{[\bar\sigma^\mu \sigma^\nu + \bar\sigma^\nu
  \sigma^\mu]^{\dot\alpha}}_{\dot \beta}= \sgnslash 2  g^{\mu\nu} {\delta^{\dot
    \alpha}}_{\dot \beta}
\ee
we check that they satisfy indeed the Clifford algebra
\be
\{\gamma^\mu,\gamma^\nu\} = \sgnslash 2 g^{\mu\nu} \left( \begin{array}{cc}
{\delta_\alpha}^\beta & 0\\
0 & {\delta^{\dot \alpha}}_{\dot \beta}  \end{array} \right)\,,
\ee
where for simplicity we omit to write the Dirac spinor indices on operators when it is obvious.

To extract chiral components, we also use the matrix
\be\label{Explicitgamma5}
\gamma^5 \equiv \frac{\ii}{4 !} \epsilon_{\mu\nu\rho\lambda}\gamma^\mu
\gamma^\nu \gamma^\rho \gamma^\lambda = \ii \gamma^0 \gamma^1 \gamma^2
\gamma^3 = \left( \begin{array}{cc}
-{\delta_\alpha}^\beta & 0 \\
0 & {\delta^{\dot \alpha}}_{\dot \beta}  \end{array}\right)\,.
\ee
We define also operators in the space of two-components spinors by
\be
\sigma^{\mu\nu} = \frac{\ii}{4}(\sigma^\mu \bar \sigma^\nu-\sigma^\nu
\bar \sigma^\mu)\,,\qquad \bar\sigma^{\mu\nu} = \frac{\ii}{4}(\bar
\sigma^\mu \sigma^\nu-\bar \sigma^\nu \sigma^\mu)\,,
\ee
such that the matrices (\ref{DefSigma}) are expressed in blocks as
\be\label{Sigmablock}
{(\Sigma^{\mu\nu})_\mathfrak{a}}^\mathfrak{b} = \left( \begin{array}{cc}
(\sigma^{\mu\nu})_\alpha^{\,\,\,\beta} & 0\\
0 & (\bar \sigma^{\mu\nu})^{\dot \alpha}_{\,\,\,\dot\beta}  \end{array} \right)\,.
\ee
Since the Dirac matrices are Hermitians, then we find that
\be
(\sigma^{\mu\nu})^\dagger = \bar \sigma^{\mu\nu}\,.
\ee
Defining the conjugation matrix $A$ and its inverse $A^{-1}$
\be
A^{\mathfrak{a}\mathfrak{b}}= \left( \begin{array}{cc}
0& {\delta^{\dot \alpha}}_{\dot \beta} \\
  {\delta_\alpha}^\beta&0\end{array} \right)\qquad A^{-1}_{\mathfrak{a}\mathfrak{b}}= \left( \begin{array}{cc}
0& {\delta_\alpha}^\beta \\
{\delta^{\dot \alpha}}_{\dot \beta}  &0\end{array}\right),
\ee
we obtain
\be\label{MagicSigma}
A \cdot \Sigma^{\mu\nu}\cdot A^{-1} = \Sigma^{\dagger \mu\nu} \,.
\ee
In components and for our chiral representation, we use the numerical
equality $A =A^{-1}= \gamma^0$ and it is customary to use the notation
$\gamma^0$ in place of $A$. For a Dirac spinor $\Psi_\mathfrak{a}$
decomposed as in Eq. (\ref{DecDiracinTCS}), the Dirac conjugate is defined as
\be\label{RigorousDiracConjugate}
 \bar \Psi \equiv \Psi^\dagger \cdot A \qquad\Leftrightarrow
 \qquad\bar \Psi ^\mathfrak{a} = \Psi_{\mathfrak{b}}^\star A^{\mathfrak{b}\mathfrak{a}}\,.
\ee
Finally, the charge conjugation operator defined in \S~\ref{SecChargeConjugation} is given by
\be
C_{\mathfrak{a}\mathfrak{b}}= \left( \begin{array}{cc}
\epsilon_{\alpha \beta}& 0 \\
  0&\epsilon^{\dot \alpha \dot \beta}\end{array} \right)\qquad (C^{-1})^{\mathfrak{a}\mathfrak{b}}= \left( \begin{array}{cc}
\epsilon^{\dot \alpha \dot \beta}& 0 \\
  0&\epsilon_{\alpha \beta}\end{array} \right)
\ee
and it satisfies the numerical equality $C = -C^{-1} = \ii \gamma^0\gamma^2$.

\subsection{Lorentz group representations}\label{AppSUdeux}

An infinitesimal Lorentz transformation is expressed as
\be
{\Lambda^\mu}_\nu = \delta^\mu_\nu +{\omega^\mu}_\nu \,,\qquad
\omega_{\mu\nu} = -\omega_{\nu\mu}\,.
\ee
A finite transformation is obtained through exponentiation of the
representation. For Dirac spinors, the representation is obtained as
\be\label{DiracRepresentation}
D(\Lambda) = {\rm exp}\left(-\frac{\ii}{2}\omega_{\mu\nu} \Sigma^{\mu\nu}\right), \qquad
\ee
where we remind the definition (\ref{DefSigma}). In explicit terms, using the two-component spinor expression
(\ref{Sigmablock}), it is given by
\be\label{DiracRepresentationM}
D(\Lambda) =  \left( \begin{array}{cc}
M & 0\\
0 & (M^\dagger)^{-1} \end{array} \right)\qquad M =
\exp(-\tfrac{\ii}{2}\omega_{\mu\nu}\sigma^{\mu\nu})\qquad (M^\dagger)^{-1} = \exp(-\tfrac{\ii}{2}\omega_{\mu\nu}\bar\sigma^{\mu\nu})\,.
\ee
The inverse representation is related thanks to the property (\ref{MagicSigma}) which gives
\be\label{MagicDinverse}
D(\Lambda^{-1}) = A^{-1} \cdot D^\dagger(\Lambda) \cdot A\,,\qquad \Leftrightarrow\quad 
{D^{\dagger \mathfrak{a}}}_\mathfrak{c}(\Lambda) A^{\mathfrak{c}\mathfrak{b}} = A^{\mathfrak{a}\mathfrak{c}} {D_\mathfrak{c}}^{\,\,\mathfrak{b}}(\Lambda^{-1}).
\ee
The representation $M$ is an element of SL(2), the covering group of the Lorentz group, and thus conservation of unit determinant implies the property
\be\label{Mmoinsun}
{M_\alpha}^\beta {M_{\alpha'}}^{\beta'} \epsilon_{\beta \beta'} =
\epsilon_{\alpha \alpha'}\qquad \Rightarrow \qquad {{M^{-1}}_\alpha}^\beta = - {M^\beta}_\alpha=- {{M^T}_\alpha}^\beta\,.
\ee
A left handed two-components spinor $\xi_\alpha$ transforms as ${M_\alpha}^\beta
\xi_\beta$ under the active transformation $\Lambda$. Using
(\ref{Mmoinsun}), a right handed two-component spinor $\xi^{\dagger\dot\alpha}$ transforms as $-{{M^\star}^{\dot
    \alpha}}_{\dot\beta} \xi^{\dagger \dot \beta} =
{({M^{\dagger\,-1})}^{\dot \alpha}}_{\dot \beta} \xi^{\dagger
  \dot\beta} $.  The representation (\ref{DiracRepresentationM}) acts
on a space which is the direct sum of left-handed and right handed
two-components spinors in agreement with the decomposition (\ref{DecDiracinTCS}).

When finding explicit expressions for the representation (that is for $M$ and $(M^\dagger)^{-1}$) we have to introduce the notation
\bea\label{sqrtpm}
{[\sqrt{\sgnslash p\cdot \sigma}]_\alpha}^\beta\equiv {[\sqrt{\sgnslash p\cdot \sigma \bar
  \sigma^0}]_\alpha}^\beta&=&\frac{\sgnslash p \cdot \sigma_{\alpha \dot
  \gamma} \bar \sigma^{0\,\dot \gamma \beta}+m\delta^{\beta}_\alpha}{\sqrt{2(E+m)}} \\
{[\sqrt{\sgnslash p\cdot \bar\sigma}]^{\dot\alpha}}_{\dot\beta}\equiv
{[\sqrt{\sgnslash p\cdot \bar \sigma 
  \sigma^0}]^{\dot\alpha}}_{\dot\beta}&=&\frac{\sgnslash p \cdot \bar\sigma^{\dot
  \alpha \gamma}
\sigma^0_{\gamma \dot \beta}+m\delta^{\dot \alpha}_{\dot \beta}}{\sqrt{2(E+m)}} \\
{[\sqrt{\sgnslash p \cdot \sigma}]^{\dot \alpha}}_{\dot \beta} \equiv
{{[\sqrt{\sgnslash \sigma^0 p\cdot
      \sigma}]}^{\dot\alpha}}_{\dot \beta}&=&\frac{\sgnslash \bar\sigma^{0\dot
  \alpha \gamma}  p \cdot \sigma_{\gamma \dot \beta}+m\delta^{\dot \alpha}_{\dot \beta}}{\sqrt{2(E+m)}} \\
{[\sqrt{\sgnslash p\cdot \bar\sigma}]_\alpha}^\beta \equiv 
{{[\sqrt{\sgnslash \sigma^0 p\cdot \bar\sigma}]}_\alpha}^\beta & =& \frac{\sgnslash
  \sigma^{0}_{\alpha \dot
    \gamma}  p \cdot \bar \sigma^{\dot
  \gamma \beta} +m\delta^{\beta}_\alpha}{\sqrt{2(E+m)}} \,.
\eea
For detailed expressions see Eqs. (2.106) and (2.107) of Ref.~\cite{BibleSpinors}. 
\com{\be
\sigma^\dagger_{\alpha \dot \beta} = \sigma_{\alpha \dot \beta}
\qquad{\rm or}\qquad
(\sigma^{\beta \dot \alpha})^\star = (\sigma^\star)^{\dot \beta \alpha}=\sigma^{\alpha \dot \beta}
\ee}
They satisfy the properties
\be\label{Magic6}
{[\sqrt{\sgnslash p\cdot \sigma}]^\alpha}_\beta =-{[\sqrt{\sgnslash p \cdot \bar
  \sigma}]_\beta}^\alpha \qquad {[\sqrt{\sgnslash p\cdot \bar\sigma}]_{\dot
    \alpha}}^{\dot \beta} =-{[\sqrt{\sgnslash p \cdot \sigma}]^{\dot\beta}}_{\dot \alpha}\,,
\ee
\com{ ${[\sqrt{\sgnzz p\cdot
  \sigma}]_\alpha}^\beta$ and raising the down and lowering the up index. The term $\sigma_{\alpha \dot
  \gamma} \bar \sigma^{0\,\dot \gamma \beta}$ becomes  
$-\sigma^{\alpha \dot  \gamma} \bar \sigma^{0}_{\,\gamma \beta}=-\bar \sigma^{\dot
  \gamma \alpha} \sigma^{0}_{\,\beta \gamma}=-\sigma^{0}_{\,\beta \gamma}\bar \sigma^{\dot
  \gamma \alpha} $.}
\be
\bar \sigma^{0\dot \alpha \alpha}{[\sqrt{\sgnslash p\cdot
    \sigma}]_\alpha}^\beta \sigma^0_{\beta \dot \beta} =
{[\sqrt{\sgnslash p \cdot \sigma}]^{\dot \alpha}}_{\dot \beta} \,,\qquad \bar \sigma^{0\dot \alpha \alpha}{[\sqrt{\sgnslash p\cdot
    \bar \sigma}]_\alpha}^\beta \sigma^0_{\beta \dot \beta} =
{[\sqrt{\sgnslash p \cdot \bar \sigma}]^{\dot \alpha}}_{\dot \beta} \,,
\ee
which can be used to change the position of indices. The definitions (\ref{sqrtpm}) are convenient since it can then be
shown that for a pure boost which transforms a momentum
from its rest-frame to a frame where it takes on-shell values $p^\mu$,
the representations are
\be\label{Boosts}
M = \sqrt{\frac{\sgnslash p\cdot \sigma}{m}}\,,\qquad (M^\dagger)^{-1} =
  \sqrt{\frac{\sgnslash p\cdot \bar \sigma}{m}}\,.
\ee

\subsection{Plane wave solutions of the Dirac equation}

In two-components form, the wave-functions $u_s$ and $v_s$ are of the form
\be\label{usvsTwoComponent}
u_{s,\,\mathfrak{a}}(\gr{p}) = \left( \begin{array}{c}
x_{s,\alpha}(\gr{p})\\
y^{\dagger \dot\alpha}_s(\gr{p})  \end{array} \right)\qquad v_{s,\,\mathfrak{a}}(\gr{p}) = \left( \begin{array}{c}
y_{s,\alpha}(\gr{p})\\
x^{\dagger \dot\alpha}_s(\gr{p})  \end{array} \right)\,.
\ee
To obtain explicit expressions, let us take a basis for the two-components spinors. This basis will be used for both the two-components spinors with down index and dotted up index. For down indices we note this basis $(\chi_s)_\alpha$ and for dotted up indices $(\eta^\dagger_s)^{\dot
  \alpha}$.  Using a direction $\hat{\gr{s}}$ having
angles $\theta,\phi$ in spherical coordinates, the components of this
basis are
\bea
\left(\chi_{\tfrac{1}{2}}\right)_\alpha &=&\left(\eta^\dagger_{\tfrac{1}{2}}\right)^{\dot \alpha}= \left(\begin{array}{c}  {\rm e}^{-\ii \phi/2} {\rm
                            cos}(\theta/2) \\ {\rm e}^{\ii\phi/2}{\rm
                            sin}(\theta/2)\end{array}\right) = {\rm
  e}^{-\tfrac{\ii}{2} \phi \sigma_z} \cdot {\rm
  e}^{-\tfrac{\ii}{2} \theta \sigma_y} \cdot \left( \begin{array}{c}
1\\
0 \end{array} \right)\\
\left(\chi_{-\tfrac{1}{2}}\right)_\alpha &=& \left(\eta^\dagger_{-\tfrac{1}{2}}\right)^{\dot \alpha}= \left(\begin{array}{c} - {\rm e}^{-\ii\phi/2}{\rm
                            sin}(\theta/2) \\  {\rm e}^{\ii\phi/2}{\rm cos}(\theta/2) \end{array}\right) = {\rm
  e}^{-\tfrac{\ii}{2} \phi \sigma_z} \cdot {\rm
  e}^{-\tfrac{\ii}{2} \theta \sigma_y} \cdot \left( \begin{array}{c}
0\\
1 \end{array} \right)\,.
\eea
The building blocks of $u_s$ and $v_s$ are then given by
\bea\label{RelativisticWeyl}
x_{s,\,\alpha}(\gr{p}) &=& {[\sqrt{p\cdot \sigma}]_\alpha}^\beta
({\chi_s})_\beta \qquad y^{\dagger \dot\alpha}_s(\gr{p}) = {[\sqrt{p\cdot
    \bar\sigma}]^{\dot\alpha}}_{\dot \beta} (\eta_s^\dagger)^{\dot\beta}\,,\\
x^\dagger_{s,\,\dot\alpha}(\gr{p}) &=& ({\chi_s^\dagger})_{\dot\beta}  {[\sqrt{p\cdot \sigma}]^{\dot
    \beta}}_{\dot \alpha}
\qquad y^{\alpha}_s(\gr{p}) =  ({\eta_s})^{\beta} {[\sqrt{p\cdot
    \bar\sigma}]_\beta}^\alpha\,.
\eea
Raising and lowering indices on these previous expressions and
using properties~(\ref{Magic6}) we find
\bea\label{RelativisticWeyl2}
x^\alpha_s(\gr{p}) &=&({\chi_s})^\beta  {[\sqrt{p\cdot \bar \sigma}]_\beta}^\alpha
\qquad y^\dagger_{s,\,\dot\alpha}(\gr{p}) =({\eta}_s^\dagger)_{\dot{\beta}} {[\sqrt{p\cdot
    \sigma}]^{\dot\beta}}_{\dot \alpha} \\
x^{\dagger \dot\alpha}_s(\gr{p}) &=&
{[\sqrt{p\cdot \bar \sigma}]^{\dot
    \alpha}}_{\dot \beta}  (\chi_s^\dagger)^{\dot\beta} 
\qquad y_{s,\,\alpha}(\gr{p}) = {[\sqrt{p\cdot
    \sigma}]_\alpha}^\beta  ({\eta_s})_{\beta}\,.
\eea
In order to compare with Eqs.~(3.1.19-3.1.22) of Ref.~\cite{BibleSpinors} we must use the relations
\bea\label{chieta}
(\chi_s)^\alpha &=& - 2 s (\eta_{-s})^{\alpha} \qquad\quad\,
(\chi_s^\dagger)^{\dot\alpha} = - 2 s (\eta_{-s}^\dagger)^{\dot\alpha}\\
{(\eta_s)}_\alpha &=&  2 s {(\chi_{-s})}_{\alpha} \qquad \quad \quad
{(\eta_s^\dagger)}_{\dot\alpha} =  2 s {(\chi_{-s}^\dagger)}_{\dot\alpha} \,.
\eea
Eqs. (\ref{RelativisticWeyl}) and (\ref{RelativisticWeyl2}) are all what is needed to get explicit expressions for $u_s$ and $v_s$ from Eqs. (\ref{usvsTwoComponent}), but also for $\bar u_s = u_s^\dagger \gamma^0$ and $\bar v_s = v_s^\dagger \gamma^0$ which are expressed in terms of two-components spinors as
\be\label{usbarvsbarTwoComponent}
\bar u_{s}^{\mathfrak{a}}(\gr{p}) = \left( \begin{array}{c}
y_{s}^{\alpha}(\gr{p})\\
x^{\dagger}_{s,\,\dot\alpha}(\gr{p})  \end{array} \right)\qquad v_{s}^{\mathfrak{a}}(\gr{p}) = \left( \begin{array}{c}
x_{s}^{\alpha}(\gr{p})\\
y^{\dagger}_{s,\,\dot\alpha}(\gr{p})  \end{array} \right)\,.
\ee
Note that the plane-wave solutions for a momentum $p^\mu$ are obtained
simply by considering the case where the momentum is at rest, for which $x_{s,\,\alpha} =
({\chi_s})_\alpha$ and $ y^{\dagger \dot\alpha}_s(\gr{p}) =
(\eta_s^\dagger)^{\dot\alpha}$ and then boosting using
Eqs.~(\ref{DiracRepresentationM}) and (\ref{Boosts}).

\subsection{Helicity basis plane wave solutions}\label{AppHelicity}

If the direction $\hat{\gr{s}}$  used to build the basis is the spatial momentum direction  $\hat{p} =
(\text{sin}(\theta)\text{cos}(\phi),\text{sin}(\theta)\text{sin}(\phi),\text{cos}(\theta))$,
then we are dealing with helicities and we have 
\be
\frac{1}{2}\hat{\gr{s}} \cdot \vec{\gr{\sigma}} \chi_s = s \chi_s
\qquad \frac{1}{2}\hat{\gr{s}} \cdot \vec{\gr{\sigma}} \eta_s = s \eta_s\,.
\ee
The relations (\ref{sqrtpm}) take in that case a very simple form, and
we obtain the relativistic two-components spinors
(\ref{RelativisticWeyl}) immediately
\bea
x_{s,\,\alpha}(p) &=& \sqrt{E - 2s |p|}
(\chi_s)_\alpha\qquad y^{\dagger\dot \alpha}_s(p) = \sqrt{E +
  2 s |p|} (\eta^\dagger_s)^{\dot \alpha}\\
  x_{s}^{\dagger \dot\alpha}(p) &=& \sqrt{E - 2s |p|}
(\chi_s^\dagger)^{\dot\alpha}\qquad y_{s,\,\alpha}(p) = \sqrt{E +
  2s |p|} (\eta_s)_{\alpha}\,\,.
\eea
Using the general form of the plane wave solutions (\ref{usvsTwoComponent}) and Eqs. (\ref{chieta}) we obtain the helicity solutions~(\ref{usvseasy}).
 
\subsection{Bilinears with two-components spinors}

Using the definitions (\ref{Explicitsphericalbasis}) and
(\ref{Explicitpolarbasis}) we obtain bilinear scalars built from two-components spinors 
\bea
X^\mu_{++} \equiv	x^{\dagger}_{+,\dot{\beta}}\bar{\sigma}^{\mu
  \dot{\beta}\alpha} x_{+,\alpha} = x^{\alpha}_{+}{\sigma}^{\mu}_{\alpha \dot \beta} x^{\dagger \dot \beta}_{+}&=& (E-|p|) (1,-\hat{p}) \\
X^\mu_{--} \equiv x^{\dagger}_{-,\dot{\beta}} \bar{\sigma}^{\mu \dot{\beta}\alpha} x_{-,\alpha}=x^{\alpha}_{-}{\sigma}^{\mu}_{\alpha \dot \beta} x^{\dagger \dot \beta}_{-}&=& (E+|p|) (1,\hat{p}) \\
X^\mu_{-+} \equiv  x^{\dagger}_{-,\dot{\beta}}\bar{\sigma}^{\mu
  \dot{\beta}\alpha} x_{+,\alpha}= x^{\alpha}_{+}{\sigma}^{\mu}_{\alpha \dot \beta} x^{\dagger \dot \beta}_{-} &=& -\sqrt{2}m(0,\epsilon_{-})\\
X^\mu_{+-} \equiv x^{\dagger}_{+,\dot{\beta}} \bar{\sigma}^{\mu \dot{\beta}\alpha} x_{-,\alpha}=x^{\alpha}_{-}{\sigma}^{\mu}_{\alpha \dot \beta} x^{\dagger \dot \beta}_{+}&=&- \sqrt{2}m(0,\epsilon_{+})  
\eea
\bea
Y^\mu_{++} \equiv y^{\alpha}_{+}{\sigma}^{\mu}_{\alpha \dot \beta}
  y^{\dagger \dot \beta}_{+}= y^{\dagger}_{+,\dot{\beta}}\bar{\sigma}^{\mu
  \dot{\beta}\alpha} y_{+,\alpha} &=& (E+|p|) (1,\hat{p}) \\
Y^\mu_{--} \equiv y^{\alpha}_{-}{\sigma}^{\mu}_{\alpha \dot \beta}
  y^{\dagger \dot \beta}_{-}= y^{\dagger}_{-,\dot{\beta}}\bar{\sigma}^{\mu
  \dot{\beta}\alpha} y_{-,\alpha} &=& (E-|p|) (1,-\hat{p}) \\
Y^\mu_{-+} \equiv y^{\alpha}_{-}{\sigma}^{\mu}_{\alpha \dot \beta}
  y^{\dagger \dot \beta}_{+}= y^{\dagger}_{+,\dot{\beta}}\bar{\sigma}^{\mu
  \dot{\beta}\alpha} y_{-,\alpha} &=& \sqrt{2}m (0,\epsilon_{-})\\
Y^\mu_{+-} \equiv y^{\alpha}_{+}{\sigma}^{\mu}_{\alpha \dot \beta}
  y^{\dagger \dot \beta}_{-}= y^{\dagger}_{-,\dot{\beta}}\bar{\sigma}^{\mu
  \dot{\beta}\alpha} y_{+,\alpha} &=& \sqrt{2}m(0,\epsilon_{+}) 
\eea
\bea
(YX)^{0i}_{++} &\equiv& y^{\alpha}_{+}[{\sigma}^{0i}]_{\alpha}^{\,\,\,\beta}
  x_{+,\beta}=  x^{\alpha}_{+}[{\sigma}^{0i}]_{\alpha}^{\,\,\,\beta}
  y_{+,\beta} = -\frac{\ii m}{2} \hat{p}^i  \\
(YX)^{ij}_{++} &\equiv& y^{\alpha}_{+}[{\sigma}^{ij}]_{\alpha}^{\,\,\,\beta}
  x_{+,\beta} = x^{\alpha}_{+}[{\sigma}^{ij}]_{\alpha}^{\,\,\,\beta}
  y_{+,\beta}={\epsilon^{ij}}_k \frac{m}{2} \hat{p}^k  \\
(YX)^{0i}_{--} &\equiv& y^{\alpha}_{-}[{\sigma}^{0i}]_{\alpha}^{\,\,\,\beta}
  x_{-,\beta}=  x^{\alpha}_{-}[{\sigma}^{0i}]_{\alpha}^{\,\,\,\beta}
  y_{-,\beta}= \frac{\ii m}{2} \hat{p}^i \\
(YX)^{ij}_{--} &\equiv& y^{\alpha}_{-}[{\sigma}^{ij}]_{\alpha}^{\,\,\,\beta}
  x_{-,\beta}=  x^{\alpha}_{-}[{\sigma}^{ij}]_{\alpha}^{\,\,\,\beta}
  y_{-,\beta} =-{\epsilon^{ij}}_k \frac{m}{2} \hat{p}^k  
\eea

\bea
(YX)^{0i}_{-+} &\equiv& y^{\alpha}_{-}[{\sigma}^{0i}]_{\alpha}^{\,\,\,\beta}
  x_{+,\beta}=x^{\alpha}_{+}[{\sigma}^{0i}]_{\alpha}^{\,\,\,\beta}
  y_{-,\beta}=- \frac{\ii}{\sqrt{2}}(E-|p|) \epsilon_-^i \\
(YX)^{ij}_{-+} &\equiv& y^{\alpha}_{-}[{\sigma}^{ij}]_{\alpha}^{\,\,\,\beta}
  x_{+,\beta}=  x^{\alpha}_{+}[{\sigma}^{ij}]_{\alpha}^{\,\,\,\beta}
  y_{-,\beta}= \frac{1}{\sqrt{2}}(E-|p|)
                 {\epsilon^{ij}}_k\epsilon_-^k  \\
(YX)^{0i}_{+-} &\equiv& y^{\alpha}_{+}[{\sigma}^{0i}]_{\alpha}^{\,\,\,\beta}
  x_{-,\beta} = x^{\alpha}_{-}[{\sigma}^{0i}]_{\alpha}^{\,\,\,\beta}
  y_{+,\beta}= -\frac{\ii}{\sqrt{2}}(E+|p|) \epsilon_+^i \\
(YX)^{ij}_{+-} &\equiv& y^{\alpha}_{+}[{\sigma}^{ij}]_{\alpha}^{\,\,\,\beta}
  x_{-,\beta} = x^{\alpha}_{-}[{\sigma}^{ij}]_{\alpha}^{\,\,\,\beta}
  y_{+,\beta}= \frac{1}{\sqrt{2}}(E+|p|) {\epsilon^{ij}}_k\epsilon_+^k  
\eea

\bea
(XY)^{0i}_{++} &\equiv&
  x^\dagger_{+,\dot\alpha}[{\bar\sigma}^{0i}]^{\dot
  \alpha}_{\,\,\,\dot\beta} y_+^{\dagger\dot\beta} = y^\dagger_{+,\dot\alpha}  [{\bar\sigma}^{0i}]^{\dot
  \alpha}_{\,\,\,\dot\beta} x^{\dagger \dot\beta}_{+}= \frac{\ii m}{2}
  \hat{p}^i\\
(XY)^{ij}_{++} &\equiv& x^\dagger_{+,\dot\alpha}[{\bar\sigma}^{ij}]^{\dot
  \alpha}_{\,\,\,\dot\beta} y_+^{\dagger\dot\beta}= y^\dagger_{+,\dot\alpha}[{\bar\sigma}^{ij}]^{\dot
  \alpha}_{\,\,\,\dot\beta} x_+^{\dagger\dot\beta} = {\epsilon^{ij}}_k \frac{m}{2} \hat{p}^k \\
(XY)^{0i}_{--} &\equiv&
  x^\dagger_{-,\dot\alpha}[{\bar\sigma}^{0i}]^{\dot
  \alpha}_{\,\,\,\dot\beta} y_-^{\dagger\dot\beta}= y^\dagger_{-,\dot\alpha}[{\bar\sigma}^{0i}]^{\dot
  \alpha}_{\,\,\,\dot\beta} x_-^{\dagger\dot\beta} = -\frac{\ii m}{2}
  \hat{p}^i  \\
(XY)^{ij}_{--} &\equiv& x^\dagger_{-,\dot\alpha}[{\bar\sigma}^{ij}]^{\dot
  \alpha}_{\,\,\,\dot\beta} y_-^{\dagger\dot\beta} = y^\dagger_{-,\dot\alpha}[{\bar\sigma}^{ij}]^{\dot
  \alpha}_{\,\,\,\dot\beta} x_-^{\dagger\dot\beta} = -{\epsilon^{ij}}_k
  \frac{m}{2} \hat{p}^k 
\eea
\bea
(XY)^{0i}_{-+} &\equiv&
  x^\dagger_{-,\dot\alpha}[{\bar\sigma}^{0i}]^{\dot
  \alpha}_{\,\,\,\dot\beta} y_+^{\dagger\dot\beta}= y^\dagger_{+,\dot\alpha}[{\bar\sigma}^{0i}]^{\dot
  \alpha}_{\,\,\,\dot\beta} x_-^{\dagger\dot\beta} = 
  \frac{\ii}{\sqrt{2}}(E+|p|) \epsilon_-^i\\
(XY)^{ij}_{-+} &\equiv& x^\dagger_{-,\dot\alpha}[{\bar\sigma}^{ij}]^{\dot
  \alpha}_{\,\,\,\dot\beta} y_+^{\dagger\dot\beta}= y^\dagger_{+,\dot\alpha}[{\bar\sigma}^{ij}]^{\dot
  \alpha}_{\,\,\,\dot\beta} x_-^{\dagger\dot\beta} = \frac{1}{\sqrt{2}}(E+|p|)
              {\epsilon^{ij}}_k   \epsilon_-^k\\
(XY)^{0i}_{+-} &\equiv&
  x^\dagger_{+,\dot\alpha}[{\bar\sigma}^{0i}]^{\dot
  \alpha}_{\,\,\,\dot\beta} y_-^{\dagger\dot\beta}= y^\dagger_{-,\dot\alpha}[{\bar\sigma}^{0i}]^{\dot
  \alpha}_{\,\,\,\dot\beta} x_+^{\dagger\dot\beta} =
  \frac{\ii}{\sqrt{2}}(E-|p|) \epsilon_+^i \\
(XY)^{ij}_{+-} &\equiv& x^\dagger_{+,\dot\alpha}[{\bar\sigma}^{ij}]^{\dot
  \alpha}_{\,\,\,\dot\beta} y_-^{\dagger\dot\beta}= y^\dagger_{-,\dot\alpha}[{\bar\sigma}^{ij}]^{\dot
  \alpha}_{\,\,\,\dot\beta} x_+^{\dagger\dot\beta} =
  \frac{1}{\sqrt{2}}(E-|p|) {\epsilon^{ij}}_k  \epsilon_+^k  \,.
\eea

\subsection{Bilinears with Dirac spinors}\label{SecuXu}

From the explicit expression of the plane wave solutions in terms of
two-components spinors (\ref{usvsTwoComponent}) and (\ref{usbarvsbarTwoComponent}), and the explicit expression of the $\gamma^\mu,\gamma^5,\Sigma^{\mu\nu}$ matrices of \S~\ref{AppClifford}, we find the needed scalar bilinear combinations
\bea\label{baruGu}
\bar{u}_r u_{s} &=& 2 m \delta^{\rm K}_{rs} \\
\bar{u}_r \gamma^5 u_{s} &=&0\\
\bar u_r \gamma^\mu u_{s} &=& X^\mu_{rs} +
Y^\mu_{rs} = 2\delta^{\rm K}_{rs}  p^{\mu}\\
\bar u_r \gamma^\mu \gamma^5 u_{s} &=&
-X^\mu_{rs} + Y^\mu_{rs}= \delta^{\rm K}_{rs}\left(4r m S^{\mu}\right)+\delta^{\rm K}_{r\,-s}\left(-\sqrt{8} m \epsilon^{\mu}_{s-r}\right)\\
\com{\bar u_r \Sigma^{0i} u_{s} &=&
(YX)^{0i}_{rs} + (XY)^{0i}_{rs}\\
\bar u_r \Sigma^{ij} u_{s} &=&
(YX)^{ij}_{rs} + (XY)^{ij}_{rs}\\}
\bar u_r \widetilde{\Sigma}^{0i} u_{s} &=&
-\ii (YX)^{0i}_{rs} +\ii (XY)^{0i}_{rs}= \delta^{\rm K}_{rs}\left(-2 r m \hat p^i \right)+\delta^{\rm K}_{r\,-s}\left(-\sqrt{2} E  \epsilon^i_{r-s}\right)\\
\bar u_r \widetilde{\Sigma}^{ij} u_{s} &=& -\ii (YX)^{ij}_{rs} +\ii (XY)^{ij}_{rs}=\delta^{\rm K}_{r\,-s}\left(\sqrt{2}\ii (s-r) |\vec{p}| {\epsilon^{i j}}_k \epsilon_{r-s}^k \right)=\delta^{\rm K}_{r\,-s}\left(-2 \sqrt{2}  p^{[i} \epsilon_{r-s}^{j]}\right)\,,
\eea
where $\delta^{\rm K}_{r\,-s}$ vanishes if $r=s$ and is unity otherwise.
When forming the similar bilinears but with negative frequency plane-wave solutions $v_s(p)$ we find
\bea\label{vXv}
\bar{v}_{r} v_{s} &=& -2 m \delta^{\rm K}_{rs}\\
\bar{v}_r \gamma^5 v_{s} &=&0\\
\bar v_r \gamma^\mu v_{s} &=& Y^\mu_{sr} +
X^\mu_{sr}= 2\delta^{\rm K}_{rs}  p^{\mu}\\
\bar v_r \gamma^\mu \gamma^5 v_{s} &=&
-Y^\mu_{sr} + X^\mu_{sr}=\delta^{\rm K}_{rs}\left(- 4 r m S^{\mu}\right)+\delta^{\rm K}_{r\,-s}\left(\sqrt{8} m \epsilon^{\mu}_{s-r}\right)\\
\com{\bar v_r \Sigma^{0i} v_{s} &=&
(XY)^{0i}_{sr} + (YX)^{0i}_{sr}\\
\bar v_r \Sigma^{ij} v_{s} &=&
(XY)^{ij}_{sr} + (YX)^{ij}_{sr}\\}
\bar v_r \widetilde{\Sigma}^{0i} v_{s} &=&
-\ii (XY)^{0i}_{sr} +\ii (YX)^{0i}_{sr} =\delta^{\rm K}_{rs}\left(2 r m \hat p^i \right)+\delta^{\rm K}_{r\,-s}\left(-\sqrt{2} E  \epsilon^i_{r-s}\right)\\
\bar v_r \widetilde{\Sigma}^{ij} v_{s} &=& -\ii
(XY)^{ij}_{sr} + \ii (YX)^{ij}_{sr} = \delta^{\rm K}_{r\,-s}\left(-2 \sqrt{2}  p^{[i} \epsilon_{r-s}^{j]}\right)\,.
\eea
\com{Explicitly we get
\bea\label{baruGu}
\bar{u}_r u_{s} &=& 2m \delta^{\rm K}_{rs} \\
\bar{u}_r \gamma^5 u_{s} &=& 0 \\
\bar{u}_r \gamma^{\mu} u_{s} &=& 2\delta^{\rm K}_{rs}  p^{\mu} \\
\bar{u}_r \gamma^{\mu} \gamma^5 u_{r} &=& 4r m S^{\mu} \\
\bar{u}_r \gamma^{\mu} \gamma^5 u_{s} &=& -\sqrt{8} m \epsilon^{\mu}_{s-r} \\
\com{\bar{u}_r  \Sigma^{0 i} u_{r} &=& 0 \\
\bar{u}_r \Sigma^{i j} u_{r} &=& 2r m {\epsilon^{i j}}_k \hat{p}^k\\
\bar{u}_r \Sigma^{0 i} u_{s} &=& i \sqrt{2}(s-r) |\vec{p}|  \epsilon^i_{r-s} \\
\bar{u}_r \Sigma^{i j} u_{s} &=&  \sqrt{2} E {\epsilon^{ij}}_k \epsilon_{r-s}^k\quad  {\rm if}\quad  r \neq s \\}
\bar{u}_r  \widetilde{\Sigma}^{0 i} u_{r} &=& -2 r m \hat p^i \\
\bar{u}_r \widetilde{\Sigma}^{0 i} u_{s} &=& -\sqrt{2} E  \epsilon^i_{r-s}\quad   {\rm if}\quad r \neq s\\
\bar{u}_r \widetilde{\Sigma}^{i j} u_{r} &=& 0 \\
\bar{u}_r \widetilde{\Sigma}^{i j} u_{s} &=&  \sqrt{2}\ii (s-r) |\vec{p}| {\epsilon^{i j}}_k    \epsilon_{r-s}^k=-2 \sqrt{2}  p^{[i} \epsilon_{r-s}^{j]} \quad {\rm if}\quad r \neq s\,.
\eea}
We realize that the bilinears built from $\bar v_r$ and $v_s$ are obtained from those built with $\bar u_r$ and $u_s$ by the simple replacement $m \to -m$.

\subsection{Covariant components of the distribution function using two-components spinors}\label{SecInterpretationImu}

The covariant components of the distribution function $f_{rs}$ can also be extracted with the two-components formalism, and it happens to lead to straightforward expressions. Let us define first the distribution functions which are $2\times2$ valued in complex spinor space by
\be
f^x_{\dot \alpha \beta}(p) \equiv \sum \limits_{rs} x^\dagger_{r,\,\dot \alpha}(p)
f_{rs}(p) x_{s,\,\beta}(p) \qquad \qquad   f_y^{\alpha \dot \beta}(p) \equiv \sum \limits_{rs} y^{\alpha}_r(p)
f_{rs}(p) y^{\dagger\dot \beta}_s(p)\,.
\ee
Due to the relations (\ref{ConnectVectorandbispinor}), we can define 4-vector distribution functions simply as
\be
f_x^{\mu}(p) = \frac{1}{2} \bar{\sigma}^{\mu \dot{\alpha}\beta}f^x_{\dot
  \alpha {\beta}}(p) \,,\qquad\qquad f_y^{\mu}(p) = \frac{1}{2} \sigma^\mu_{\alpha \dot\beta}f_y^{\alpha \dot{\beta}}(p)\,.
\ee
Using the definitions (\ref{DefIVQ}) and (\ref{DefcalQ}) we get
\be
f_x^\mu =\frac{1}{2}(I p^\mu - m {\cal Q}^\mu)  =  \begin{cases}{\cal I}^{-\,\mu}\quad{\rm particles}\\{\cal I}^{+\,\mu}\quad{\rm antiparticles,}\end{cases}\,\qquad \qquad f_y^\mu =\frac{1}{2}( I
p^\mu + m {\cal Q}^\mu) =  \begin{cases}{\cal I}^{+\,\mu}\quad{\rm particles}\\{\cal I}^{-\,\mu}\quad{\rm antiparticles.}\end{cases}
\ee
The interpretation is thus that ${\cal I}^{\pm}_\mu$ are vectors of chiral parts, with ${\cal I}^{-}_\mu$ (resp. ${\cal I}^{+}_\mu$) being a the left chiral part (resp. right chiral part) vector for particles or the right chiral part (resp. left chiral part) vector for antiparticles.

\end{document}